\documentclass[fleqn,usenatbib]{mnras}

\usepackage{newtxtext,newtxmath}
\usepackage[T1]{fontenc}

\DeclareRobustCommand{\VAN}[3]{#2}
\let\VANthebibliography\thebibliography
\def\thebibliography{\DeclareRobustCommand{\VAN}[3]{##3}\VANthebibliography}

\usepackage{graphicx}	% Including figure files
\usepackage{amsmath}	% Advanced maths commands
\usepackage{tablefootnote}  %Footnotes in tables

\usepackage{soul} %For strikethrough \st{}

\usepackage{threeparttable} %Table notes package

\title[CN/CO in U/LIRGs]{Stored in the archives: Uncovering the CN/CO intensity ratio with ALMA in nearby U/LIRGs}

\author[B. Ledger et al.]{
B. Ledger,$^{1}$\thanks{E-mail: ledgeb1@mcmaster.}
T. Saito,$^{2}$
D. Iono,$^{2,3}$
C. D. Wilson$^{1}$
\\
$^{1}$Department of Physics and Astronomy, McMaster University, 1280 Main St. W., Hamilton, Ontario L8S 4M1, Canada\\
$^{2}$National Astronomical Observatory of Japan, National Institutes of Natural Sciences, 2-21-1 Osawa, Mitaka, Tokyo, 181-8588\\
$^{3}$Department of Astronomical Science, The Graduate University for Advanced Studies, SOKENDAI, 2-21-1 Osawa, Mitaka, Tokyo
181-8588\\
}

\date{Accepted XXX. Received YYY; in original form ZZZ}

\pubyear{2023}

\begin{document}
\label{firstpage}
\pagerange{\pageref{firstpage}--\pageref{lastpage}}
\maketitle

\begin{abstract}
We present an archival Atacama Large Millimeter/submillimeter Array (ALMA) study of the CN $N = 1-0$ / CO $J = 1-0$ intensity ratio in nearby ($z < 0.05$) Ultra Luminous and Luminous Infrared Galaxies (U/LIRGs). We identify sixteen U/LIRGs that have been observed in both CN and CO lines at $\sim500$ pc resolution based on sixteen different ALMA projects. We measure the (CN bright)/CO and (CN bright)/(CN faint) intensity ratios at an ensemble of molecular clouds scales (CN bright = CN $N = 1-0$, $J = 3/2-1/2$; CN faint = CN $N = 1-0$, $J = 1/2-1/2$ hyperfine groupings). Our global measured (CN bright)/CO ratios range from 0.02-0.15 in LIRGs and 0.08-0.17 in ULIRGs. We attribute the larger spread in LIRGs to the variety of galaxy environments included in our sample. Overall, we find that the (CN bright)/CO ratio is higher in nuclear regions, where the physical and excitation conditions favour increased CN emission relative to the disk regions. 10 out of 11 galaxies which contain well-documented active galactic nuclei show higher ratios in the nucleus compared to the disk. Finally, we measure the median resolved (CN bright)/(CN faint) ratio and use it to estimate the total integrated CN line optical depth in ULIRGs ($\tau \sim 0.96$) and LIRGs ($\tau \sim 0.23$). The optical depth difference is likely due to the higher molecular gas surface densities found in the more compact ULIRG systems.%, and suggests that CN optical depth is not responsible for the variations in the CN bright / CO intensity ratio seen in LIRGs.

\end{abstract}


%%%%%%%%%%%%%%%%%%%%%%%%%%%%%%%%%%%%%%%%%%%%%%%%%%

%%%%%%%%%%%%%%%%% BODY OF PAPER %%%%%%%%%%%%%%%%%%

\section{Introduction}
\label{sec:introduction}

%Paragraph describing molecular gas and its importance in galaxy evolution, star formation, etc.
The processes of galaxy formation and evolution are guided by the rates at which stars form in galaxies, which makes understanding the connection between the interstellar medium (ISM) and star formation a major area of research in modern astronomy (see the review by \citealt{Kennicutt2012} and references therein). The molecular gas content of galaxies plays an important role in regulating star formation, as it is the fuel from which future stars are formed. Observers typically use carbon monoxide, CO, and a corresponding conversion factor, $\alpha$\textsubscript{CO}, to estimate the cold molecular gas content of galaxies (see the review by \citealt{Bolatto2013} and references therein). The lowest energy rotational line, $^{12}$CO ($J = 1-0$) $-$ hereafter referred to as simply ``CO'' unless otherwise specified $-$ is often found to be optically thick and can thus be excited and observed even in the coldest parts of molecular clouds with a low excitation energy ($\sim5$ K) and gas densities of $n<10^{2}$ cm$^{-3}$\citep{Shirley2015}. CO can also be excited in the diffuse molecular gas when there is sufficient column density for self-shielding to occur \citep{Bolatto2013}. Thus, CO is a good tracer for the bulk cold gas content in the ISM. However, stars form in the densest regions of molecular clouds \citep{Lada1991} and to target this dense gas ($n$\textsubscript{H\textsubscript{2}} $\gtrapprox10^{4}$ cm$^{-3}$), molecules with higher critical densities need to be used. Molecules like hydrogen cyanide (HCN), HCO$^{+}$, and CS are the common tracers used by observers \citep{Wu2010, Garcia2012, Kennicutt2012}, with the HCN (1-0) line representing a well-established empirical tracer of dense gas in galaxies. Dense molecular gas has been found to correlate with the star formation rate (SFR) in galaxies on global scales with a nearly linear power law slope \citep{Gao2004, Usero2015AJ, Gallagher2018}, and the correlation extends to individual molecular clouds on Galactic scales \citep{Wu2005, Shimajiri2017}.

The HCN line luminosity converts to a dense gas mass through a conversion factor ($\alpha$\textsubscript{HCN}; \citealt{Gao2004}). Variations in this conversion factor have been observed or predicted in the Milky Way and nearby galaxies \citep{Gracia2008b, Vega2008, Garcia2012, Usero2015AJ, Shimajiri2017, Onus2018, Jones2023}. Additionally, dense gas will be exposed to strong stellar feedback and ultra-violet (UV) radiation fields from newly formed stars. \cite{Shimajiri2017} demonstrate that the HCN conversion factor depends on the strength of this UV radiation field. HCN emission has also been found to be optically thick when observed in the centres of galaxies (\citealt{Jimenez2017} and references therein), with optical depths in the range of $\tau=2-11$. Since the link between HCN and dense gas is complicated due to its high optical depth and the variations in $\alpha$\textsubscript{HCN}, it is useful to explore other bright dense gas tracers; in this work, we use an astrochemically related molecule to HCN, the cyanide radical (CN), as our dense gas tracer ($n$\textsubscript{crit} $\gtrapprox10^{5}$ cm$^{-3}$; \citealt{Shirley2015}).

CN can be formed through photodissociation of HCN, neutral-neutral reactions with various intermediary species involving carbon and nitrogen, and reactions with ionized carbon \citep{Aalto2002, Boger2005, Chapillon2012}. Models of photon-dominated regions (PDRs) predict the abundance of CN to increase on the surfaces of molecular clouds, where the UV radiation field is strongest, with the abundance of HCN higher in the more shielded centres of molecular clouds \citep{Boger2005}. Somewhat surprisingly, however, \cite{Wilson2023} found that the CN/HCN intensity ratio is relatively constant when compared to star formation rate surface density ($\Sigma$\textsubscript{SFR}) and gas surface density ($\Sigma$\textsubscript{gas}) in galaxies over a range of sub-kiloparsec spatial scales. Furthermore, the CN/CO and HCN/CO intensity ratios show a tight correlation, implying that the CN/CO intensity ratio can also be used to trace the dense gas fraction in galaxies \citep{Wilson2023}. Additional detailed studies of CN and HCN intensities and abundances are required to test the predictions of PDR models in galaxies.

 There have been a large number of projects which have used the Atacama Large Millimeter/Submillimeter Array (ALMA) to observe CN ($N = 1-0$), hereafter CN (1-0), in spectral line surveys of starbursts and galaxies with active galactic nuclei (AGN; e.g., \citealt{Sakamoto2014, Harada2018, Martin2021, Saito2022b}). CN (1-0) has been observed in local mergers and merger remnants (e.g., \citealt{Konig2016, Sakamoto2017, Ueda2017, Ueda2021, Ledger2021}), the circumnuclear ring of the spiral galaxy M83 \citep{Harada2019}, and in starburst and AGN outflows (e.g., \citealt{Sakamoto2014, Meier2015, Walter2017, Lutz2020, Saito2022b}). Gas-rich early type galaxies have been detected in CN (1-0) \citep{Young2021, Young2022}, and the line has been seen in absorption towards radio nuclei, AGN, and bright cluster galaxies \citep{Rose2019, Kameno2020, Baek2022}. Observations of CN (1-0) with ALMA in multiple bright starbursts and AGN have been compared with HCN (1-0) and CO (1-0) (\citealt{Wilson2018, Ledger2021, Wilson2023}).

Additional observations of CN (1-0) include spectral line surveys with the Nobeyama 45 m telescope (e.g., \citealt{Nakajima2018, Takano2019}) and the Large Millimeter Telescope \citep{Cruz2020}. CN (1-0) was detected in molecular gas outflows in Mrk 231 \citep{Cicone2020} with the IRAM PdBI telescope, and in nearby galaxies with the IRAM 30 m telescope \citep{Henkel1998, Watanabe2014, Aladro2015}. Higher transition lines of CN have been studied with ALMA in nearby galaxies (e.g., \citealt{Saito2015, Nakajima2015, Rose2020, Ledger2021}) and galaxies at redshifts 2.5-3.5 \citep{Geach2018, Canameras2021}. For more discussion of previous observations of CN with single-dish and/or non-ALMA data (e.g., \citealt{Aalto2002, Costagliola2011}), we refer the reader to \cite{Wilson2018}.

\cite{Wilson2018} explored the CN/CO ratio in nearby galaxies using archival Cycle 0 ALMA data and found significant CN/CO spatial variations within some galaxies, as well as global differences between individual galaxies. The galaxies ranged from nearby starbursts to Ultra-Luminous and Luminous Infrared Galaxies, or U/LIRGs, and the line ratios were measured on kiloparsec scales. \cite{Wilson2023} measured the CN/CO line ratio in a sample of nearby galaxies, covering starbursts, Seyfert 2 galaxies with AGN, and U/LIRGs. They found the CN/CO ratio also demonstrated variations between the galaxies in their sample and spatially within their galaxies. Both \cite{Wilson2018} and \cite{Wilson2023} found that some ratio variations correlated spatially with regions of higher star formation rates. Variations in CN/CO ratios were also found in a gas-rich elliptical \citep{Young2022} and bright cluster galaxies \citep{Rose2019}. On more resolved scales in galaxies, \citet{Meier2015} found an enhanced CN/CO ratio in the inner nuclear region of NGC 253 compared to the outer disk. \citet{Cicone2020} measured total CN/CO line luminosity ratios in Mrk 231 and found they are enhanced by a factor of $\sim3$ in the line wings of the outflow. They attribute this enhancement in the CN/CO ratio in the outflow to stronger UV radiation fields present in the gas, perhaps from the formation of massive stars in the outflow.

 A major challenge of quantifying any trend in the CN/CO ratio in X-ray dominated regions (XDRs) is the limited spatial scales of most extragalactic observations in galaxies with an AGN. XDRs most strongly affect the gas properties in the innermost $<100$ pc of the galaxy nuclei \citep{Wolfire2022}. \cite{Wilson2018} found that the CN/CO ratio decreases in the vicinity of 3 of 4 AGN in her galaxy sample. In contrast, \cite{Wilson2023} found an increase in the CN/CO ratio in the centres of NGC 7469 and NGC 1808, two galaxies with strong Seyfert nuclei. NGC 1068, a nearby Seyfert 2 galaxy, has also been shown to have an increasing CN/CO ratio near the AGN and the jet-driven molecular outflow \citep{Saito2022b}.  It is clear that more work needs to be done to fully understand any trends in the CN/CO ratio in XDRs in galaxies.

U/LIRGs are ideal laboratories for studying the impact of radiation fields on the properties of molecular gas in galaxies (see e.g., \citealt{Lonsdale2006} and \citealt{Perez2021}). These galaxies have high infrared luminosity ($L$\textsubscript{IR}) and are usually mergers of galaxies, starbursts, and contain AGN \citep{Sanders2003, Armus2009, Pearson2016}. U/LIRGs have large fractions of dense molecular gas \citep{Solomon1992, Gao2004, Privon2017, Sliwa2017}, and the correlation between HCN line luminosity and SFR extends into this high-IR regime \citep{Gao2004, Wu2005}. \cite{Gracia2008b} found that the star formation efficiency of dense gas, SFE\textsubscript{dense}, is higher in U/LIRGs than normal galaxies, indicating that these galaxies are star forming engines that efficiently convert their large gas reservoirs into stars at extremely high rates. U/LIRGs can contain PDRs, XDRs, or a combination of both, and therefore are prime targets to study the CN/CO ratio in external galaxies.

The goal of this paper is to present an observational picture of the CN/CO intensity ratio in a comprehensive sample of U/LIRGs. We focus on galaxies which are nearby ($z < 0.05$), have L\textsubscript{IR} $> 10^{11}$ L\textsubscript{$\odot$}, have starbursts, AGN, or some mixture of the two, and have been observed with ALMA. Section \ref{sec:sample_selection} describes our sample selection and data processing procedures, including calibration, imaging, and analysis of 16 different archival ALMA projects. In Section \ref{sec:results}, we discuss our measured line intensity ratios and compare them between a subset of ULIRGs and LIRGs. We summarize our conclusions in Section \ref{sec:conclusions}. We defer any comparisons of this ratio with $\Sigma$\textsubscript{SFR} or PDR and XDR models to a future paper.

\section{Sample Selection and Data Analysis}
\label{sec:sample_selection}

\begin{figure*}
	\includegraphics[width=\textwidth]{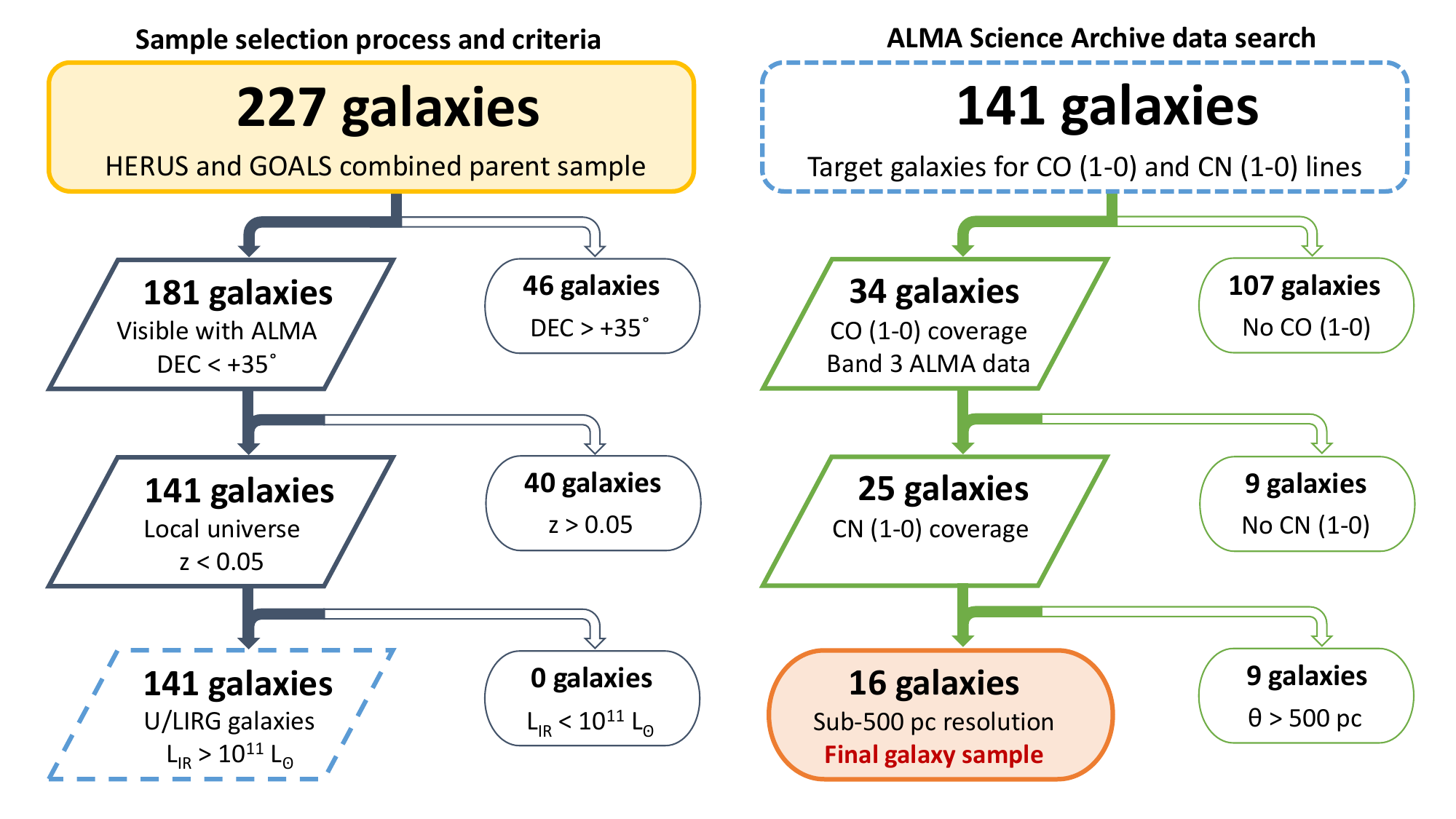}
    \caption{A graphical representation of the selection process to obtain our CN/CO ratio galaxy sample. Starting with the HERUS and GOALS parent surveys, we narrowed our sample based on the criteria of DEC $> +35^{\circ}$, $z < 0.05$, and L\textsubscript{IR} $> 10^{11}$ L\textsubscript{$\odot$} (left column in figure). We then searched the ALMA Science Archive on June 20, 2022, for potential galaxy targets that had both CO (1-0) and CN (1-0) lines with a physical resolution of $\theta < 500$ pc (right column in figure). From the 227 initial galaxy candidates, we obtained a final sample of 16 galaxies previously observed with ALMA.}
    \label{fig:selection}
\end{figure*}

In this section, we describe the methods by which we selected our sample of U/LIRGs and measured our intensity ratios. We describe the data calibration and image processing methods and discuss how we obtained our different data products. Finally, we summarize the various ways we measure our intensity ratios, which lead to the main results of this paper. 

The lowest rotational line of CN, CN ($N = 1-0$), has two main hyperfine groupings with rest frequencies of $\sim113.49$ GHz and $\sim113.17$ GHz and quantum number designations of CN ($N = 1-0, J = 3/2-1/2$) and CN ($N = 1-0, J = 1/2-1/2$), respectively. The hyperfine groupings of CN ($N = 1-0$) are made up of 9 individual hyperfine spectral lines \citep{Skatrud1983}, but at the turbulent line widths and velocity and spatial resolutions of extragalactic observations, these are typically blended into two larger groupings of five (brighter) and four (fainter) spectral lines\footnote{https://home.strw.leidenuniv.nl/~moldata/datafiles/cn-hfs.dat}. Throughout this paper, we will denote the CN ($N = 1-0, J = 3/2-1/2$) and CN ($N = 1-0, J = 1/2-1/2$) lines as CN bright and CN faint, respectively.

\subsection{The HERUS and GOALS U/LIRG surveys}
\label{subsec:HERUS_and_GOALS}
For this archival project, we wanted to find as many nearby U/LIRGs as possible that were observed in both the CO (1-0) and CN (1-0) lines at sub-kiloparsec scales in public ALMA data. As our parent sample, we compiled a master list of galaxies combined from the HERschel Ultra Luminous InfraRed Galaxy Survey (HERUS, \citealt{Pearson2016}) and The Great Observatories All-Sky LIRG Survey (GOALS, \citealt{Armus2009}). The HERUS survey is a flux-limited sample of low-redshift ($z \leq 0.2$) ULIRGs identified from the \textit{IRAS} PSC-z survey \citep{Saunders2000} with 60 $\mu$m fluxes greater than 1.8 Jy \citep{Pearson2016}. The \textit{Herschel} observations recovered nearly complete CO ladders of 43 ULIRGs, with the science goal of tracing different temperature components of the gas within PDR and XDR regions \citep{Pearson2016}. Of the 43 ULIRGs in our parent sample, 25 are unique to HERUS, while 18 are also covered in GOALS. The GOALS survey is a large collection of data from the \textit{Spitzer Space Telescope}, the \textit{Chandra X-Ray Observatory}, the \textit{Hubble Space Telescope}, and the \textit{Galaxy Evolution Explorer} \citep{Armus2009}. A main science objective of GOALS is to describe the mechanisms responsible for the enhanced IR power generation in an unbiased representation (e.g., starbursts, AGN, mergers) of LIRGs \citep{Armus2009}. GOALS compiled observations of 202 low-redshift ($z < 0.088$) LIRGs, of which 18 are also covered in the HERUS survey. The U/LIRGs observed in HERUS and GOALS are, by definition, galaxies with L\textsubscript{IR} $> 10^{11}$ L\textsubscript{$\odot$}. We therefore start with a sample of galaxies with L\textsubscript{IR} $> 10^{11}$ L\textsubscript{$\odot$} (Figure \ref{fig:selection}).

\subsection{Sample selection criteria}
\label{subsec:selection_criteria}
From the HERUS and GOALS surveys, we obtained a combined parent sample of 227 potential U/LIRG targets for our CN/CO ratio study. In this section, we describe the criteria by which we narrowed down our target sample to 141 galaxies. Figure \ref{fig:selection} demonstrates the selection process visually.

We required our targets be visible with ALMA, which has an upper Declination (DEC) limit of +47$^{\circ}$ (see the ALMA Proposer's Guide \footnote{G. Privon et al. 2022, ALMA Cycle 9 Proposer’s Guide, ALMA Doc. 9.2 v1.4; https://almascience.eso.org/documents-and-tools/cycle9/alma-proposers-guide}). To ensure that we targeted galaxies with optimal observing conditions and to limit the possibility of shadowing effects, we imposed a stricter requirement of DEC $< +35^{\circ}$. We wanted to study the CN/CO ratio in nearby galaxies with sub-kiloparsec resolution (corresponding to a physical resolution representative of an ensemble of molecular clouds). To accomplish this resolution goal with ALMA, we imposed an upper redshift limit of $z = 0.05$, which corresponds to D $\sim200$ Mpc\footnote{The redshift cut removed 40 galaxies from our sample. 3 galaxies near the threshold redshift value with $z \gtrapprox 0.05$ (D $\gtrapprox 200$ Mpc) were found to have archival CO and CN data with sub-kiloparsec resolution (IRAS F14378-3651, IRAS 19542+1110, and IRAS F01364-1042); however, these galaxies were removed from the sample list due to the CO and CN coverage criteria described in Section \ref{subsec:ALMA_data}.}. At this distance, a 1 arcsec ALMA beam corresponds to $\sim1$ kiloparsec physical resolution.

%We note, however, that the CN/CO ratio has been observed with ALMA in nearby starbursts, Seyfert galaxies, early-type galaxies, and molecular outflows with L\textsubscript{IR} $< 10^{11}$ L\textsubscript{$\odot$} (\citealt{Meier2015, Wilson2018, Rose2019, Cicone2020, Young2021, Wilson2023}). The CN/CO ratios in these galaxies will be compared to our results in Section \ref{subsubsec:cn_co_prev_obs}.

\subsection{The ALMA CO and CN galaxy sample}
\label{subsec:ALMA_data}
A list of the 141 galaxy targets was run through the ALMA Science Archive query tool on June 20, 2022. 46 of the 141 galaxies had public Band 3 data available, which were taken over ALMA Cycles 0-7. Of these 46 galaxies, 12 did not have spectral window coverage of the redshifted CO (1-0) line. This further reduced our potential candidate galaxies to 34 (Figure \ref{fig:selection}). Furthermore, 9 of the 34 galaxies did not have spectral window coverage of the redshifted brightest CN (1-0) line, leaving 25 galaxies which had both CO and CN covered in Band 3 with ALMA (Figure \ref{fig:selection}). We imposed an angular resolution limit to correspond to a physical scale of $500$ pc. This requirement further reduced our potential candidates by 9 galaxies (Figure \ref{fig:selection}). Of the 9 galaxies that were removed, 5 had CO and CN observed with $500$ pc $<\theta<1$ kiloparsec resolution, and 4 had CO and CN observed with $\theta>1$ kiloparsec resolution.

After our extended sample selection process, we found 16 U/LIRGs with both CO (1-0) and CN (1-0) observed at $\theta<500$ pc resolution in the ALMA archive. 8 of the 16 galaxies are covered in the HERUS survey, while all 16 galaxies are covered in the GOALS surveys. Table \ref{tab:projects} lists the project codes which correspond to the archival ALMA data.

Our galaxy sample is comprised of twelve LIRGs but only four ULIRGs, which is likely not a statistically significant and representative sample of ULIRGs in the local Universe. We have compared our small sample of ULIRGs with those in the HERUS survey \citep{Pearson2016} in terms of redshift and $L$\textsubscript{IR}. We are not representative of the HERUS redshift distribution as we have selected only those ULIRGs within $z<0.05$. However, our population of ULIRGs is representative of the lower half of infrared luminosities covered by the HERUS sample of ULIRGs (e.g., log($L$\textsubscript{IR}) $< 12.3$).

In our results discussion, we compare our population of LIRGs with our population of ULIRGs as independent samples, acknowledging that future work comparing galaxies with their distribution of infrared luminosities is forthcoming and needed for our small statistical sample of galaxies (Ledger et al. \textit{in prep.}).

\subsubsection{Excluded ALMA data sets}
\label{subsubsec:excluded_ALMA}

We only consider data from ALMA Cycles 1-6. There were no corresponding projects which met our criteria with ALMA Cycle 7 or later dates at the time of searching the archive. We removed the only Cycle 0 project which met our selection criteria, project code 2011.0.00525.S, with observations of NGC 3256. The corresponding Cycle 6 observations for NGC 3256 are of higher spatial resolution and sensitivity and can be better calibrated than the Cycle 0 data. \cite{Wilson2018} measured the (CN bright)/CO ratio for NGC 3256 using the archival Cycle 0 data and we compare with her results in Section \ref{subsubsec:cn_co_prev_obs}.

We use only project code 2015.1.00167.S for Arp 220, although two other projects exist in the ALMA archive (2015.1.00113.S and 2017.1.00042.S). We ignore these two projects as they are of significantly higher angular resolution (physical scales of $\sim 10$ pc), and therefore we would need significant degradation of the data to match our target scale of 500 pc.

\tabcolsep=0.02cm
\begin{table}
\centering
    \caption{Archival ALMA Cycle 1-6 projects with both CO (1-0) and CN (1-0) observations in nearby U/LIRGs.} %\color{red}{For now, of our original 21 projects only 2 needed to be removed from this table since we did not include them in data reduction. These are Arp 220 (2015.1.00113.S) and NGC 3256 (2011.0.00525.S). 2019.1.01664.S still has no publication as of Feb 20 2023.}}
    \begin{tabular}{ccc}
    \hline
        Galaxy\textsuperscript{\textit{a}} & CO and CN project code & ALMA data reference \\
        \hline
        IRAS 13120-5453 & 2015.1.00287.S & \cite{Sliwa2017} \\
        Arp 220 & 2015.1.00167.S & \cite{Brown2019} \\
        %& \color{red}{2015.1.00113.S} & \cite{Scoville2017} \\
        IRAS F05189-2524 & 2012.1.00306.S & \cite{Lutz2020} \\
        IRAS F10565+2448 & 2019.1.01664.S & PI: J. Wang \\
        NGC 6240 & 2015.1.00003.S & \cite{Saito2018b} \\
        IRAS F18293-3413 & 2015.1.01191.S & \cite{Saito2020} \\
        NGC 3256 & 2018.1.00223.S & \cite{Ueda2021} \\
        %& \color{red}{2011.0.00525.S} & \cite{Sakamoto2014} \\
        NGC 1614 & 2013.1.00991.S & \cite{Konig2016} \\
         & 2013.1.01172.S & \cite{Saito2016} \\
        NGC 7469 & 2013.1.00218.S & \cite{Wilson2019} \\
         & 2017.1.00078.S & \cite{Izumi2020} \\
        NGC 2623 & 2015.1.01191.S & \cite{Brown2019} \\
        NGC 3110 & 2013.1.01172.S & \cite{Kawana2022} \\
        ESO 320-G030 & 2016.1.00263.S & \cite{Pereira2020} \\
        NGC 1068 & 2012.1.00657.S & PI: S. Takano \\
                 & 2018.1.01684.S & \cite{Saito2022b} \\
        NGC 5104 & 2015.1.01191.S & PI: Z. Zhang \\
        NGC 4418 & 2016.1.00177.S & \cite{Lutz2020} \\
        NGC 1365 & 2015.1.01135.S & \cite{Zabel2019} \\
    \hline
    \end{tabular}
    \label{tab:projects}
    \begin{tablenotes}
        \item \textit{Notes:} \textsuperscript{\textit{a}}The galaxies are listed in order of decreasing infrared luminosity.
    \end{tablenotes}
\end{table}

\subsection{ALMA data calibration and imaging}
\label{subsec:data_reduction}
In this section, we present an overview of the calibration, reduction, and imaging processes for our ALMA data. Data reduction and calibration were performed using Common Astronomy Software Applications (\texttt{CASA}; \citealt{McMullin2007}) and the PHANGS-ALMA pipeline v3 \citep{Leroy2021}. Many of the relevant imaging parameters are provided in Table \ref{tab:imaging_params}.

\subsubsection{Calibration}
\label{subsubsec:calibration}
We were generously provided with calibrated \textit{uv} measurement sets for NGC 7469 (only project code 2013.1.00218.S; \citealt{Wilson2023}) and Arp 220 \citep{Brown2019}. Similarly, we already had calibrated data for IRAS 13120-5453 \citep{Ledger2021}, NGC 1068 \citep{Saito2022a, Saito2022b}, and NGC 6240 (O. Klimi, private communication).

To calibrate the remaining ALMA data, relevant projects were downloaded using scripts generated from the ALMA Science Archive\footnote{https://almascience.nao.ac.jp/aq/} in late June 2022 (data for IRAS F10565+2448 was downloaded and calibrated at a later date, after its public release). The individual project sets were calibrated using the ``ScriptForPI.py'' \texttt{PYTHON} script and specificed \texttt{CASA} version in the calibration script (except for projects 2013.1.00991.S and 2013.1.01172.S, where version 4.7.2 was used for calibration instead of the recommended \texttt{CASA} version).

\tabcolsep=0.2cm
\begin{table*}
\centering
    \caption{Data reduction and imaging parameters.}% \color{red}Should recheck that all the values here match the data from NAOJ before submitting, but I trust that I added everything correctly last summer. Once I have transferred the data over with Osvald, then I can check things locally on astro.}
    \begin{tabular}{cccccccc}
    \hline
        & & & \multicolumn{2}{c|}{Channel width ($\Delta V$)} & \multicolumn{2}{c|}{Sensitivity per channel} & \\
        Galaxy\textsuperscript{\textit{a}} & Native CO (1-0) beam & Smoothed round beam\textsuperscript{\textit{b}} & CO (1-0) & CN (1-0)  & CO (1-0) & CN (1-0) & Pixel size\textsuperscript{\textit{b}} \\
         & (arcsec $\times$ arcsec, PA in degrees) & (arcsec) &  \multicolumn{2}{c|}{(km s$^{-1}$)} &  \multicolumn{2}{c|}{(mJy beam$^{-1}$)} & (arcsec) \\
        \hline
        IRAS 13120-5453 & $0.7\times0.64, 1$ & $0.72$ & $20.0$ & $20.0$ & $1.06$ & $1.06$ & $0.36$ \\
        Arp 220 & $0.79\times0.47, 12$ & $1.17$ & $20.32$ & $20.3$ & $1.21$ & $1.50$ & $0.58$ \\
        IRAS F05189-2524 & $0.45\times0.42, -56$ & $0.55$ & $20.32$ & $20.64$ & $0.36$ & $0.72$ & $0.28$ \\
        IRAS F10565+2448 & $0.36\times0.26, -37$ & $0.52$ & $20.32$ & $20.3$ & $0.38$ & $0.76$ & $0.26$ \\
        NGC 6240 & $0.58\times0.57, -44$ & $0.89$ & $20.0$ & $20.0$ & $1.32$ & $1.24$ & $0.44$ \\
        IRAS F18293-3413 & $0.81\times0.57, -85$ & $1.2$ & $24.14$ & $24.51$ & $2.88$ & $4.31$ & $0.6$ \\
        NGC 3256 & $1.63\times1.45, 78$ & $2.34$ & $19.69$ & $23.22$ & $2.55$ & $1.15$ & $1.17$ \\
        NGC 1614 & $0.85\times0.53, 90$ & $1.49$ & $17.78$ & $23.22$ & $1.9$ & $2.97$ & $0.74$ \\
        NGC 7469 & $0.44\times0.30, -55$ & $1.46$ & $15.24$ & $20.64$ & $1.98$ & $2.55$ & $0.73$ \\
        NGC 2623 & $1.04\times0.86, 1$ & $1.34$ & $19.05$ & $24.51$ & $1.9$ & $3.37$ & $0.67$ \\
        NGC 3110 & $1.72\times1.35, -86$ & $1.8$ & $17.78$ & $23.22$ & $1.38$ & $2.0$ & $0.9$ \\
        ESO 320-G030 & $0.39\times0.34, -29$ & $2.5$ & $17.78$ & $23.22$ & $2.52$ & $3.86$ & $1.25$ \\
        NGC 1068\textsuperscript{\textit{c}} & $3.92\times3.67, -74$ & $7.38$ & $22.23$ & $23.23$ & $10.17$ & $3.05$ & $3.69$ \\
        NGC 5104 & $0.78\times0.58, 61$ & $1.14$ & $19.05$ & $24.51$ & $2.73$ & $5.45$ & $0.57$ \\
        NGC 4418 & $1.67\times1.28, 81$ & $2.83$ & $15.24$ & $20.64$ & $1.13$ & $1.21$ & $1.42$ \\
        NGC 1365 & $1.85\times1.43, 84$ & $5.76$ & $19.05$ & $24.51$ & $10.71$ & $11.18$ & $2.88$ \\
    \hline
    \end{tabular}
    \label{tab:imaging_params}
    \begin{tablenotes}
        \item \textit{Notes:} \textsuperscript{\textit{a}}The galaxies are listed in order of decreasing infrared luminosity.
        \item \textsuperscript{\textit{b}}Both the pixel and smoothed beam size for each galaxy are set by the target resolution of 500 pc.
        \item \textsuperscript{\textit{c}}We note that the resolution of NGC 1068 was tapered to a larger beamsize during cleaning to reduce the amount of post-process smoothing required. The original synthesized resolution of the dataset is $0.43$ arcsec $\times$ $0.38$ arcsec \citep{Saito2022a, Saito2022b}.
    \end{tablenotes}
\end{table*}

\subsubsection{Imaging}
\label{subsubsec:imaging}

Following calibration, we performed continuum subtraction and imaging using the PHANGS-ALMA pipeline v3 \citep{Leroy2021}. We already had images for IRAS 13120-5453 \citep{Ledger2021} and NGC 6240 (O. Klimi, private communication) at the target angular resolution, and as a result these galaxies were not reduced with the pipeline. Processed data cubes (produced using the PHANGS-ALMA pipeline v3) for NGC 1068 were kindly provided by the project PIs at the target resolution \citep{Saito2022a, Saito2022b}.

The PHANGS-ALMA pipeline reduces the calibrated data by joining together data for each galaxy from multiple measurement sets and project codes before performing continuum subtraction with \texttt{CASA}'s \texttt{uvcontsub} task on line-free channels. To obtain the best continuum subtraction, we specified the CO (1-0) line and both CN (1-0) hyperfine groupings to be excluded as channels with line emission. The spectral resolution of each datacube was determined as an integer multiplication of the original channel width. We targeted a channel width of $\sim20$ km s$^{-1}$ for all galaxies (Table \ref{tab:imaging_params}) to increase the sensitivity for the often weak emission from the CN (1-0) lines.

Most of the default PHANGS-ALMA pipeline v3 parameters were used for imaging the calibrated and continuum subtracted data. For the cleaning procedure, we set the default weighting to Brigg's \citep{Briggs1995} with robust = 0.5, and the primary beam was limited to $>20$\%. For the CN (1-0) line, we produced data cubes with a large spectral width to simultaneously image the two main hyperfine groupings around rest frequencies of $113.490970$ GHz and $113.170492$ GHz in the same data cube. Therefore, our final image products consist of a CO (1-0) cube and a CN (1-0) cube which contains both hyperfine groupings. We note that we did not recover both CN (1-0) lines in NGC 1365, as the faint line was not covered in the spectral window set up.

Although the PHANGS-ALMA pipeline v3 has the ability to produce a variety of further data products, we did not use the pipeline for data processing after producing the initial data cubes. Instead, any additional reduction processes were performed on the data cubes using \texttt{CASA} version $6.5.0.15$ and the \texttt{CASA} tasks \texttt{imsmooth}, \texttt{imsubimage}, \texttt{imregrid}, \texttt{imrebin}, \texttt{imstat}, and \texttt{exportfits}. These steps included: smoothing the cubes to the target beam resolution in arcsec (corresponding to $500$ pc), extracting a primary beam slice at the channel with peak CO intensity for correction of the moment 0 maps, and rebinning or regridding the images to minimum Nyquist sampled pixel size of 250 pc (Table \ref{tab:imaging_params}).

\subsection{Data products}
\label{subsec:data_products}
\subsubsection{CO (1-0) line products}
\label{subsubsec:co_products}
Integrated intensity maps for the CO (1-0) line were created using the Sun moment map method \citep{Sun2018, Sun2020}. The Sun method uses an expanding mask technique to find signal in the calibrated data cubes. Three-dimensional noise cubes are created and used to measure the RMS noise ($\sigma$\textsubscript{RMS}) for signal-free parts of the CO (1-0) line cubes in each pixel; the mean $\sigma$\textsubscript{RMS} for each galaxy is listed in Table \ref{tab:imaging_params}. Initial signal masks are then defined by selecting pixels which have positive CO (1-0) line emission and which have a signal-to-noise (S/N) ratio of at least 4 over two neighbouring channels. Finally, the mask expands to include pixels out to a lower S/N threshold of 2 \citep{Rosolowsky2006, Sun2018, Sun2020}.

We produced integrated intensity (moment 0) maps in K km s$^{-1}$ units for the CO (1-0) line cubes and corrected these by the primary beam using the Astropy Spectral Cube package \citep{Ginsburg2019}. We also made moment 1 (intensity-weighted velocity) and moment 2 (spectral line width) maps for the CO (1-0) line to use in CN processing (Section \ref{subsubsec:cn_products}). Uncertainties on these maps were calculated by taking the mean RMS noise in each pixel from the three-dimensional noise cubes ($\sigma$\textsubscript{RMS}) and multiplying this by the number of channels ($N$\textsubscript{chan}) with signal and the channel width ($\Delta V$), using
\begin{equation}
    \sigma\textsubscript{pix} = \sigma\textsubscript{RMS} \times \sqrt{N_{\text{chan}}} \times \Delta V.
    \label{eqn:uncertainty}
\end{equation}
 The resulting maps were used to estimate the uncertainty on the measured CO (1-0) fluxes and luminosities for each galaxy (Section \ref{subsubsec:fluxes_and_ratios}; Table \ref{tab:fluxes}).

In this paper, we do not convert the CO intensities to molecular gas surface densities ($\Sigma$\textsubscript{mol}) or masses (M\textsubscript{H\textsubscript{2}}), as is often done using an $\alpha$\textsubscript{CO} conversion factor \citep{Bolatto2013}. The typical conversion factor that is used for U/LIRGs is lower by roughly a factor of 5 compared to normal spirals, disks of galaxies, or the Milky Way, and would hold especially true in the CO bright central regions \citep{Downes1998}. We expect that the commonly accepted U/LIRG conversion factor, $\alpha$\textsubscript{CO} $= 1.088$ M\textsubscript{$\odot$} (K km s$^{-1}$ pc$^{2}$)$^{-1}$ (which includes a factor of 1.36 to account for Helium), is appropriate for the galaxies in our sample. U/LIRGs have high star formation rate surface densities, molecular gas surface densities, and are highly centrally concentrated \citep{Solomon1992, Downes1998, Lonsdale2006, Perez2021}. If some portion of the disks of our U/LIRG sample have a different $\alpha$\textsubscript{CO}, this would account for a small fraction of the total percentage of CO flux and produce a small effect in global measurements.

\subsubsection{CN (1-0) line products}
\label{subsubsec:cn_products}

\begin{figure}
	\includegraphics[width=0.99\columnwidth]{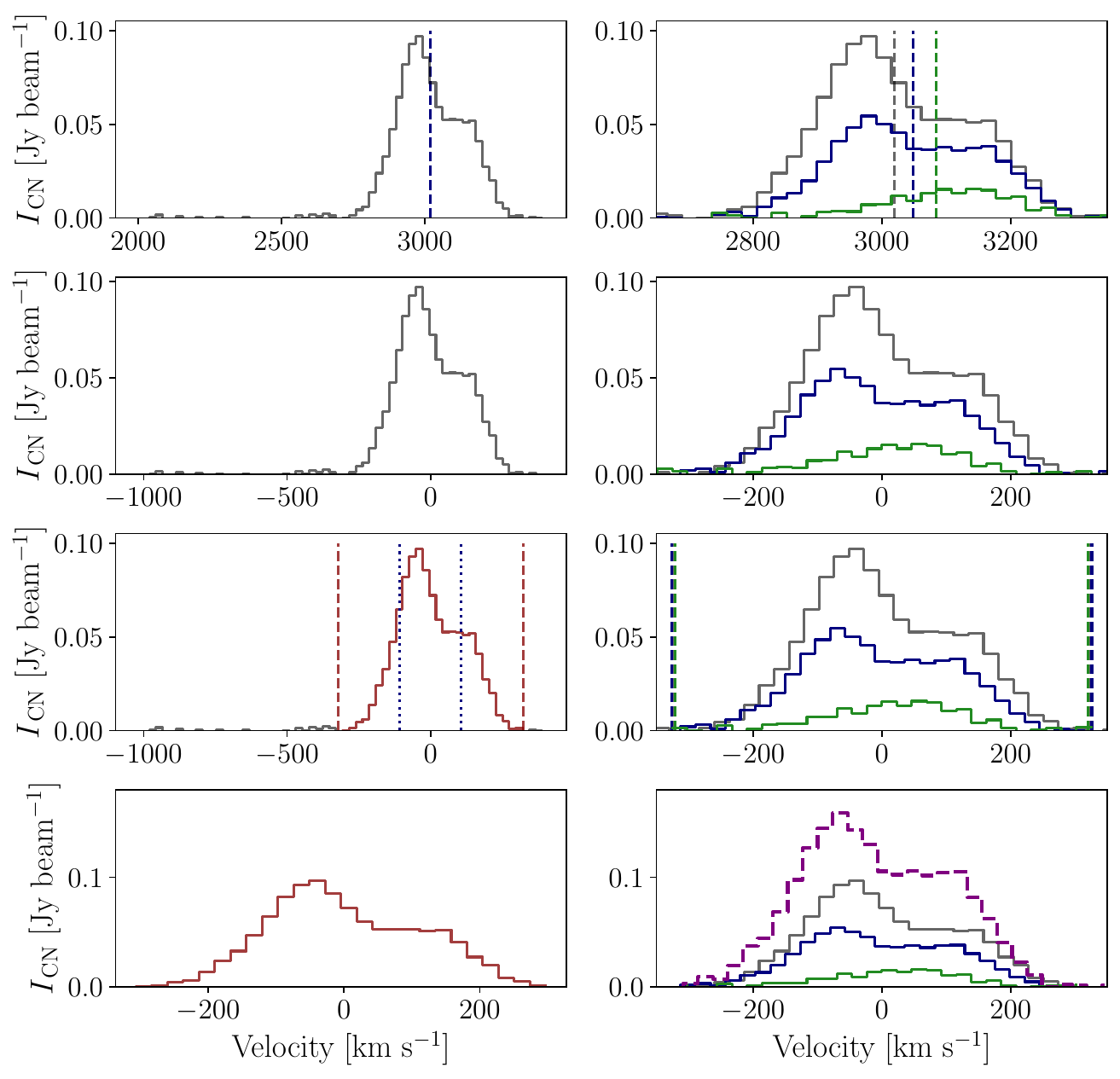}
    \caption{This figure illustrates an example of the shuffle-stack process for obtaining the moment 0 maps and the total integrated spectra for the CN bright and CN faint lines in ESO 320-G030. \textit{Left column}: This column shows an example of the shuffle and integration of the CN bright line in the strongest pixel. From top to bottom, we show the native CN velocity spectrum, the spectrum that has been shifted by the peak central velocity of the CO line, the spectrum overplotted by red dashed lines that show the width of integration ($3\times$ the CO line width, which is shown as the blue dotted lines), and the spectrum to be integrated in this pixel. \textit{Right column}: This column shows the shuffle and stack method for 3 example pixels. The top 3 figures are the same as the left column, but for the 3 pixels. The bottom plot shows the sum of the three stacked individual spectra as the magenta dashed line, which are integrated as part of the global spectrum.}
    \label{fig:stacking}
\end{figure}

\begin{figure*}
	\includegraphics[width=\textwidth]{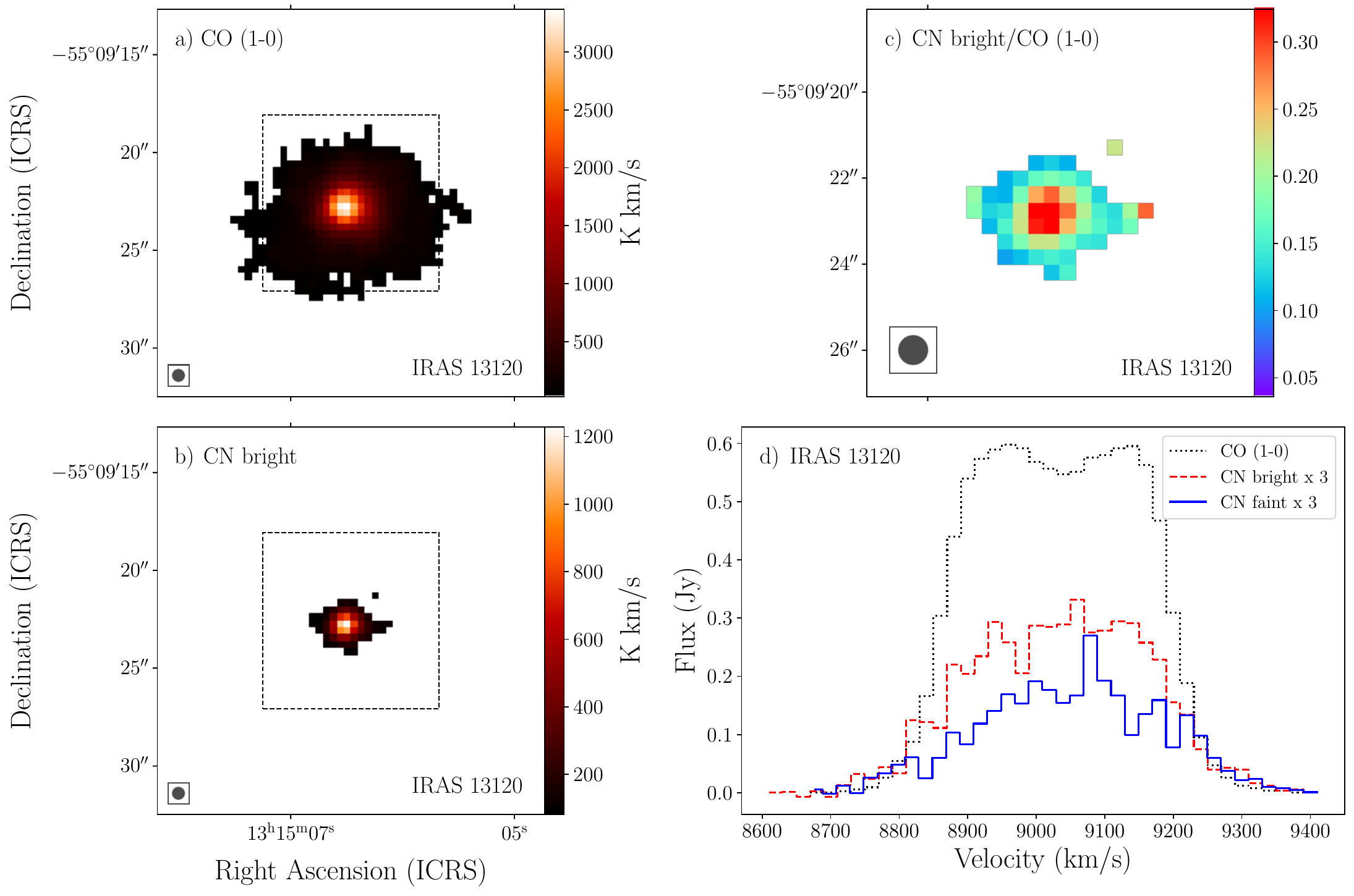}
    \caption{This figure shows the moment 0 and (CN bright)/CO intensity ratio maps in K km s$^{-1}$ units in IRAS 13120, as well as the global spectra. The circle in the bottom left corner of each panel denotes the size of the beam smoothed to 500 pc. \textit{a)} The total integrated intensity of the CO (1-0) line. The dashed square indicates the region in \textit{(c)}. \textit{b)} The total integrated intensity of the CN bright line. The pixels included here have a S/N of $>6\sigma$ and $>3\sigma$ in the CN bright and CN faint lines, respectively. \textit{c)} The (CN bright)/CO intensity ratio. The colour bar is clipped at the mean value plus or minus 80\%. \textit{d)} The total integrated spectra of the CO line (black dotted), CN bright line (red dashed), and CN faint line (blue). Both CN lines have been multiplied by a factor of 3 for demonstration purposes.}
    \label{fig:iras13120_4panel}
\end{figure*}

To measure the intensities of the weaker CN (1-0) lines, we used a spectral shuffle and stack method, similar to that described in e.g., \citet{Schruba2011} and \citet{Leroy2016}. Figure \ref{fig:stacking} demonstrates the applied shuffle-stack method. We spatially masked the CN image cubes to include only pixels where we have emission in the CO moment 0 maps, since we do not expect any weaker CN emission where there is no CO emission. The CN spectrum in each pixel was then shifted (the shuffe method) by the CO velocity from the moment 1 map and masked to only include signal within a spectral width of $3\times$ the CO linewidth ($\sigma$\textsubscript{CO}) from the moment 2 map. In each pixel, the CN emission was then integrated to obtain an intensity in K km s$^{-1}$ units. After completing this process in each pixel, we had an integrated intensity map for each of the CN bright and CN faint lines. These maps include pixels with weak emission, as our shuffle method includes any CN emission (regardless of S/N) found within the $3\sigma$\textsubscript{CO} linewidth. Both the CN bright and CN faint lines in our data become comparable to the total linewidth of the stronger CO emission after blending the individual hyperfine structure lines within each grouping (e.g., Figure \ref{fig:iras13120_4panel}d). We also create uncertainty maps for the CN lines using Equation \ref{eqn:uncertainty}, and use these values for estimating the uncertainty on the measured CN (1-0) fluxes and luminosities (Section \ref{subsubsec:fluxes_and_ratios}; Table \ref{tab:fluxes}) and applying any subsequent S/N cuts. The CN uncertainty is also used for determining upper limits on the (CN bright)/CO ratio when binning by CO intensity (Section \ref{subsubsec:CO_binning_method}).

Our final data products include moment 0 integrated intensity maps, uncertainty maps, and total integrated spectra for the CN bright, CN faint, and CO lines in each galaxy (see e.g., Figure \ref{fig:iras13120_4panel} and Appendix \ref{append:galaxy_images}).

\subsubsection{Measured fluxes, luminosities, and ratio maps}
\label{subsubsec:fluxes_and_ratios}

\begin{table*}
\centering
    \caption{CN (1-0) and CO (1-0) total integrated intensities and luminosities calculated using Equation \ref{eqn:lum}.}
    \begin{tabular}{ccccccccc}
    \hline
        Galaxy\textsuperscript{\textit{a}} & $S$\textsubscript{CO (1-0)} & $L$\textsubscript{CO (1-0)}\textsuperscript{\textit{b}} & $S$\textsubscript{CN (1-0)} bright\textsuperscript{\textit{c}} & $L$\textsubscript{CN (1-0)} bright\textsuperscript{\textit{b}} & S\textsubscript{CN (1-0)} faint\textsuperscript{\textit{c}} & $L$\textsubscript{CN (1-0)} faint\textsuperscript{\textit{b}} & N\textsuperscript{\textit{d}} & Distance\textsuperscript{\textit{e}} \\
         & (Jy km s$^{-1}$) & (K km s$^{-1}$ pc$^{2}$) & (Jy km s$^{-1}$)& (K km s$^{-1}$ pc$^{2}$) & (Jy km s$^{-1}$)& (K km s$^{-1}$ pc$^{2}$) & (pixels) & (Mpc) \\
        \hline
        IRAS 13120-5453  & $206\pm10$   & $86\pm4$ & $35\pm2$    & $15\pm1$ & $21\pm2$    & $9.3\pm0.9$ & $506$ & $137$ \\
        Arp 220          & $569\pm28$   & $87\pm5$ & $51\pm3$    & $8.0\pm0.5$ & $25\pm2$    & $3.9\pm0.3$ & $489$ & $81.1$ \\
        IRAS F05189-2524 & $23\pm1$     & $18\pm1$ & $2.0\pm0.3$ & $1.8\pm0.3$ & $1.0\pm0.3$ & $1.1\pm0.3$ & $174$ & $188$ \\ 
        IRAS F10565+2448 & $86\pm4$     & $70\pm4$ & $7\pm1$     & $6.1\pm0.5$ & $4.0\pm0.5$ & $3.6\pm0.5$ & $656$ & $194$ \\ 
        NGC 6240         & $236\pm12$   & $60\pm3$ & $20\pm2$    & $5.3\pm0.6$ & $12\pm2$    & $3.3\pm0.6$ & $396$ & $106$ \\ 
        IRAS F18293-3413 & $902\pm45$   & $125\pm6$ & $48\pm6$    & $6.9\pm0.8$ & $23\pm5$    & $3.3\pm0.7$ & $954$ & $77.2$ \\ 
        NGC 3256         & $1686\pm84$  & $79\pm4$ & $41\pm2$    & $2.0\pm0.1$ & $21\pm1$    & $1.0\pm0.1$ & $883$ & $44.3$ \\        
        NGC 1614         & $357\pm18$   & $38\pm2$ & $54\pm4$    & $6.0\pm0.4$ & $31\pm3$    & $3.4\pm0.3$ & $608$ & $67.9$ \\ 
        NGC 7469         & $359\pm18$   & $37\pm2$ & $36\pm3$    & $3.7\pm0.3$ & $19\pm2$    & $2.0\pm0.2$ & $766$ & $66$ \\
        NGC 2623         & $129\pm6$    & $21\pm1$ & $9\pm2$     & $1.5\pm0.3$ & $4\pm2$     & $0.8\pm0.3$ & $186$ & $83.4$ \\ 
        NGC 3110         & $271\pm14$   & $38\pm2$ & $7\pm2$     & $1.0\pm0.3$ & $3\pm2$     & $0.4\pm0.3$ & $790$ & $77.8$ \\
        ESO 320-G030     & $292\pm15$   & $18\pm1$ & $34\pm3$    & $2.1\pm0.2$ & $17\pm2$    & $1.1\pm0.1$ & $225$ & $50.7$ \\         
        NGC 1068         & $2768\pm138$ & $13\pm1$ & $174\pm9$   & $0.85\pm0.04$ & $72\pm4$    & $0.35\pm0.02$ & $209$ & $13.97$ \\        
        NGC 5104         & $172\pm9$    & $28\pm1$ & $12\pm3$    & $2.0\pm0.6$ & $5\pm4$     & $0.9\pm0.6$ & $324$ & $84.6$ \\ 
        NGC 4418         & $121\pm6$    & $3.6\pm0.2$ & $10\pm1$    & $0.31\pm0.02$ & $4.0\pm0.5$ & $0.11\pm0.02$ & $140$ & $35.3$ \\ 
        NGC 1365         & $3399\pm170$ & $31\pm2$ & $116\pm16$  & $0.25\pm0.01$ & -           & -     & $786$ & $19.57$ \\ %
    \hline
    \end{tabular}
    \label{tab:fluxes}
    \begin{tablenotes}
        \item \textit{Notes:} \textsuperscript{\textit{a}}The galaxies are listed in order of decreasing infrared luminosity.
        \item \textsuperscript{\textit{b}}All given luminosity values have been divided by a factor of $1\times10^{8}$.
        \item \textsuperscript{\textit{c}}The CN ($N = 1-0, J = 3/2-1/2$) and CN ($N = 1-0, J = 1/2-1/2$) lines are denoted as CN bright and CN faint, respectively
        \item \textsuperscript{\textit{d}}The number of pixels with CO (1-0) line detections.
        \item \textsuperscript{\textit{e}}All luminosity distances were converted from redshifts with Ned Wright's (Updated) Cosmology Calculator adopting WMAP 5-year cosmology with $H_{0} = 70.5$ km s$^{-1}$ Mpc$^{-1}$, $\Omega = 1$, $\Omega_{\rm m} = 0.27$ in the 3K CMB frame, except for NGC 7469 (SN type Ia distance is from \citealt{Ganeshalingam2013}) and NGC 1068 and NGC 1365 (tip of the red giant branch distance measures are from \citealt{Anand2021}). We note that any uncertainties from distance estimates are less relevant when considering line ratios.
    \end{tablenotes}
\end{table*}

The flux values for the CO (1-0) line were calculated directly from the integrated intensity maps. In each galaxy, we summed the intensities from all pixels with CO emission. To measure the fluxes for the CN (1-0) lines, we stacked the spectra from each pixel and obtained a total integrated spectrum, which we then integrated to obtain the CN bright and CN faint line fluxes (Figure \ref{fig:stacking}). This method significantly improved our S/N for the CN lines, and allowed us to detect weaker CN emission and better obtain the true global values of our measured intensity ratios.

Line luminosities were subsequently calculated using Equation 3 from \citet{Solomon2005}:
\begin{equation}
    L\textsubscript{mol} = 3.25\times10^{7} \frac{D\textsubscript{L}^{2}}{(1 + z)^{3}} \frac{1}{\nu\textsubscript{mol}^{2}} S\textsubscript{mol}\Delta v.
    \label{eqn:lum}
\end{equation}
We used the measured flux value and its uncertainty for each molecule as S\textsubscript{mol}$\Delta v$ [Jy km s$^{-1}$], with the corresponding line frequency, $\nu$\textsubscript{mol} [GHz]. Galaxy distances, $D$\textsubscript{L} [Mpc], are listed in Table \ref{tab:fluxes}. We adopted WMAP 5-year cosmology ($H_{0} = 70.5$ km s$^{-1}$ Mpc$^{-1}$, $\Omega = 1$, $\Omega_{\rm m} = 0.27$) in the CMB frame. Uncertainties were estimated on the line luminosities assuming all uncertainty comes from the fluxes (no uncertainty in $D$\textsubscript{L}, $z$, and $\nu$\textsubscript{mol}). Any distance uncertainties are less relevant when considering line ratios.

Ratio maps were created for (CN bright)/CO and (CN bright)/(CN faint) lines by dividing the individual moment maps. To avoid any low S/N pixels that were still included in the shuffle-stack creation of the CN moment maps, we spatially masked the CN bright and CN faint moment maps to include pixels with $>6\sigma$ and $>3\sigma$ emission in each spectral feature, respectively. This S/N threshold was chosen for each feature from the expected 2-to-1 (CN bright)/(CN faint) ratio that would come from optically thin CN emission in Local Thermodynamic Equilibrium \citep{Skatrud1983, Wang2004, Shirley2015, Tang2019}. Our main observational results, e.g., the CN and CO moment 0 maps, (CN bright)/CO ratio map, and total integrated spectra for all three lines, are combined and presented in four panel plots (Figure \ref{fig:iras13120_4panel} and Appendix \ref{append:galaxy_images}), where only pixels with emission from all three lines have been included. (CN bright)/(CN faint) ratio maps for all galaxies are shown in Appendix \ref{append:cn_bright_faint_ratio_maps}.

We do not apply any correction for inclination angle of individual galaxies to our measured intensities. 13 of the 16 U/LIRGs in our sample are in some merger stage (see e.g., \citealt{Stierwalt2013}), and therefore an inclination angle is difficult to interpret (a comparison of our measured ratios with merger stage is deferred to a future paper; Ledger et al. \textit{in prep.}). The galaxies which are not mergers are NGC 4418, ESO 320-G030, and NGC 1365 \citep{Stierwalt2013}. For reference, the inclination angles of these three galaxies are $i=62^{\circ}$ for NGC 4418 \citep{Sakamoto2013}, $i=43^{\circ}$ for ESO 320-G030 \citep{Pereira2016}, and $i=40^{\circ}$ for NGC 1365 \citep{Sakamoto2007}.

\subsubsection{CO (1-0) intensity binning}
\label{subsubsec:CO_binning_method}

We also measure the (CN bright)/CO intensity ratios by binning pixels by the intensity of the CO (1-0) line. We split the range of CO intensities in logarithmic space into 15 CO intensity bins, spaced by $\sim10^{0.25}$ K km s$^{-1}$. Choosing 15 bins allows for a minimum of 8 bins in each galaxy. The first bin is equal to or less than $10^{0.5}$ K km s$^{-1}$, and the final bin is equal to or greater than $10^{3.75}$ K km s$^{-1}$. For the pixels in each bin, we measure the total integrated CO (1-0) intensity and shuffle-stack to get the total integrated CN (1-0) intensity.

 \citet{Beslic2021} and \citet{denBrok2021, denBrok2022} perform a similar binning analysis when stacking into galactocentric radial bins and CO (2-1) intensity bins. Similar to \citet{denBrok2022}, we also identify upper limits on the binned intensities of the CN bright line if they are detected below a 3$\sigma$ level. The results from binning and plots of the CO intensity binned ratio for each galaxy are discussed in Section \ref{subsec:cn_co_nuclei}.

 Although we do not explicitly convert to molecular gas surface densities in this work, the reader can convert our CO intensity bins to $\Sigma$\textsubscript{mol} bins [M\textsubscript{$\odot$} pc$^{-2}$] by multiplying by the ULIRG conversion factor: $\alpha$\textsubscript{CO} $= 1.088$ M\textsubscript{$\odot$} (K km s$^{-1}$ pc$^{2}$)$^{-1}$.

\subsection{Measuring the CN and CO intensity ratios}
\label{subsec:ratio_summary}

\begin{table*}
\centering
    \caption{Intensity ratios in K km s$^{-1}$ of CN and CO in 16 U/LIRGs.}
    \begin{tabular}{cccccccc}
    \hline
        Galaxy\textsuperscript{\textit{a}} & (CN bright)/CO & (CN bright)/CO & (CN bright)/CO & (CN bright)/(CN faint) & (CN bright)/(CN faint)  & log($L$\textsubscript{IR})\textsuperscript{\textit{c}} & Type of AGN\textsuperscript{\textit{d}} \\
         & peak ratio & global ratio & spatial average\textsuperscript{\textit{b}} &  global ratio & spatial average\textsuperscript{\textit{b}} & ($L_{\odot}$) & \\
        \hline
        IRAS 13120-5453  & $0.35\pm0.01$ & $0.17\pm0.01$ & $0.18\pm0.01$ & $1.64\pm0.01$ & $1.55\pm0.03$ & 12.28 & Obscured \\
        Arp 220          & $0.07\pm0.01$ & $0.09\pm0.01$ & $0.26\pm0.02$ & $2.07\pm0.01$ & $2.16\pm0.08$ & 12.21 & Obscured \\
        IRAS F05189-2524 & $0.26\pm0.01$ & $0.10\pm0.01$ & $0.20\pm0.01$ & $1.74\pm0.01$ & $1.73\pm0.06$ & 12.16 & Seyfert 2 \\
        IRAS F10565+2448 & $0.22\pm0.01$ & $0.08\pm0.01$ & $0.15\pm0.01$ & $1.70\pm0.01$ & $1.63\pm0.03$ & 12.07 & None \\
        NGC 6240         & $0.10\pm0.01$ & $0.08\pm0.01$ & $0.16\pm0.02$ & $1.62\pm0.01$ & $1.78\pm0.07$ & 11.85 & Obscured \\
        IRAS F18293-3413 & $0.13\pm0.01$ & $0.05\pm0.01$ & $0.12\pm0.01$ & $2.14\pm0.01$ & $1.80\pm0.05$ & 11.79 & None \\
        NGC 3256         & $0.06\pm0.01$ & $0.02\pm0.01$ & $0.05\pm0.01$ & $1.99\pm0.01$ & $2.07\pm0.04$ & 11.75 & Obscured \\       
        NGC 1614         & $0.17\pm0.01$ & $0.15\pm0.01$ & $0.20\pm0.01$ & $1.76\pm0.01$ & $1.72\pm0.03$ & 11.65 & LINER \\
        NGC 7469         & $0.28\pm0.01$ & $0.10\pm0.01$ & $0.27\pm0.02$ & $1.88\pm0.01$ & $1.73\pm0.02$ & 11.59 & Seyfert 1 \\
        NGC 2623         & $0.13\pm0.01$ & $0.07\pm0.01$ & $0.13\pm0.01$ & $2.06\pm0.01$ & $1.88\pm0.06$ & 11.59 & LINER \\
        NGC 3110         & $0.07\pm0.01$ & $0.02\pm0.01$ & $0.07\pm0.01$ & $2.32\pm0.01$ & $1.64\pm0.05$ & 11.35 & None \\
        ESO 320-G030     & $0.17\pm0.01$ & $0.11\pm0.01$ & $0.24\pm0.03$ & $2.02\pm0.01$ & $2.00\pm0.04$ & 11.35 & None \\
        NGC 1068         & $0.08\pm0.01$ & $0.06\pm0.01$ & $0.07\pm0.01$ & $2.44\pm0.01$ & $2.75\pm0.07$ & 11.29 & Seyfert 2 \\        
        NGC 5104         & $0.09\pm0.01$ & $0.07\pm0.02$ & $0.11\pm0.01$ & $2.15\pm0.02$ & $1.99\pm0.07$ & 11.21 & None \\
        NGC 4418         & $0.12\pm0.01$ & $0.08\pm0.01$ & $0.09\pm0.01$ & $2.81\pm0.01$ & $2.58\pm0.08$ & 11.16 & Obscured \\
        NGC 1365\textsuperscript{\textit{e}} & $0.05\pm0.01$ & $0.03\pm0.01$ & $0.05\pm0.01$ & - & - & 11.08 & Obscured \\
    \hline
    \end{tabular}
    \label{tab:ratios}
    \begin{tablenotes}
        \item \textit{Notes:} \textsuperscript{\textit{a}}The galaxies are listed in order of decreasing infrared luminosity.
        \item \textsuperscript{\textit{b}}Spatial averages are calculated from the intensity ratio maps as the mean pixel value and the uncertainties are given by jackknife sampling of the mean. The spatial averages are measured on maps with the pixels masked with $>6\sigma$ and $>3\sigma$ detections in the CN bright and CN faint lines, respectively.
        \item \textsuperscript{\textit{c}}The log($L$\textsubscript{IR}) values are taken from the GOALS survey \citep{Armus2009} and we corrected them to our luminosity distances (as listed in Table \ref{tab:fluxes}).
        \item \textsuperscript{\textit{d}}We classify the type of AGN based on previous literature as an obscured AGN (Obscured), an optically identified AGN (Seyfert 1, Seyfert 2, or LINER), or no documented evidence for an AGN (None). For further discussion see Section \ref{subsubsec:cn_co_agn_position}. Galaxies with an obscured AGN: IRAS 13120 \citep{Teng2015}, Arp 220 \citep{Sakamoto2017}, NGC 6240 \citep{Iwasawa2011}, NGC 3256 southern nucleus \citep{Sakamoto2014}, NGC 4418 \citep{Ohyama2019}, and NGC 1365 \citep{Swain2023}. Galaxies with an optically classified AGN: IRAS F05189 \citep{Smith2019}, NGC 1614 \citep{Konig2013}, NGC 7469 \citep{Liu2014}, NGC 2623 \citep{Aalto2002}, and NGC 1068 \citep{Saito2022a, Saito2022b}. Galaxies with no conclusive evidence for an AGN: IRAS F10565 \citep{Iwasawa2011}, IRAS F18293, NGC 3110 \citep{Espada2018}, NGC 5104, and ESO 320 \citep{Gonzalez2021}.
        \item \textsuperscript{\textit{e}}We only report (CN bright)/CO ratios because the spectral window for the NGC 1365 ALMA data did not cover the CN faint line.
    \end{tablenotes}
\end{table*}

We measured global, peak, and spatially averaged (CN bright)/CO and (CN bright)/(CN faint) intensity ratios for each galaxy. The measured ratios are listed in Table \ref{tab:ratios}.

\begin{itemize}
    \item \textit{(CN bright)/CO global ratios:} Global ratio values are measured using the total integrated intensities in Jy km s$^{-1}$ of each spectral line and then converting to K km s$^{-1}$. For CO, we sum the detected pixels in the integrated intensity map. For CN bright, we integrate the shuffle-stacked spectral line. This approach takes advantage of the better S/N of the CN lines obtained through our shuffle-stacking method.
    \item \textit{(CN bright)/CO peak ratios:} To obtain the (CN bright)/CO peak ratio values, we measure the integrated intensity of each line using only the pixels in the strongest CO (1-0) intensity bin (see the binning method described in Section \ref{subsubsec:CO_binning_method}). Note this is the ratio measured at the peak CO emission, and not necessarily the maximum (CN bright)/CO ratio in a given galaxy.
    \item \textit{(CN bright)/CO spatial averages:} The spatially averaged ratio values are obtained from averaging the pixels in the S/N matched moment maps, which may include uncertainties due to the S/N matching process (e.g., CO is masked using the Sun method and CN bright is masked at $>6\sigma$).
    \item \textit{(CN bright)/(CN faint) global ratios and spatial averages}: The (CN bright)/(CN faint) global and spatial average ratios are measured in the same way as the (CN bright)/CO ratios, with CN bright and CN faint masked at $>6\sigma$ and $>3\sigma$, respectively. The spatially averaged ratios use the (CN bright)/(CN faint) S/N matched maps, which are shown in Figure \ref{fig:cn_bright_faint_maps} in Appendix \ref{append:cn_bright_faint_ratio_maps}.
\end{itemize}

\textit{A brief note on Arp 220}: The individual pixels of Arp 220 are not highly reliable because of varying RMS values across the data cube. However, we anticipate that the global values obtained will be trustworthy as they are calculated from multiple pixels and will average out the noise fluctuations. We do not consider absorption effects in our analysis, which have been seen previously in Arp 220, see e.g., \citet{Ueda2022}, although we do find subtle absorption signatures in the spectra of the CO and CN lines. The individual pixel data points for Arp 220 are not included in Figures \ref{fig:cn_optical_depth}, \ref{fig:cn_hist_violin}, or \ref{fig:cn_co_hist}, but we have quoted their values in Tables \ref{tab:ratios}, \ref{tab:CN_optical_histogram}, and \ref{tab:CN_CO_histogram_values}. We recreated these three figures including the pixels in Arp 220 and provide them in Appendix \ref{append:arp220}.

\section{The CN and CO Intensity Ratios in U/LIRGs}
\label{sec:results}

\subsection{CN is more optically thin in LIRGs than ULIRGs}
\label{subsec:cn_optical_depth_results}

\begin{figure}
	\includegraphics[width=\columnwidth]{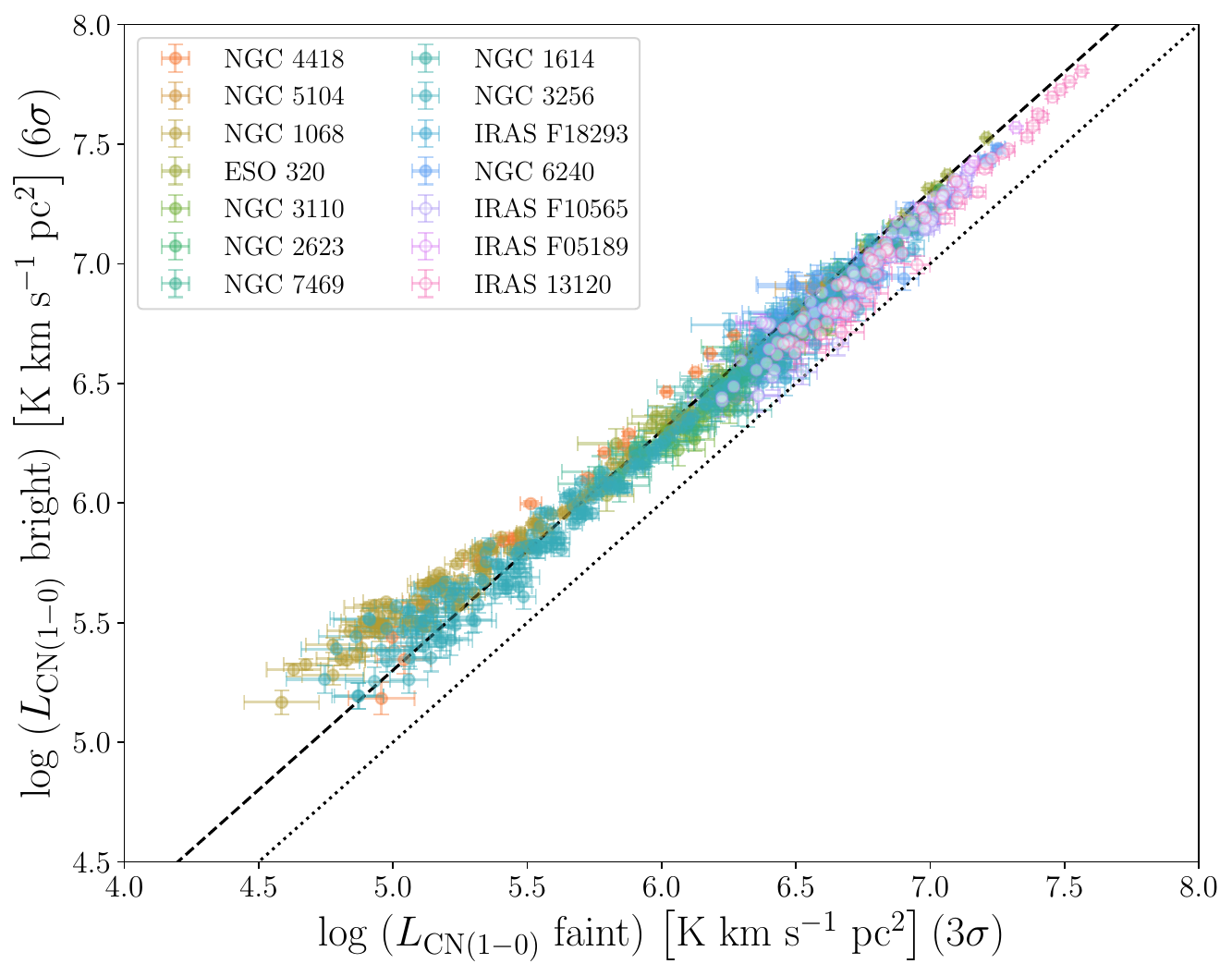}
    \caption{This figure compares the CN bright and CN faint lines on a pixel-by-pixel basis for 14 of the 16 galaxies in our sample. The (CN bright)/(CN faint) intensity ratio is lower in ULIRGs than LIRGs, implying that CN emission in more optically thick in ULIRGs. NGC 1365 is not included because the CN faint line is not covered by the spectral windows in the ALMA data. Data points from Arp 220 are not included due to uncertain pixels, but an example including this galaxy can be found in Figure \ref{fig:cn_optical_depth_arp220} in Appendix \ref{append:arp220}. Uncertainties on individual pixels are calculated from Equation \ref{eqn:uncertainty}. Both pixel values and their uncertainties have been converted from fluxes to luminosities using Equation \ref{eqn:lum}. The pixels for the CN bright and CN faint lines are show with S/N cuts of $>6\sigma$ and $>3\sigma$, respectively. Pixels have been colourized by galaxy. Open circles are ULIRG galaxies. Closed circles are LIRG galaxies. The black-dotted and black-dashed lines represent 1:1 and 2:1 luminosity ratios, respectively. The colours for individual galaxies match those in Figures \ref{fig:cn_co_ratio_violin_all} and \ref{fig:ratio_comparison}.}
    \label{fig:cn_optical_depth}
\end{figure}

\begin{figure*}
	\includegraphics[width=\textwidth]{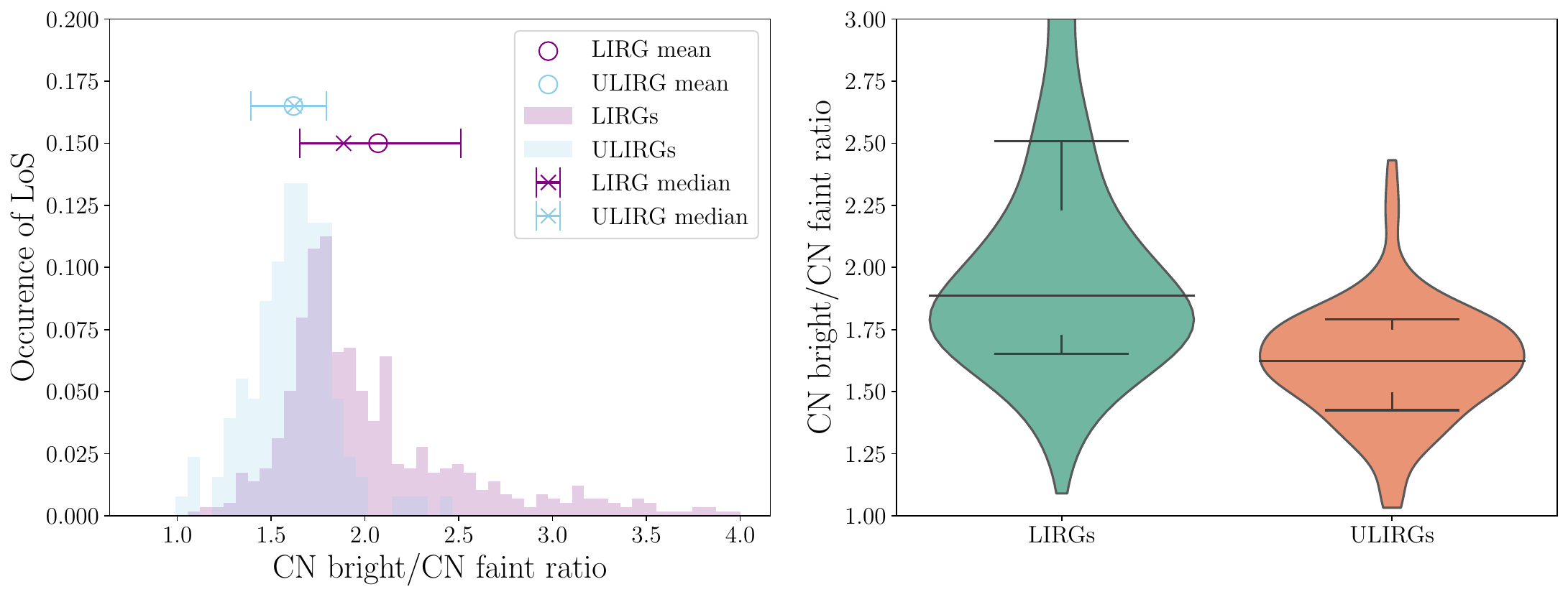}
    \caption{This figure shows the histograms of the (CN bright)/(CN faint) ratio after splitting the sample into pixels from ULIRGs and LIRGs. The median and mean (CN bright)/(CN faint) intensity ratios are lower in ULIRGs than LIRGs. Data points from Arp 220 are not included, but an example including this galaxy can be found in Figure \ref{fig:cn_hist_violin_arp220} in Appendix \ref{append:arp220}. \textit{Left:} The y-axis shows the number of pixels with the specific ratio value, and the x-axis gives the ratio in a linear scale. The blue and violet histograms correspond to the ULIRG and LIRG data points, respectively. The open circles are the mean values of each distribution. The cross represents the median value, while the error bars extend to the 16\textsuperscript{th} and 84\textsuperscript{th} percentiles. A decreasing (CN bright)/(CN faint) ratio corresponds to an increasing optical depth. \textit{Right:} Violin plots of the (CN bright)/(CN faint) intensity ratio in ULIRGs (orange) compared to LIRGs (green). The black lines correspond to the 16\textsuperscript{th}, 50\textsuperscript{th}, and 84\textsuperscript{th} percentiles. The y-axis gives the (CN bright)/(CN faint) ratio in a linear scale.}
    \label{fig:cn_hist_violin}
\end{figure*}

\tabcolsep=0.15cm
\begin{table}
\centering
    \caption{(CN bright)/(CN faint) intensity ratio distributions.}
    \begin{tabular}{ccccc}
    \hline
        & \multicolumn{3}{c|}{(CN bright)/(CN faint) ratio} & \\
        Galaxy\textsuperscript{\textit{a,b}} & Mean\textsuperscript{\textit{c}} & 16\textsuperscript{th}-50\textsuperscript{th}-84th\textsuperscript{th} & Mode\textsuperscript{\textit{d}} & $N$\textsuperscript{\textit{e}} \\
        \hline
        All ULIRGs & $1.62\pm0.02$ & $1.39-1.62-1.79$ & $1.5$ & $127$ \\
        \hline
        IRAS 13120-5453 & $1.56\pm0.03$ & $1.36-1.58-1.75$ & $1.5$ & $53$ \\
        Arp 220 & $2.17\pm0.08$ & $1.38-2.02-3.01$ & $1.8$ & $114$ \\
        IRAS F05189-2524 & $1.74\pm0.06$ & $1.56-1.72-1.87$ & $1.5$ & $16$ \\
        IRAS F10565+2448 & $1.64\pm0.03$ & $1.43-1.64-1.81$ & $1.5$ & $58$ \\
        \hline
        All LIRGs & $2.07\pm0.02$ & $1.65-1.89-2.51$ & $1.7$ & $587$ \\
        \hline
        NGC 6240 & $1.79\pm0.07$ & $1.54-1.71-2.12$ & $1.6$ & $27$ \\
        IRAS F18293-3413 & $1.81\pm0.05$ & $1.38-1.87-2.14$ & $1.3$ & $53$ \\
        NGC 3256 & $2.08\pm0.04$ & $1.74-1.89-2.48$ & $1.7$ & $175$ \\
        NGC 1614 & $1.73\pm0.03$ & $1.57-1.69-1.91$ & $1.6$ & $77$ \\
        NGC 7469 & $1.74\pm0.02$ & $1.61-1.72-1.84$ & $1.7$ & $67$ \\
        NGC 2623 & $1.89\pm0.06$ & $1.72-1.82-2.11$ & $1.7$ & $13$ \\
        NGC 3110 & $1.65\pm0.05$ & $1.44-1.65-1.9$ & $1.6$ & $19$ \\
        ESO 320-G030 & $2.01\pm0.04$ & $1.85-2.0-2.15$ & $1.9$ & $32$ \\
        NGC 1068 & $2.76\pm0.07$ & $2.02-2.55-3.49$ & $2.1$ & $103$ \\
        NGC 5104 & $2.00\pm0.07$ & $1.89-2.0-2.11$ & $1.8$ & $4$ \\
        NGC 4418 & $2.59\pm0.08$ & $2.42-2.66-2.8$ & $2.7$ & $17$ \\
    \hline
    \end{tabular}
    \label{tab:CN_optical_histogram}
    \begin{tablenotes}
        \item \textit{Notes:} \textsuperscript{\textit{a}}The galaxies are listed in order of decreasing infrared luminosity.
        \item \textsuperscript{\textit{b}}NGC 1365 is not included in this table because the CN faint line is not covered by the spectral windows in the ALMA data. Arp 220 is not included in the calculation of the "All ULIRGs" values because the individual scatter points are not trustworthy.
        \item \textsuperscript{\textit{c}}The uncertainty on the mean was calculated using jackknife resampling.
        \item \textsuperscript{\textit{d}}The mode is the most common value in each histogram when dispersed in 50 bins ranging from 0.0 to 5.0.
        \item \textsuperscript{\textit{e}}The number of points in the histogram for each galaxy as defined by the number of pixels seen in the spatial (CN bright)/(CN faint) ratio maps in Appendix \ref{append:cn_bright_faint_ratio_maps}. These ratio maps show pixels with $>6\sigma$ and $>3\sigma$ detections in the CN bright and CN faint lines, respectively.
    \end{tablenotes}
\end{table}

\subsubsection{The (CN bright)/(CN faint) intensity ratio}
\label{subsubsec:cn_bright_faint_ratio}

We measure the (CN bright)/(CN faint) ratio in 16 U/LIRG galaxies using a global ratio calculated from comparing the total integrated intensities of the CN ($N = 1-0, J = 3/2-1/2$) and CN ($N = 1-0, J = 1/2-1/2$) hyperfine groupings. Additionally, we calculate a spatially averaged ratio using S/N matched ratio maps (Figure \ref{fig:cn_bright_faint_maps}). The ratios are presented in Table \ref{tab:ratios}. On average, the (CN bright)/(CN faint) ratio we measure is lower when using the spatial averaging method than when calculating a global ratio from the shuffle-stacked spectra. This comparison implies we are picking up more bright relative to faint CN emission when using the stacking process. The uncertainty when calculating the spatially averaged ratio is higher than for the global ratios, in part because we have reduced the number of pixels used in calculating the spatial averages due to the S/N matching of the CN bright ($>6\sigma$ cut) and CN faint ($>3\sigma$ cut) lines. Using the integrated global stacked spectra, we measure (CN bright)/(CN faint) ratios ranging from $1.61-2.80$ in our galaxy sample. Using spatial averaging, we measure (CN bright)/(CN faint) ratios ranging from $1.55-2.75$. The overall trend of ULIRGs having lower (CN bright)/(CN faint) ratios than LIRGs and the range of ratio values are similar between the two methods.
%Since our S/N matched maps use a $>6 \sigma$ cut for CN bright and a $> 3\sigma$ cut for CN faint, we are preferentially biased to lower (CN bright)/(CN faint) ratios as a result of cutting more CN bright emission with the higher threshold.

Figure \ref{fig:cn_optical_depth} compares the luminosity of the CN bright line with the luminosity of the CN faint line in 14 of our 16 galaxies. NGC 1365 is not included, since the CN faint line was not observed in this galaxy. Arp 220 was excluded because we do not trust individual pixels because of varying RMS values across the data cube due to poor quality data (see Section \ref{subsec:ratio_summary}); a version including Arp 220 is given in Figure \ref{fig:cn_optical_depth_arp220} in Appendix \ref{append:arp220}. We have used line luminosities in Figure \ref{fig:cn_optical_depth} instead of intensities to remove distance dependencies when comparing the ratio between galaxies. 

The CN bright and CN faint lines should have a 2:1 brightness ratio for optically thin emission in LTE conditions (see e.g., \citealt{Skatrud1983}). To adjust for this, we use $>6\sigma$ and $>3\sigma$ S/N cuts for the CN bright and CN faint lines, respectively, in Figure \ref{fig:cn_optical_depth}. The majority of individual pixels scatter around the 2:1 line; however, at higher line luminosities, there are a significant number of pixels that lie between the 1:1 dotted line and 2:1 dashed lines, indicating CN is becoming more optically thick in these regions.

Figure \ref{fig:cn_hist_violin} shows histograms of the (CN bright)/(CN faint) ratio after grouping the galaxies into ULIRG and LIRG samples. In Figure \ref{fig:cn_hist_violin}, lower (CN bright)/(CN faint) ratios tend to be found in ULIRGs toward the left in the figure, matching the results from Figure \ref{fig:cn_optical_depth}. Table \ref{tab:CN_optical_histogram} provides the mean, mode, and 16th, 50th and 84th quartiles of the histogram values for each galaxy and the combined values for the LIRG and ULIRG distributions.

\subsubsection{Estimating CN optical depths}
\label{subsubsec:cn_optical_depth}

The (CN bright)/(CN faint) intensity ratio can be used to estimate CN optical depth \citep{Skatrud1983, Wang2004, Tang2019}. We observe CN (1-0) line widths in excess of 100 km s$^{-1}$, which is sufficient to blend the hyperfine components of the nine individual CN lines which make up the two larger hyperfine groups. As a result, the optical depths we estimate are average $\tau$ values representative of the total integrated CN lines, and not opacities of individual hyperfine components or line peak opacities. We use the median ratio values from Table \ref{tab:CN_optical_histogram} to estimate the CN optical depth in ULIRGs compared to LIRGs. Under the initial assumption that the CN lines are optically thin and in LTE, we use Equation (1) from \cite{Tang2019} to estimate the optical depth as
\begin{equation}
    \frac{I\textsubscript{CN bright}}{I\textsubscript{CN faint}} = \frac{1 - e^{-\tau_1}}{1 - e^{-\tau_2}}.
    \label{eqn:cn_optical_depth}
\end{equation}

\noindent{$\tau_1$ and $\tau_2$ are the optical depths of the CN bright and CN faint lines, respectively, and $\tau_1$ = $2\tau_2$ \citep{Skatrud1983}. Using the median (CN bright)/(CN faint) ratios from Table \ref{tab:CN_optical_histogram}, we estimate the average CN optical depth to be $\tau = 0.96$ and $\tau = 0.23$ in ULIRGs and LIRGs, respectively. We note that CN is more optically thick in ULIRGs than LIRGs for our sample, although we only have 3 ULIRGs included in our analysis. Looking at Table \ref{tab:CN_optical_histogram}, the trend is that galaxies with lower infrared luminosities ($L\textsubscript{IR}$) have lower CN optical depths, with (CN bright)/(CN faint) ratios closer to 2.}
%Solving Equation \ref{eqn:cn_optical_depth}, we find that the CN optical depth can be calculated as $\tau = -2 \ \text{ln}( \frac{I\textsubscript{CN bright}}{I\textsubscript{CN faint}} - 1)$. 

There are two galaxies with (CN bright)/(CN faint) $> 2$, NGC 1068 and NGC 4418, for which using this calculation method would mean $\tau\ll1$. We attribute these high ratios to possible S/N matching issues, underestimated emission from the CN faint line, non-LTE effects in the excitation of the CN fine structure distribution, possible line blending effects, or the complicated influence of an AGN, which can be found in both galaxies.

%Optical depth effects can have a major impact on the emission seen from molecular lines in extragalactic system, particularly U/LIRGs which have large molecular gas reservoirs in their ISM. Optically thin emission indicates that the line intensity we are observing directly corresponds to molecular abundance, e.g., higher intensity means higher abundance. When a line becomes optically thick, we tend to underestimate the true intensity and as such, our line ratios and interpretations can be skewed. It is typically assumed that the CO (1-0) line is optically thick at ensemble of cloud scales in extragalactic systems (see e.g., \citealt{Bolatto2013} and references therein). To better interpret whether CN optical depth effects will impact our measured CN/CO intensity ratios, we have estimated the CN optical depth using the (CN bright)/(CN faint) intensity ratio to be $\tau = 0.96$ and $\tau = 0.23$ in ULIRGs and LIRGs, respectively.

CN opacities have been previously estimated in four nearby starburst galaxies. \cite{Henkel1998} observed CN in M82 using the IRAM 30m telescope and measured the (CN bright)/(CN faint) ratio to be $\sim2.33$, implying optically thin CN emission. IC 342 was found to have optically thick CN emission, with a ratio of $\sim 1.45$ \citep{Henkel1998} and an estimated optical depth of $4.09$ \citep{Nakajima2018}. \cite{Wang2004} observed NGC 4945 with the Swedish-ESO Submillimeter Telescope. The authors found a (CN bright)/(CN faint) ratio of $\sim1.58$ and a moderate CN optical depth of $1.09$. Similarly, \citet{Tang2019} used ALMA to measure a ratio of $1.68$ and an optical depth of $0.8$ in NGC 4945.

\cite{Henkel2014} measure the (CN bright)/(CN faint) ratio to be $1.55$ in the starburst nucleus of NGC 253 using the IRAM 30m (with a peak temperature ratio of $\sim1.96$). From the ratio, \cite{Henkel2014} calculate a moderate optical depth of $0.5$. \citet{Tang2019} measure a (CN bright)/(CN faint) ratio of $1.58$ with ALMA in NGC 253, corresponding to an optical depth of $1.1$. Their results suggest that the CN emission is becoming optically thick in NGC 253. In contrast, \citet{Meier2015} use ALMA to measure a (CN bright)/(CN faint) ratio of $\sim2$ in the inner disk of NGC 253, concluding that CN is optically thin. They found that their observed CN emission is strongest near the star forming regions of NGC 253 and the origin of the molecular outflow. \cite{Nakajima2018} quote an optical depth $0.39$ in NGC 253, also suggesting a moderate to low optical depth in the central starburst, in agreement with \cite{Meier2015}. 

CN opacities have also been measured in three U/LIRGs. \cite{Henkel2014} observed Mrk 231 and found an integrated intensity ratio for (CN bright)/(CN faint) of $\sim2$, but do not quote an optical depth. \cite{Cicone2020} measure the CN (1-0) line in the molecular outflows of Mrk 231 and found a (CN bright)/(CN faint) intensity ratio of $\sim1.83$. The measured (CN bright)/(CN faint) ratios in Mrk 231 are higher than the average values we find for the ULIRGs in our sample, suggesting this galaxy has more optically thin CN emission. \cite{Tang2019} used ALMA to observe CN (1-0) in NGC 1068 and found a (CN bright)/(CN faint) ratio of $1.7$ and an optical depth of $0.7$. \citet{Nakajima2018} estimated an optical depth of $1.42$ in NGC 1068. In contrast, we measure a (CN bright)/(CN faint) ratio of $2.44$ in NGC 1068, suggesting optically thin emission. Differences in our methods for recovering CN emission using the shuffle-stack method could lead to this discrepancy for NGC 1068. \cite{Konig2016} measured a (CN bright)/(CN faint) integrated intensity ratio of $1.89$ in NGC 1614, while we measure a ratio of $1.75$ in NGC 1614 using the shuffle-stack method. Our conclusions agree with \cite{Konig2016} that CN emission will be moderately optically thin in this galaxy.

In summary, previous measures of CN optical depth in galaxies suggest that CN (1-0) can indeed be used as an optically thin gas tracer for most cases, although moderate optical depths have been measured in some systems. Our analysis indicates a trend toward higher CN optical depths in ULIRGs relative to LIRGs, but we also observe variations between individual galaxies.

\subsection{(CN bright)/CO is higher in ULIRGs compared to LIRGs}
\label{subsec:cn_co_ratio_results}

\begin{figure*}
	\includegraphics[width=\textwidth]{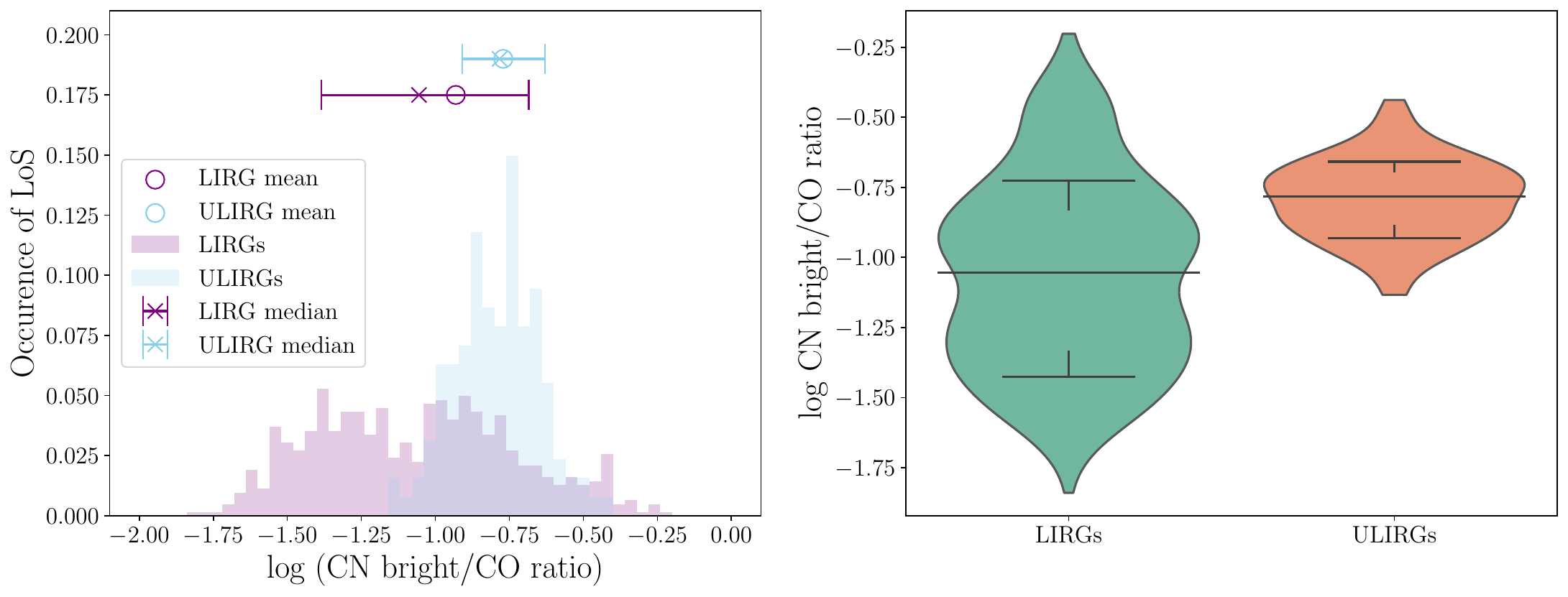}
    \caption{This figure shows the histograms of the (CN bright)/CO ratio after splitting the sample into pixels from ULIRGs and LIRGs. The (CN bright)/CO intensity is higher in ULIRGs than LIRGs; however, there is more spread in the ratio in LIRGs compared to ULIRGs. Data points from Arp 220 are not included, but an example including this galaxy can be found as Figure \ref{fig:cn_co_hist_arp220} in Appendix \ref{append:arp220}. \textit{Left:} The y-axis shows the number of pixels with the specific ratio value, and the x-axis gives the (CN bright)/CO ratio on a logarithmic scale. The blue and violet histograms correspond to the ULIRG and LIRG data points, respectively. The open circles are the mean values of each distribution. The cross represents the median value, while the error bars extend to the 16\textsuperscript{th} and 84\textsuperscript{th} percentiles. \textit{Right:} Violin plots of the (CN bright)/CO intensity ratio in ULIRGs (orange) compared to LIRGs (green). The black lines correspond to the 16\textsuperscript{th}, 50\textsuperscript{th}, and 84\textsuperscript{th} percentiles. The y-axis gives the (CN bright)/CO ratio on a logarithmic scale.}
    \label{fig:cn_co_hist}
\end{figure*}

\tabcolsep=0.15cm
\begin{table}
\centering
    \caption{(CN bright)/CO intensity ratio distributions.}
    \begin{tabular}{ccccc}
    \hline
        & \multicolumn{3}{c|}{(CN bright)/CO ratio} & \\
        Galaxy\textsuperscript{\textit{a}} & Mean\textsuperscript{\textit{b}} & 16\textsuperscript{th}-50\textsuperscript{th}-84th\textsuperscript{th} & Mode\textsuperscript{\textit{c}} & $N$\textsuperscript{\textit{d}} \\
        \hline
        All ULIRGs\textsuperscript{\textit{e}} & $0.17\pm0.01$ & $0.12-0.16-0.22$ & $0.17$ & $127$ \\
        \hline
        IRAS 13120-5453 & $0.17\pm0.01$ & $0.12-0.16-0.23$ & $0.12$ & $53$ \\
        Arp 220 & $0.22\pm0.02$ & $0.07-0.14-0.42$ & $0.06$ & $114$ \\
        IRAS F05189-2524 & $0.21\pm0.01$ & $0.17-0.21-0.24$ & $0.18$ & $16$ \\
        IRAS F10565+2448 & $0.15\pm0.01$ & $0.1-0.15-0.19$ & $0.18$ & $58$ \\
        \hline
        All LIRGs & $0.12\pm0.01$ & $0.04-0.09-0.19$ & $ 0.04$ & $623$ \\
        \hline
        NGC 6240 & $0.15\pm0.02$ & $0.1-0.12-0.14$ & $0.09$ & $27$ \\
        IRAS F18293-3413 & $0.11\pm0.01$ & $0.09-0.11-0.14$ & $0.09$ & $53$ \\
        NGC 3256 & $0.05\pm0.01$ & $0.03-0.04-0.06$ & $0.03$ & $175$ \\
        NGC 1614 & $0.19\pm0.01$ & $0.14-0.17-0.25$ & $0.15$ & $77$ \\
        NGC 7469 & $0.25\pm0.01$ & $0.11-0.24-0.38$ & $0.36$ & $67$ \\
        NGC 2623 & $0.13\pm0.01$ & $0.12-0.13-0.14$ & $0.12$ & $13$ \\
        NGC 3110 & $0.07\pm0.01$ & $0.06-0.07-0.1$ & $0.06$ & $19$ \\
        ESO 320-G030 & $0.24\pm0.03$ & $0.12-0.19-0.38$ & $0.09$ & $32$ \\
        NGC 1068 & $0.08\pm0.01$ & $0.04-0.06-0.11$ & $0.03$ & $103$ \\
        NGC 5104 & $0.11\pm0.01$ & $0.09-0.11-0.13$ & $0.09$ & $4$ \\
        NGC 4418 & $0.09\pm0.01$ & $0.05-0.08-0.11$ & $0.06$ & $17$ \\
        NGC 1365 & $0.05\pm0.01$ & $0.04-0.06-0.07$ & $0.03$ & $36$ \\
    \hline
    \end{tabular}
    \label{tab:CN_CO_histogram_values}
    \begin{tablenotes}
        \item \textit{Notes:} \textsuperscript{\textit{a}}The galaxies are listed in order of decreasing infrared luminosity.
        \item \textsuperscript{\textit{b}}The uncertainty on the mean was calculated using jackknife resampling.
        \item \textsuperscript{\textit{c}}The mode is the most common value in each histogram when dispersed in 50 bins ranging from 0.0 to 1.5.
        \item \textsuperscript{\textit{d}}The number of pixels that are detected in all three spectral features, CN bright ($>6\sigma$), CN faint ($>3\sigma$), and CO.
        \item \textsuperscript{\textit{e}}Arp 220 is not included in the calculation of the "All ULIRGs" values because the individual scatter points are not trustworthy.
    \end{tablenotes}
\end{table}

\begin{figure*}
	\includegraphics[width=\textwidth]{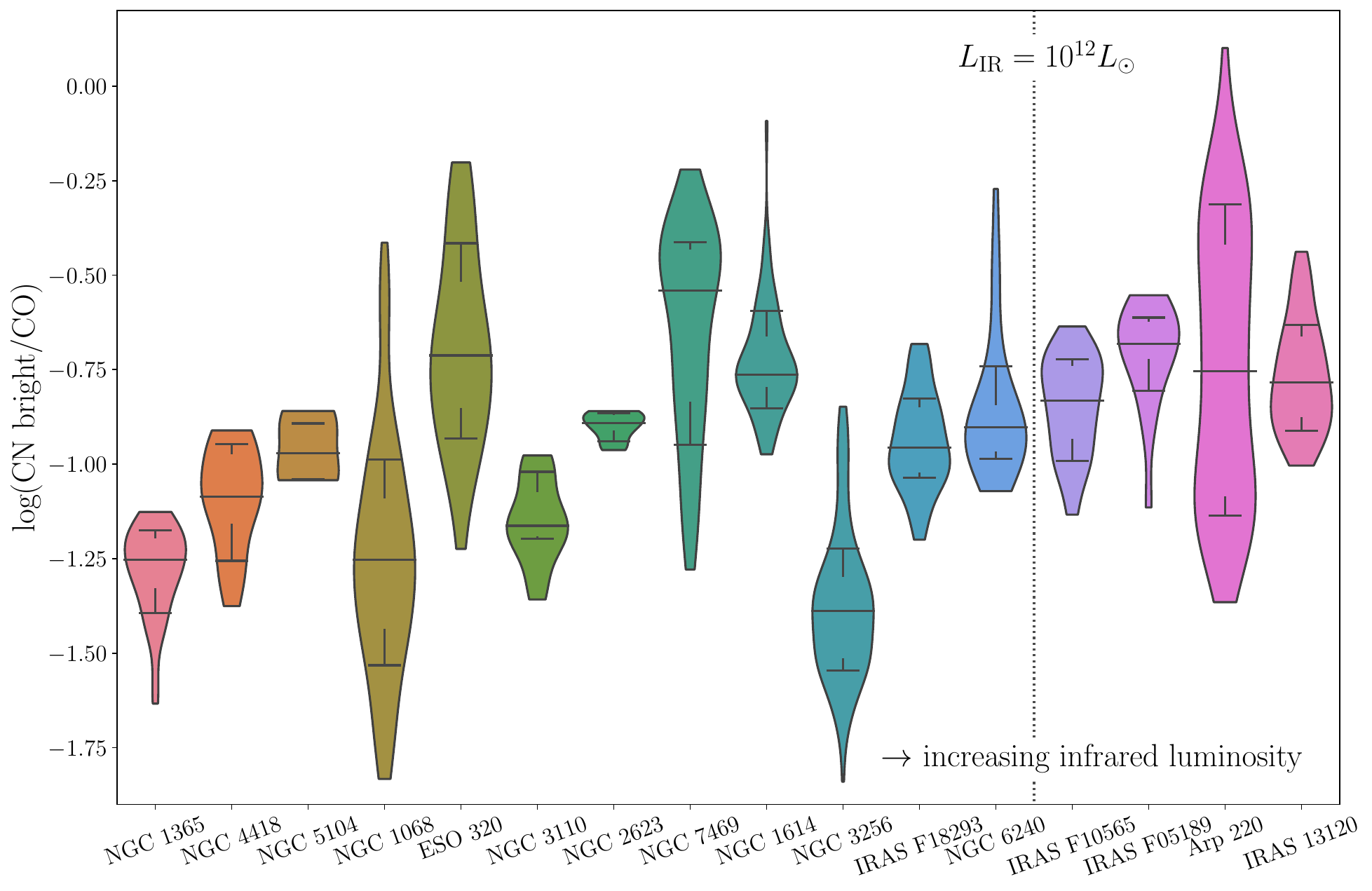}
    \caption{This figure shows the log-scale histograms of the CN bright/CO ratio as violin plots for each galaxy in our sample. There is a large spread in the (CN bright)/CO intensity ratio in our sample of LIRGs due to varying galaxy morphologies and types. The (CN bright)/CO intensity ratio is higher and has less spread on average in our ULIRG sample compared to the LIRGs. The violin plots are a visual way of representing the area-weighted averages from Table \ref{tab:ratios}. The galaxies are organized in increasing order of infrared luminosity from left to right. The black lines correspond to the 16\textsuperscript{th}, 50\textsuperscript{th}, and 84\textsuperscript{th} percentiles. Only pixels detected in the CO, CN bright $>6\sigma$, and CN faint $>3\sigma$ lines are considered. Arp 220 is included but it should be noted that the individual scatter points are less trustworthy. The colours for individual galaxies match those in Figures \ref{fig:cn_optical_depth} and \ref{fig:ratio_comparison}.}
    \label{fig:cn_co_ratio_violin_all}
\end{figure*}

\subsubsection{The (CN bright)/CO intensity ratio}
\label{subsubsec:cn_bright_co_ratio}

Using the integrated global stacked spectra, we measure (CN bright)/CO ratios ranging from $0.02-0.17$ in our galaxy sample. Using spatial averaging, we measure (CN bright)/CO ratios ranging from $0.047-0.26$. For both methods, we find that ULIRGs have higher (CN bright)/CO ratios than LIRGs. The spatially averaged (CN bright)/CO ratios tend to be higher than the global ratios. We attribute these higher ratios to the $>6\sigma$ S/N cut used for the CN bright line when creating the (CN bright)/CO ratio maps, which removes some of the pixels which have weaker CN bright emission relative to the CO emission. The fact that CN emission is weaker relative to CO was also considered in \cite{Wilson2018}, who used S/N matching between the two lines to avoid misinterpreting the (CN bright)/CO ratio because of the different line strengths and detection effects. We argue that our global ratio is a better representation of the true global (CN bright)/CO ratio than the spatially averaged ratios in our galaxies.

Figure \ref{fig:cn_co_hist} shows histograms of the (CN bright)/CO ratio after grouping the galaxies into ULIRG and LIRG samples. Table \ref{tab:CN_CO_histogram_values} provides the mean, mode, and 16th, 50th and 84th quartiles of the histogram values for each galaxy and the combined values for the LIRG and ULIRG distributions (measured without Arp 220). By showing the distribution of pixel values rather than a single global ratio value for each galaxy, Figure \ref{fig:cn_co_hist} demonstrates the complicated nature of the (CN bright)/CO ratio in galaxies. We conclude that the (CN bright)/CO ratio tends to be higher on average in ULIRGs (median value of 0.16) compared to LIRGs (median value of 0.09); however, the spread in the ratio values from the $16$\textsuperscript{th}-$84$\textsuperscript{th} percentiles is significantly larger in LIRGs ($0.04-0.19$) compared to ULIRGs ($0.12-0.22$).

Figure \ref{fig:cn_co_ratio_violin_all} shows the (CN bright)/CO ratio histogram distributions for the individual galaxies in our sample, organized by increasing infrared luminosity. Figure \ref{fig:cn_co_ratio_violin_all} demonstrates that the ULIRGs have more compact distributions than the LIRGs. The larger scatter in the LIRGs can be attributed to differences in galaxy morphologies, less extreme ratio conditions found in galaxy disks which we can identify in certain systems, and the presence of a starburst/AGN in some galaxies. LIRGs with a smaller range of (CN bright)/CO ratios are found to have accompanying compact ratio maps (compact in an absolute physical sense relative to the matched 500 pc resolution; Appendix \ref{append:galaxy_images}). 3 of our 4 ULIRGs (IRAS 13120, IRAS F05189, and IRAS F10565) also have compact ratio maps.

%\color{red}{In the compact U/LIRGs, we would expect $\alpha$\textsubscript{CO} to be lower than the value in normal spirals and the Milky Way across the entire galaxy, indicating that a lower H\textsubscript{2} column density would be sufficient to give our measured CO intensities. The measured ratios can therefore be partially accounted for by the anomalously bright CN emission we observe in these systems. For example, the brightest ULIRG in our sample, IRAS 13120, has a CN luminosity which is nearly double the next brightest galaxy in our sample, Arp 220 (Table \ref{tab:fluxes}).}\color{black}{}

Our galaxy sample is quite heterogeneous, with many galaxies in various merger stages and hosting AGN, starbursts, or some combination of the two. There is a range of $\sim1.5$ dex in infrared luminosity between our galaxies, and they span distances between $10 - 200$ Mpc. As such, it can be complicated to disentangle the physical origins of the variations observed in our (CN bright)/CO intensity ratios (Table \ref{tab:ratios}). We plan to explore the variations in our observed (CN bright)/CO intensity ratio as a function of e.g., infrared luminosity, C[II] luminosity, merger stage, AGN fraction, in future work (Ledger et al. \textit{in prep.}).

\subsubsection{(CN bright)/CO ratios in previous observations of comparable galaxies}
\label{subsubsec:cn_co_prev_obs}

In general, our measured (CN bright)/CO intensity ratios and their variations agree with those previously observed in U/LIRG or starburst systems. \cite{Meier2015} measured a (CN bright)/CO ratio of $0.11$ in the inner nuclear disk of NGC 253 compared to $0.035$ in the outer nuclear disk. This result matches what we have seen for our starburst systems, where (CN bright)/CO appears to be stronger in the nuclear regions of galaxies and decreases in the less extreme disk regions. However, as NGC 253 is not a LIRG, we can only roughly compare the trends of our ratio values with what we might expect in the starburst nucleus of this galaxy. \citet{Wilson2018} measured a global ratio of $0.05$ in NGC 253 at kiloparsec scales, which is likely averaging out the variations seen on resolved scales.

\cite{Wilson2018} carried out a study of the (CN bright)/CO intensity ratio using Cycle 0 ALMA data in a sample of eight galaxies with starbursts and AGN, including four LIRGs and one ULIRG. The four LIRGs (AM 2246-490, NGC 3256, VV 114 and AM 1300-233) have (CN bright)/CO ratios ranging from $0.02 - 0.14$ when using S/N matching between the lines\footnote{The S/N matching method used in \citet{Wilson2018} involved iterating on different S/N cuts for the CN and CO lines to minimize the impact of the different line strengths on the measured line ratios. For exact details on the S/N matching performed, we refer the reader to \citet{Wilson2018}.}. This wide range of values is comparable with the global ratio range we find for our LIRG sample, which spans from $0.02$ to $0.15$. The one shared LIRG between our samples is NGC 3256, and we both find a global (CN bright)/CO ratio of $\sim 0.02$. The ULIRG (AM 2055-425) has a ratio of $0.07$, which is comparable to the global ratios found in Arp 220 ($0.09\pm0.02$), IRAS F05189-2524 ($0.10\pm0.04$), and IRAS F10565+2448 ($0.08\pm0.02$). Our work builds on the observed CN/CO ratio study performed in \citet{Wilson2018} by adding four additional ULIRGs and eleven additional LIRGs. Further, we use newer ALMA data (Cycles 1-6) to measure the CN/CO intensity ratios on resolved as well as global scales, and include the CN faint line in our analysis.

\cite{Cicone2020} observed both CN and CO emission in the molecular outflow of Mrk 231. The authors measured the total CN/CO line luminosity ratio to be $\sim0.21$, but found that the ratio is enhanced by a factor of $\sim3$ to $0.70$ and $0.9$ in the blue and redshifted line wings of the outflow, respectively. They attribute this enhancement of the CN/CO ratio to stronger UV radiation fields present in the gas, perhaps from the formation of massive stars in the outflow. \citet{Wilson2023} measured the global (CN bright)/CO intensity ratio in a sample of four LIRGs, three of which overlap with our galaxy sample (IRAS 13120, NGC 7469, and NGC 3256). The global ratios measured by \citet{Wilson2023} are $0.176$, $0.103$, and $0.045$ in IRAS 13120, NGC 7469, and NGC 3256, respectively. Our global ratios of $0.17$ and $0.10$ agree with their results for IRAS 13120 and NGC 7469, but our global ratio in NGC 3256 is lower ($0.02$). However, our spatially averaged (CN bright)/CO intensity ratio of $0.05$ is in agreement for NGC 3256. \cite{Wilson2018}, \cite{Cicone2020} and \cite{Wilson2023} found higher (CN bright)/CO ratios toward regions with higher SFRs in U/LIRGs. Future work measuring the star formation rate surface densities in our galaxy sample will allow us to directly compare any trends of CN/CO with SFR and UV radiation field.

\begin{figure*}
	\includegraphics[width=\textwidth]{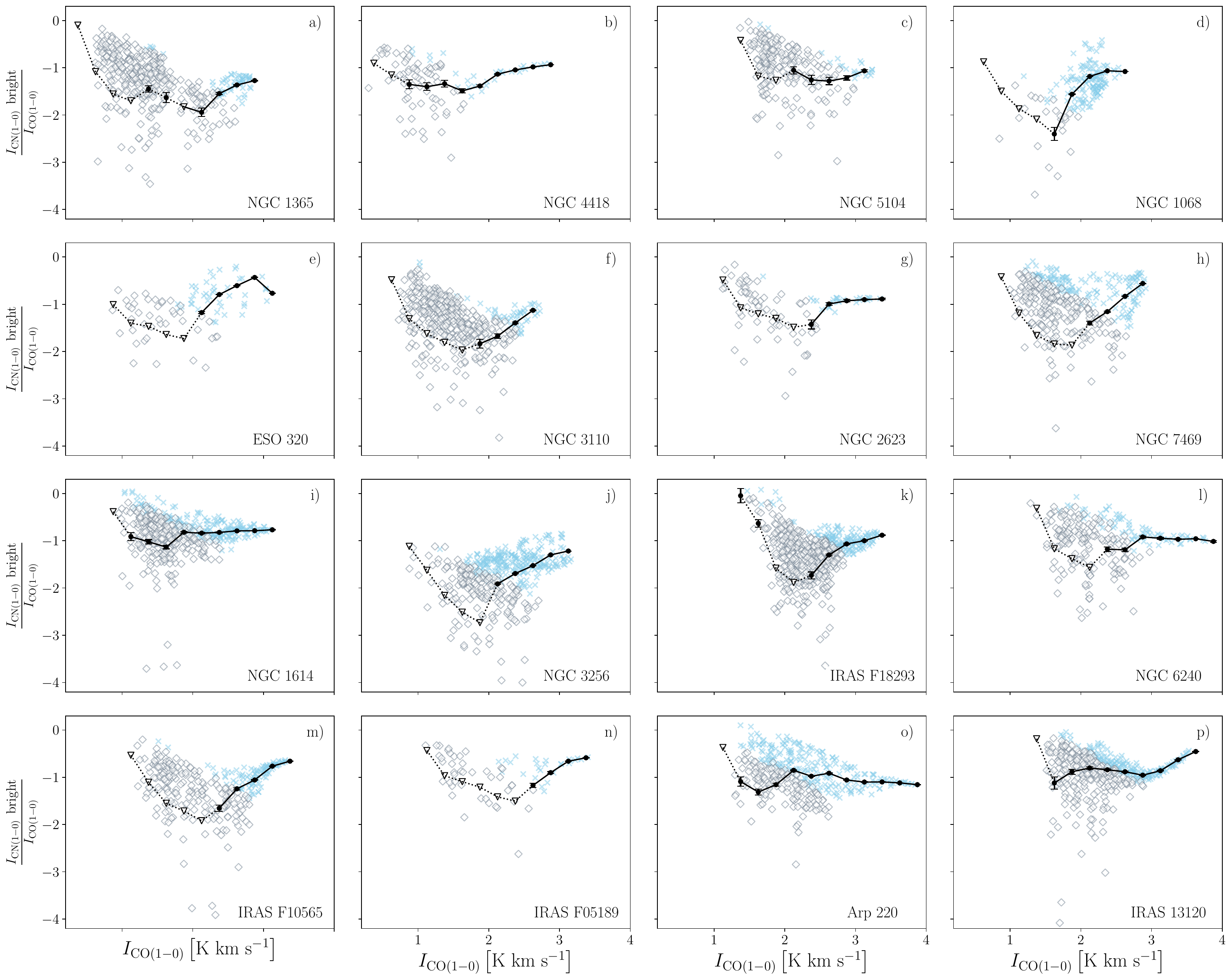}
    \caption{This figure shows the flat or gently increasing trend in the log-scale (CN bright)/CO intensity ratios with CO intensity (K km s$^{-1}$ units). The large round connected symbols represent the binned values in the CO intensity bins with $>3\sigma$ detections in the CN stacked spectra (see Section \ref{subsubsec:CO_binning_method}). The open triangles represent the CO intensity binned pixels which correspond to $<3\sigma$ detections in the stacked CN bright spectra. The detected bins are connected by solid lines, while the non-detected binned data points are connected by dotted lines. The uncertainties on the binned data points are from Equation \ref{eqn:uncertainty}. The faint blue scattered crosses represent the individual pixels detected in CN with $>3\sigma$, while the grey open diamonds are the pixels with $<3\sigma$.}
    \label{fig:cn_co_bins_all_gals}
\end{figure*}

\subsection{CN emission is stronger in galaxy nuclei than extended disks}
\label{subsec:cn_co_nuclei}

\begin{figure*}
	\includegraphics[width=\textwidth]{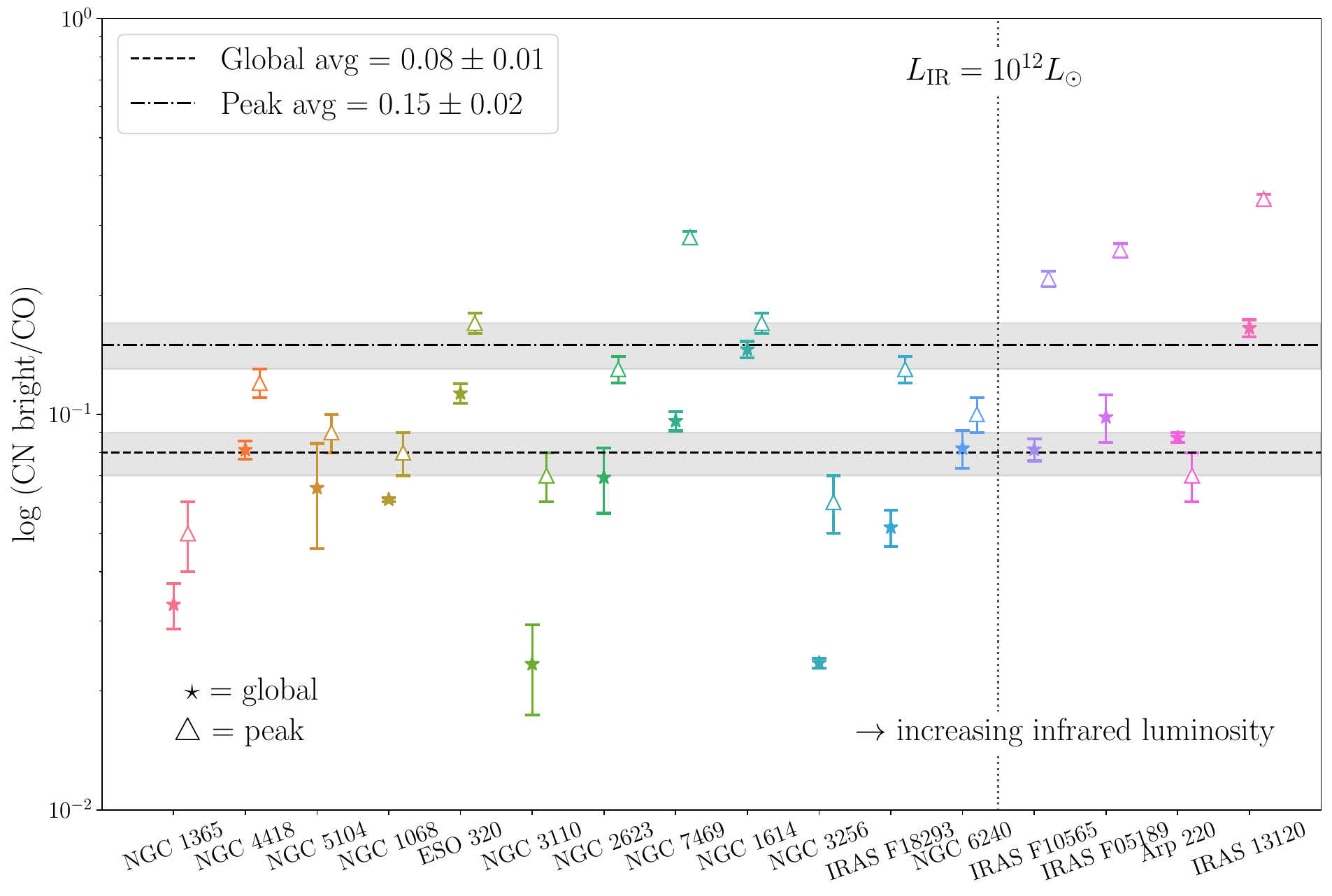}
    \caption{This figure shows the log-scale global and peak intensity (CN bright)/CO ratios in our sample of galaxies. The (CN bright)/CO intensity ratio at the CO intensity peak is higher in each galaxy compared to the global value. The CO intensity peak, and therefore peak ratio, roughly lines up with the nuclear region of each galaxy.The galaxies are organized in increasing order of infrared luminosity from left to right. Closed star symbols give the global intensity ratio as measured from the total CN bright and CO integrated spectra (Table \ref{tab:ratios}). Closed triangles give the peak intensity ratio as measured in the highest CO intensity bin (Section \ref{subsubsec:CO_binning_method}; Figure \ref{fig:cn_co_bins_all_gals}. The dotted blue and dot-dashed indigo lines gives the mean values for the global and peak intensity ratios, respectively. The uncertainties include the 5\% flux uncertainty on Band 3 observations with ALMA.}
    \label{fig:ratio_comparison}
\end{figure*}

We look for resolved variations in the CN/CO intensity ratio in individual galaxies by binning the pixels in each galaxy into 15 CO intensity bins (Section \ref{subsubsec:CO_binning_method}). Figure \ref{fig:cn_co_bins_all_gals} shows a compilation of the (CN bright)/CO intensity ratios versus CO intensity in each galaxy in our sample, distinguishing detected and non-detected pixels using a $>3\sigma$ detection limit for the CN bright line. The CO intensity binning is effectively a rough ``gas surface density binning'', since CO intensity can be converted to a gas surface density (for Figure \ref{fig:cn_co_bins_all_gals}, use an $\alpha$\textsubscript{CO} = 1.088 M\textsubscript{$\odot$} (km s$^{-1}$ pc$^2$)$^{-1}$; \citealt{Bolatto2013}). Binning by CO intensity also mimics a rough radial trend, with the higher intensity bins corresponding to nuclear regions of galaxies and the lower bins being in the extended disk regions. We see that there is a flat or gently rising trend of (CN bright)/CO for higher CO intensity bins which holds for most galaxies in our sample. In some galaxies, particularly the ULIRGs, we see a factor of 2 or 3 increase in the (CN bright)/CO intensity ratio from the lowest CO intensity bin to the highest.

The trends seen in Figure \ref{fig:cn_co_bins_all_gals} can be compared to the (CN bright)/CO global and peak ratios in Table \ref{tab:ratios}. The peak ratio represents the (CN bright)/CO ratio in the pixels found in the highest CO intensity bin, which roughly corresponds to the nuclear region of each galaxy. For most galaxies which exhibit higher peak ratios than global ratios, the (CN bright)/CO ratio gently rises with increasing CO intensity. When we compare these galaxies with the corresponding (CN bright)/CO ratio maps, we see that CN bright and CO are both spatially extended. The extended emission from both lines indicates that we are able to compare the ratio in the nuclear region of the galaxy to the extended disk, where the CN bright emission is weaker and the (CN bright)/CO ratio is lower. Galaxies that have (CN bright)/CO global ratios comparable with their peak ratios (e.g., NGC 1614 and NGC 5104), are those where the CN bright emission is compact but the CO is extended. This trend may indicate that the galaxies have a compact, dense core and CN bright emission is concentrated in the galaxy nucleus and therefore we do not resolve any trends with CO intensity.

Figure \ref{fig:ratio_comparison} summarizes our comparison of the (CN bright)/CO ratio in the peak CO intensity bin compared to the global value in our entire galaxy sample. For every galaxy in our sample except Arp 220, the global (CN bright)/CO intensity ratio is lower than the ratio measured in the peak CO intensity bin. The average global (CN bright)/CO intensity ratio in our entire sample is $0.08\pm0.01$, a factor of 2 lower than the peak average of $0.15\pm0.02$.

%Figure \ref{fig:cn_co_bins_example} (left) shows the results of binning in IRAS 13120, distinguishing detected and non-detected pixels using a $>3\sigma$ detection limit for CN bright. The pixels in IRAS 13120 span 10 CO intensity bins, with an increase in the (CN bright)/CO intensity ratio for higher CO intensities. The right side of Figure \ref{fig:cn_co_bins_example} compares binned (CN bright)/CO intensity ratio for all galaxies. We can see that the trend of (CN bright)/CO increasing for higher CO intensity bins holds for most galaxies in our sample. Trends within individual galaxies are shown in Figure \ref{fig:cn_co_bins_all_gals} in Appendix \ref{append:cn_co_bins}. The CO intensity binning is effectively a rough ``gas surface density binning'', since CO intensity can be converted through various factors to a gas surface density \citep{Bolatto2013}. Binning by CO intensity also mimics a rough radial trend, with the higher intensity bins corresponding to nuclear regions of galaxies and the lower bins being in the extended disk regions.

\subsubsection{Does (CN bright)/CO trace the position of a nuclear starburst or AGN?}
\label{subsubsec:cn_co_agn_position}

The trend of increasing (CN bright)/CO intensity ratio in the nuclear regions of our galaxy sample matches the prediction that CN emission should increase in a PDR \citep{Boger2005}. Galactic nuclei of U/LIRGs tend to be starburst dominant \citep{Lonsdale2006}, indicating that most nuclear regions in these galaxies are large PDRs. \citet{Meier2015} also found a higher (CN bright)/CO intensity ratio in the starburst centre of NGC 253 compared to the surrounding disk. For galaxies in our sample which do not host an AGN, we argue that the increase in the peak (CN bright)/CO ratio is a result of increased starburst-driven PDRs. Future work will compare the (CN bright)/CO intensity ratio maps to maps of surface density of star formation rates, so we can more directly compare the (CN bright)/CO ratio with the physical driver of star formation.

A more complicated influence in the nuclear regions of some of our galaxy sample is the presence of an active galactic nucleus. An AGN will create a localized region of enhanced X-ray emission (an XDR) which will significantly impact the physical and chemical properties of the molecular gas \citep{Meijerink2007}. The typical scale of influence of an AGN is roughly $50-100$ pc (see e.g., \citealt{Izumi2020}). The prediction from XDR models in \citet{Meijerink2007} is that there will be an increased CN/HCN abundance ratio due to the enhanced X-ray emission, and so we might expect a similar (CN bright)/CO intensity ratio increase in an XDR as in a PDR. \citet{Wilson2023} found evidence which supports this claim, with an increased ratio in the centres of two Seyfert nuclei galaxies NGC 7469 and NGC 1808. Additionally, \citet{Saito2022b} show an increasing CN/CO ratio near the AGN in NGC 1068 and the jet-driven molecular outflow. In contrast, \cite{Wilson2018} found that the global (CN bright)/CO ratio decreased in the vicinity of 3 of the 4 AGN in her galaxy sample, which is contrary to XDR model predictions \citep{Meijerink2007}.

In Table \ref{tab:ratios}, we classify each of our 16 galaxies according to the type of AGN documented in the literature. There is no conclusive evidence for an optically identified or obscured AGN in 5 of our 16 galaxies: IRAS F10565, IRAS F18293, NGC 3110, NGC 5104, and ESO 320. Five galaxies in our sample have strong, optically defined AGN: IRAS F05189, NGC 1614, NGC 7469, NGC 2623, and NGC 1068. IRAS F05189 is optically classified as a Seyfert 2 \citep{Smith2019}. NGC 1068 is a Seyfert 2 AGN with a jet-driven molecular outflow \citep{Saito2022a, Saito2022b}. NGC 7469 has a strong nuclear type-1 AGN \citep{Liu2014}, with a sphere of influence of $\sim3$ pc with $M$\textsubscript{BH} = $1.06\times10^{7} M$\textsubscript{$\odot$} \citep{Peterson2014}, and a central XDR \citep{Izumi2020}. NGC 1614 \citep{Konig2013} and NGC 2623 \citep{Aalto2002} are both classified as LINERs.

Six galaxies in our sample have obscured or embedded AGN: IRAS 13120, Arp 220, NGC 6240, the southern nucleus of NGC 3256, NGC 4418, and NGC 1365. IRAS 13120 has been optically classified as a Seyfert 2 AGN \citep{Veron2010}; however, it is likely that the AGN is inactive, heavily obscured, and Compton-thick ($N$\textsubscript{H}$>10^{24}$ cm$^{-2}$, \citealt{Teng2015}). The two nuclei of Arp 220 are heavily obscured \citep{Sakamoto2017, Scoville2017}, but the presence of an AGN has been inferred from X-ray \citep{Paggi2017} and gamma-ray \citep{Yoast2017} observations. NGC 6240 is heavily obscured \citep{Iwasawa2011} and has two separated AGNs \citep{Saito2018b}. The southern nucleus of NGC 3256 has an embedded, dormant AGN \citep{Sakamoto2014}, with evidence from both infrared and X-ray emission \citep{Ohyama2015} and observations of a jet-driven outflow \citep{Sakamoto2014, Brunetti2021}. An optical spectroscopic study of NGC 4418 found an enshrouded compact core with no luminous AGN \citep{Ohyama2019}. Finally, \citet{Swain2023} confirmed the presence of an obscured AGN in NGC 1365 using multiwavelength observations.

We find that 10 out of 11 galaxies in our sample that have well-documented AGN show an increase in the peak (CN bright)/CO intensity ratio relative to the global ratio (six of these galaxies show a significant increase within our uncertainties). The one galaxy of 11 which does not show an increase is Arp 220, which is thought to host two heavily obscured AGN in its merging nuclei, detected in X-rays \citep{Paggi2017} and gamma rays \citep{Yoast2017}. The high column densities seen in this system \citep{Sakamoto2017, Scoville2017} combined with our inability to resolve the two individual nuclear centres make this a challenging result to interpret in the general context of the impact of AGN on the (CN bright)/CO ratio.

Our work has offered significant evidence for the enhancement of the (CN bright)/CO intensity ratio in the galaxies with an AGN, which may be related to the increased X-ray emission dominating the nuclear regions of these galaxies. However, for most of our systems, the 500 pc resolution is not sufficient to disentangle competing AGN and starburst effects in the nuclear region and it may be difficult to identify any trends in (CN bright)/CO within an XDR. More highly resolved studies of individual systems with known AGN are required to conclusively compare our results with models of XDRs and PDRs in U/LIRGs. Future line ratio studies on 50-100 pc scales near known AGN would be beneficial for a more direct comparison with XDR models.

\section{Conclusions}
\label{sec:conclusions}

 We have observationally quantified the CN ($N = 1-0$) / CO ($J = 1-0$) intensity ratio in a large selection of nearby U/LIRGs using the power of the ALMA archive. We measured the (CN bright)/CO and (CN bright)/(CN faint) intensity ratios in four ULIRGs and twelve LIRGs, matching calibration, imaging, and data analysis techniques between galaxies. We have quantitatively and qualitatively compared our ULIRG and LIRG samples. Our main conclusions are as follows:
\begin{enumerate}
    \item Globally measured ratios using spectral stacking methods offer insight into the effect that spatial averaging has on intensity ratios due to S/N differences. We argue that our shuffle-stack method allows us to recover more weak CN emission and better recover the ``true'' CN/CO intensity ratio in our galaxies.
    \item The (CN bright)/(CN faint) intensity ratio is higher in LIRGs compared to ULIRGs. Converting this ratio into a CN optical depth indicates that CN is more optically thick in ULIRGs than LIRGs, although CN is optically thin or moderately thick in most cases. We measure the average optical depth to be $\tau\sim0.96$ in ULIRGs and $\tau\sim0.23$ in LIRGs.
    \item Our measured (CN bright)/CO ratios are higher in ULIRGs than LIRGs. As we have a heterogenous sample of 16 U/LIRG galaxies spanning various merger stages, AGN, and starburst contributions, we are unable to disentangle the exact physical origin of our observed line ratio variations. We plan to explore this physical origin and compare with global galactic properties in forthcoming work (Ledger et al. \textit{in prep.}).
    \item The (CN bright)/CO ratio shows more spread in LIRGs than ULIRGs, with the resolved ratio spanning a $16$\textsuperscript{th}-$84$\textsuperscript{th} percentile range of $0.15$ in LIRGs compared to $0.1$ in ULIRGs. The global ratio values range from $0.02-0.15$ in LIRGs and $0.08-0.17$ in ULIRGs. This difference is likely due to the larger range of galaxy types, morphologies, and components probed in LIRGs compared to the more compact ULIRGs.
    \item In 15 of our 16 galaxies, the (CN bright)/CO intensity ratio is higher in the peak CO intensity bin than the global value (only Arp 220 demonstrates a decrease in the peak ratio, and this may result from untrustworthy pixel effects when binning this galaxy). The average peak ratio is $0.15\pm0.02$, while the average global ratio is $0.08\pm0.01$. We argue that the increase in the (CN bright)/CO intensity ratio is a complicated combination of the presence of starbursts and/or AGN. In particular, 6 out of 11 galaxies which have well-documented AGN show a statistically significant increase in the (CN bright)/CO intensity ratio in the peak CO intensity bin relative to the global emission. Both optically defined AGN and obscured AGN have a similar impact on the ratio. This may be significant evidence that CN emission and/or abundance is enhanced by AGN. Future work on intensity ratios measured at $<100$ pc is necessary to explore the impact an AGN on resolved scales.
\end{enumerate}

We plan to compare our CN/CO intensity ratios star formation rate surface densities derived using radio continuum data in future work. We also plan to compare the global ratios with various parameters compiled in the GOALs survey for each galaxy, e.g., $L$\textsubscript{IR}, merger stage, AGN fraction, and [CII] luminosity (Ledger et al. \textit{in prep.}).

\section*{Acknowledgements}

We thank the anonymous referee for their detailed comments and revisions which improved the quality and content of this work.

This paper makes use of the following ALMA data:

\noindent{ADS/JAO.ALMA\#2012.1.00306.S,}
ADS/JAO.ALMA\#2012.1.00657.S,
ADS/JAO.ALMA\#2013.1.00218.S,
ADS/JAO.ALMA\#2013.1.00991.S,
ADS/JAO.ALMA\#2013.1.01172.S,
ADS/JAO.ALMA\#2015.1.00003.S,
ADS/JAO.ALMA\#2015.1.00167.S,
ADS/JAO.ALMA\#2015.1.00287.S,
ADS/JAO.ALMA\#2015.1.01135.S,
ADS/JAO.ALMA\#2015.1.01191.S,
ADS/JAO.ALMA\#2016.1.00177.S,
ADS/JAO.ALMA\#2016.1.00263.S,
ADS/JAO.ALMA\#2017.1.00078.S,
ADS/JAO.ALMA\#2018.1.00223.S, 
ADS/JAO.ALMA\#2018.1.01684.S, and
ADS/JAO.ALMA\#2019.1.01664.S.

ALMA is a partnership of ESO (representing its member states), NSF (USA), and NINS (Japan), together with NRC (Canada), MOST and ASIAA (Taiwan), and KASI (Republic of Korea), in cooperation with the Republic of Chile. The Joint ALMA Observatory is operated by ESO, AUI/NRAO, and NAOJ. The National Radio Astronomy Observatory is a facility of the National Science Foundation operated under cooperative agreement by Associated Universities, Inc. Any ALMA specification details in this research has made use of G. Privon et al. 2022, ALMA Cycle 9 Proposer’s Guide, ALMA Doc. 9.2 v1.4.

This research has made use of the NASA/IPAC Extragalactic Database (NED) which is operated by the Jet Propulsion Laboratory, California Institute of Technology, under contract with the National Aeronautics and Space Administration. The computing resources available at NAOJ were essential for this project. Data analysis was in part carried out on the Multi-wavelength Data Analysis System (MDAS) operated by the Astronomy Data Center (ADC), National Astronomical Observatory of Japan.

BL acknowledges partial support from an NSERC Canada Graduate Scholarship-Doctoral and an Ontario Graduate Scholarship. BL acknowledges funding for a 10-week Japan Society for the Promotion of Science (JSPS)-Mitacs summer research program which made this work possible. BL would like to thank Drs. Daisuke Iono and Toshiki Saito for hosting him at NAOJ and their guidance during the first months of this project. CDW acknowledges financial support from the Canada Council for the Arts through a Killam Research Fellowship. The research of CDW is supported by grants from the Natural Sciences and Engineering Research Council of Canada (NSERC) and the Canada Research Chairs program.

\textit{Software:} This research has made use of the following software packages: \texttt{ASTROPY}, a community-developed core PYTHON package for astronomy \citep{astropy2013, astropy2018, astropy2022}, \texttt{CASA} \citep{McMullin2007}, \texttt{SciPy} \citep{scipy2020}, \texttt{MATPLOTLIB} \citep{Hunter2007}, \texttt{NUMPY} \citep{harris2020}, and \texttt{SPECTRAL-CUBE} \citep{Ginsburg2019}.

%%%%%%%%%%%%%%%%%%%%%%%%%%%%%%%%%%%%%%%%%%%%%%%%%%
\section*{Data Availability}
The derived data generated in this research will be shared on reasonable request to the corresponding author.



\bibliographystyle{mnras}
%\bibitem[\protect\citeauthoryear{Privon}{2022}]{Privon2022}
%G. Privon et al. 2022, ALMA Cycle 9 Proposer’s Guide, ALMA Doc. 9.2 v1.4; https://almascience.eso.org/documents-and-tools/cycle9/alma-proposers-guide
\bibliography{paper_CO_CN_ratio_biblio.bib} % if your bibtex file is called example.bib

%\begin{thebibliography}{99}
%\end{thebibliography}

%%%%%%%%%%%%%%%%%%%%%%%%%%%%%%%%%%%%%%%%%%%%%%%%%%

%%%%%%%%%%%%%%%%% APPENDICES %%%%%%%%%%%%%%%%%%%%%
\newpage

\appendix
\section{Galaxy images}
\label{append:galaxy_images}
The 4 panel plots of each galaxy are shown in decreasing order of the galaxy's infrared luminosity. The four panels represent: \textit{(a)} the moment 0 map for the CO line; \textit{(b)} the moment 0 map for the CN bright line; \textit{(c)} the (CN bright)/CO intensity ratio map; \textit{(d)} the total integrated spectra for each of the CN bright, CN faint, and CO lines. Refer to the discussion in Section \ref{subsec:data_products} for more details and interpretation of individual plots. The maps included in this appendix section were used when measuring the spatially averaged (CN bright)/CO intensity ratios (Table \ref{tab:ratios}).

\begin{figure*}
	\includegraphics[width=0.9\textwidth]{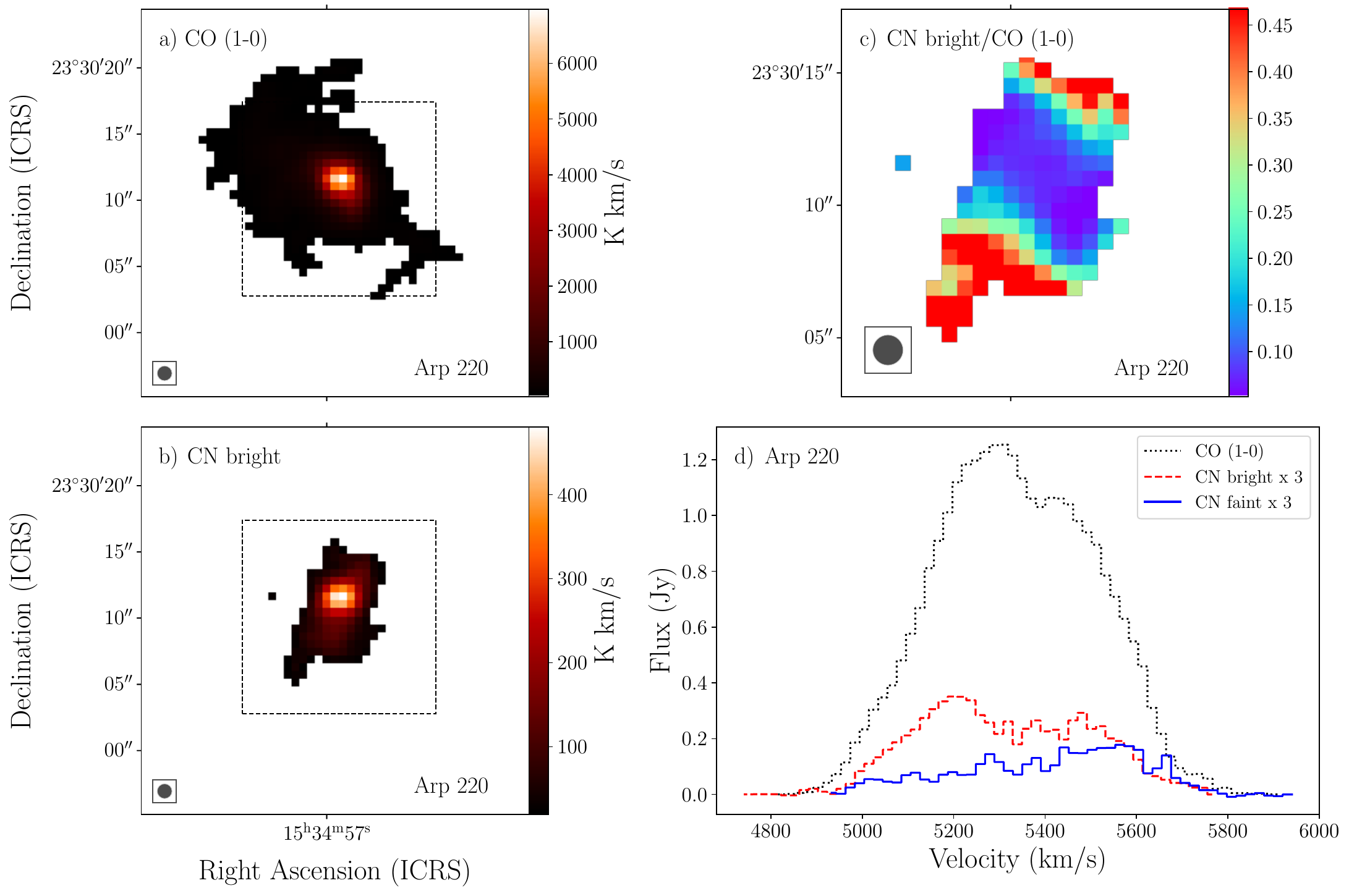}
    \caption{This figure shows an example of the moment 0 maps and (CN bright)/CO intensity ratio in K km s$^{-1}$ units in Arp 220. The circle in the bottom left corner is the size of the beam smoothed to 500 pc. \textit{a)} The total integrated intensity of the CO (1-0) line. The dashed square indicates the region in \textit{(c)}. \textit{b)} The total integrated intensity of the CN bright line. The pixels included here have a S/N of $>6\sigma$ and $>3\sigma$ in the CN bright and CN faint lines, respectively. \textit{c)} The (CN bright)/CO intensity ratio. The colour bar is clipped at the mean value plus or minus 80\%. \textit{d)} The total integrated spectra of the CO line (black dotted), CN bright line (red dashed), and CN faint line (blue). Both CN lines have been multiplied by a factor of 3 for demonstration purposes.}
    \label{fig:arp220_4panel}
\end{figure*}

\begin{figure*}
	\includegraphics[width=0.9\textwidth]{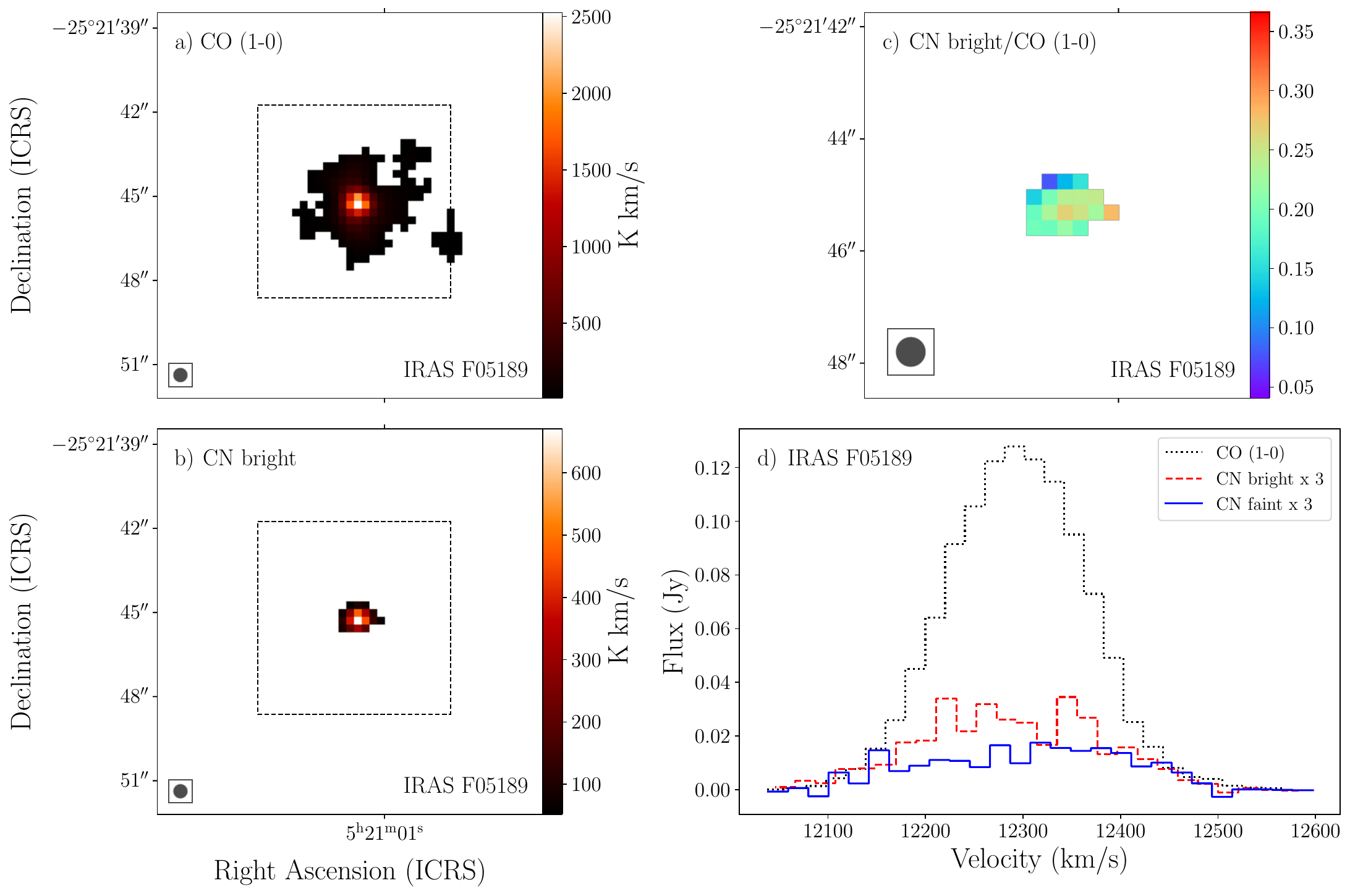}
    \caption{Moment maps, ratio maps, and spectra for IRAS F05189. See Figure \ref{fig:arp220_4panel} for more details.}
    \label{fig:irasf05189_4panel}
\end{figure*}

\begin{figure*}
	\includegraphics[width=0.9\textwidth]{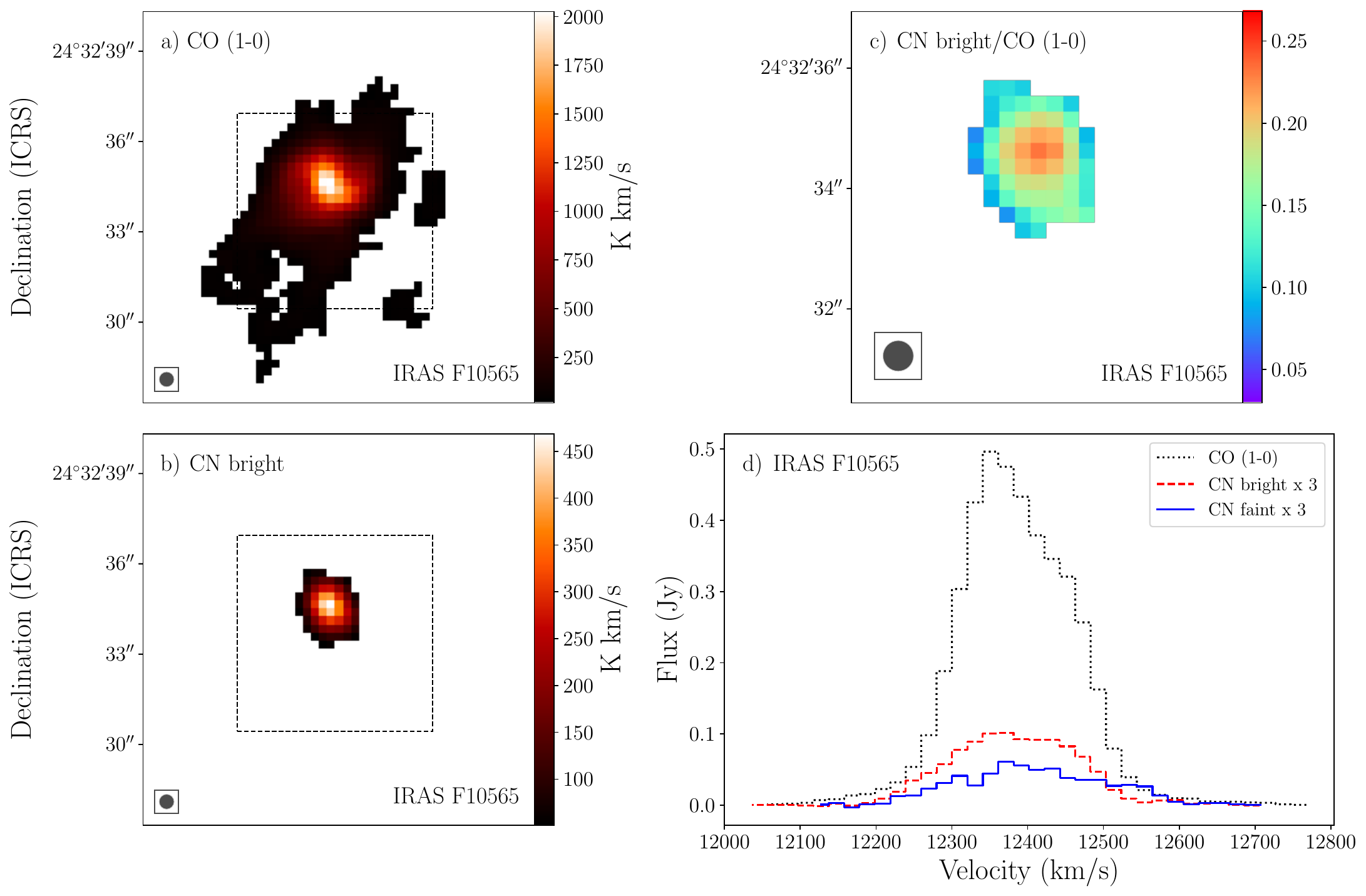}
    \caption{Moment maps, ratio maps, and spectra for IRAS F10565. See Figure \ref{fig:arp220_4panel} for more details.}
    \label{fig:irasf10565_4panel}
\end{figure*}

\begin{figure*}
	\includegraphics[width=0.9\textwidth]{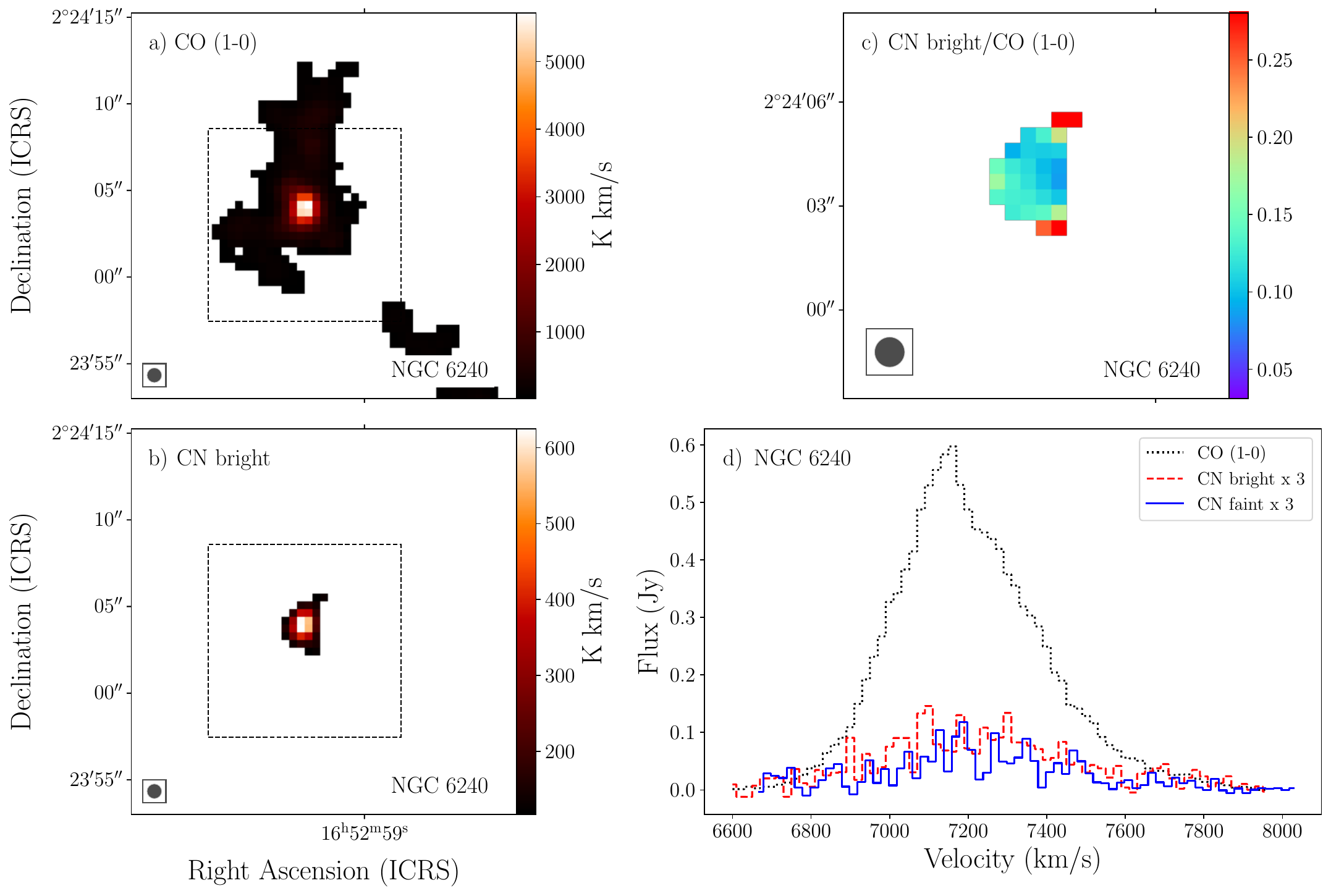}
    \caption{Moment maps, ratio maps, and spectra for NGC 6240. See Figure \ref{fig:arp220_4panel} for more details.}
    \label{fig:ngc6240_4panel}
\end{figure*}

\begin{figure*}
	\includegraphics[width=0.9\textwidth]{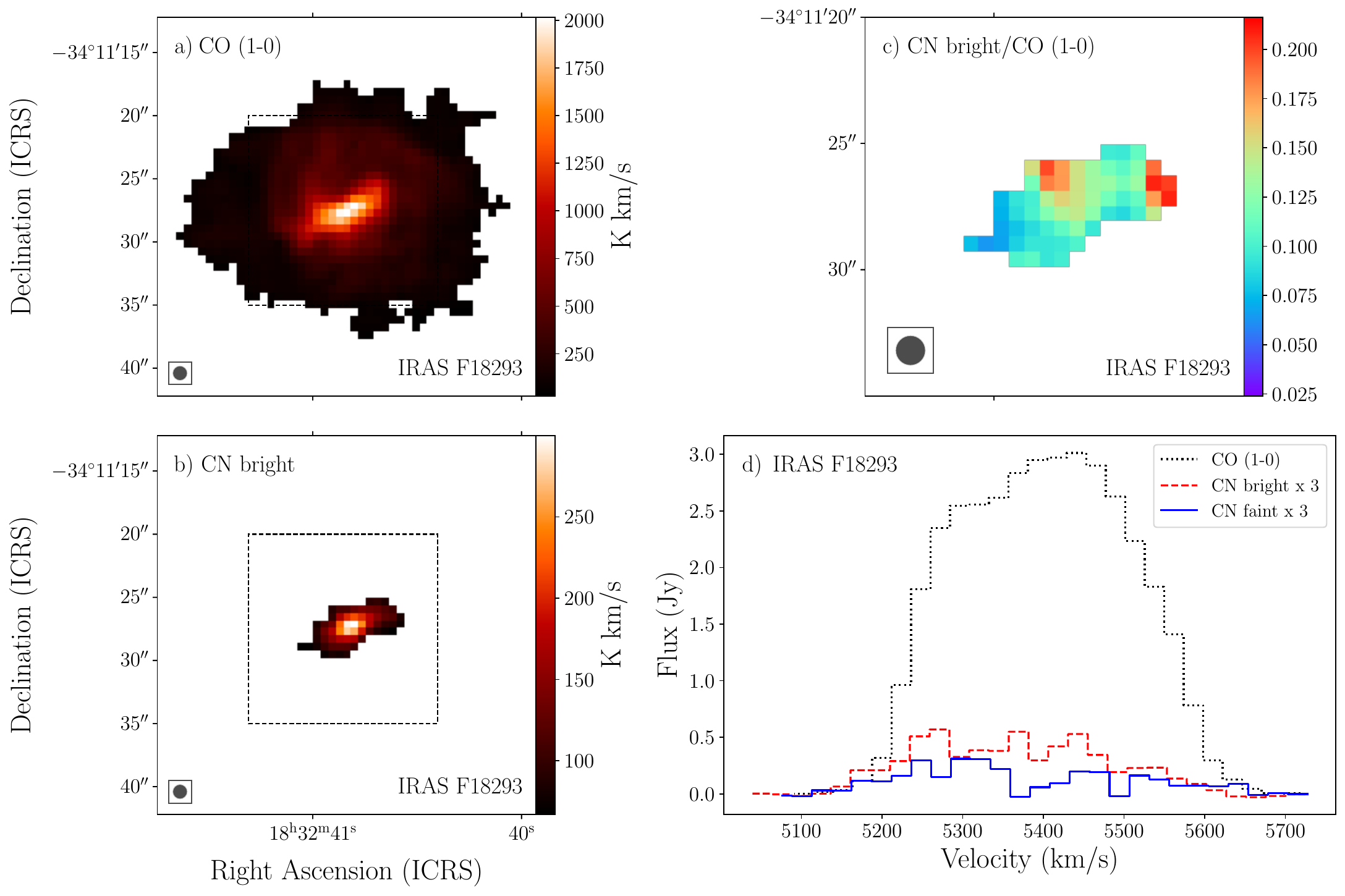}
    \caption{Moment maps, ratio maps, and spectra for IRAS F18293. See Figure \ref{fig:arp220_4panel} for more details.}
    \label{fig:irasf18293_4panel}
\end{figure*}

\begin{figure*}
	\includegraphics[width=0.9\textwidth]{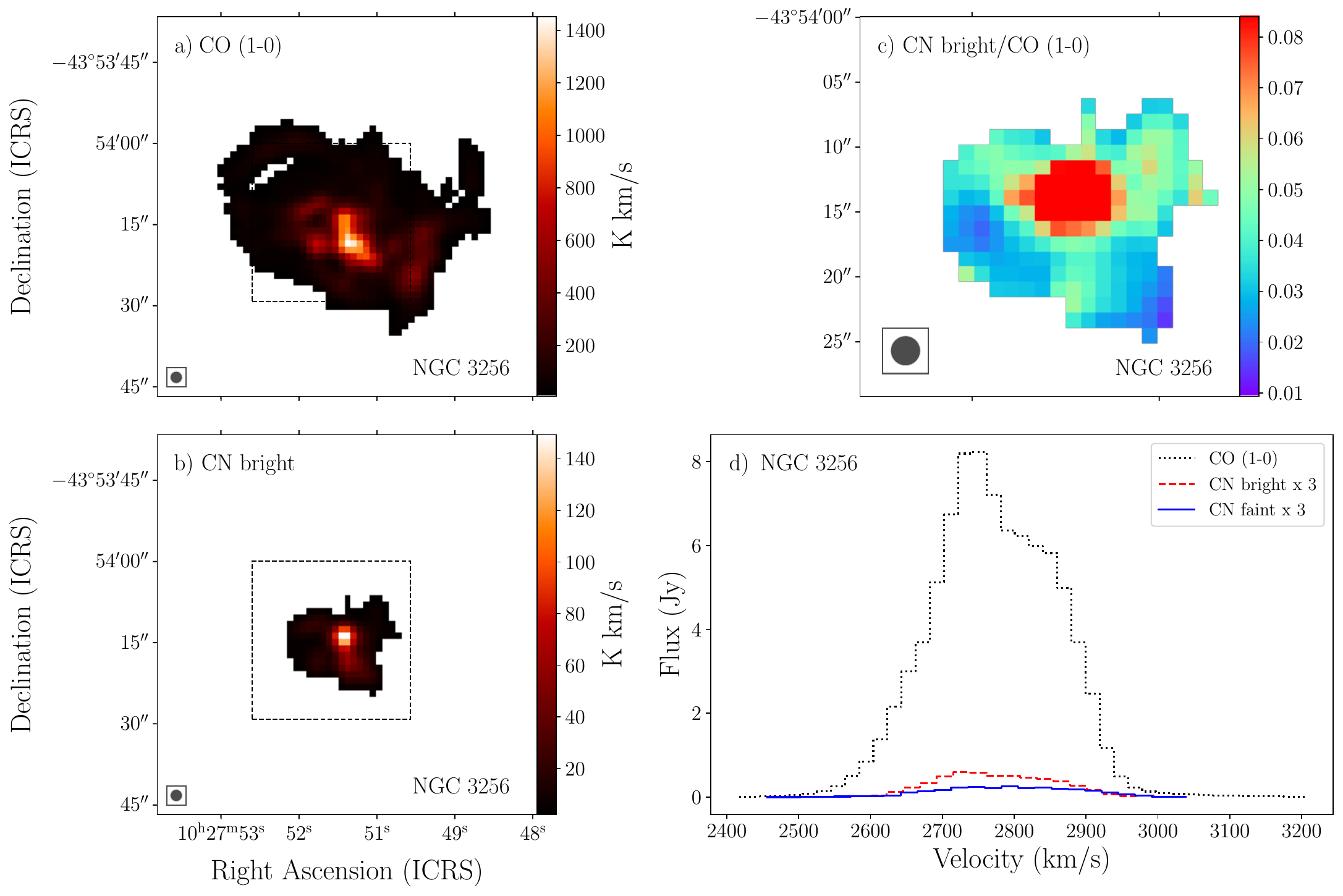}
    \caption{Moment maps, ratio maps, and spectra for NGC 3256. See Figure \ref{fig:arp220_4panel} for more details.}
    \label{fig:ngc3256_4panel}
\end{figure*}

\begin{figure*}
	\includegraphics[width=0.9\textwidth]{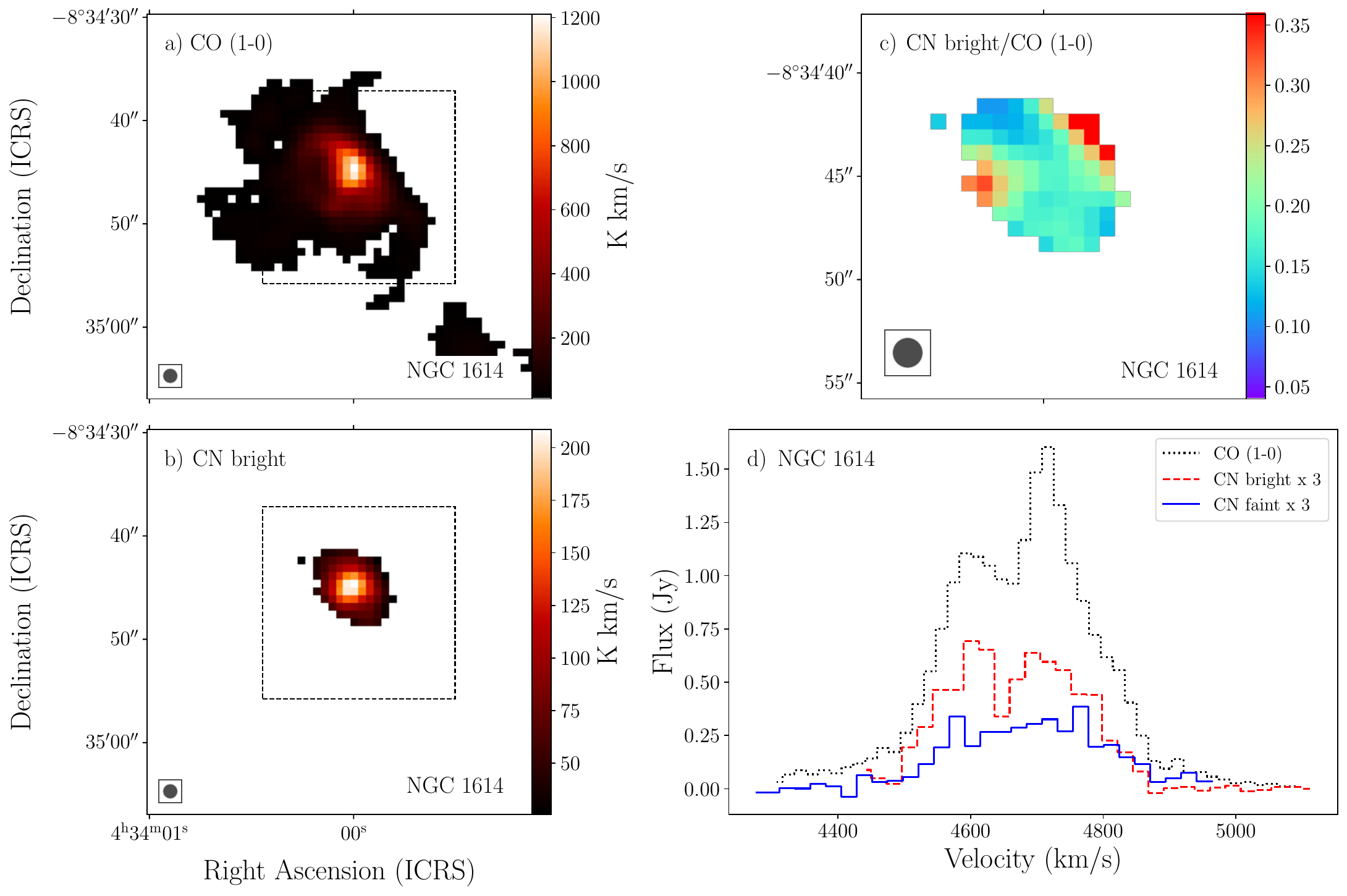}
    \caption{Moment maps, ratio maps, and spectra for NGC 1614. See Figure \ref{fig:arp220_4panel} for more details.}
    \label{fig:ngc1614_4panel}
\end{figure*}

\begin{figure*}
	\includegraphics[width=0.9\textwidth]{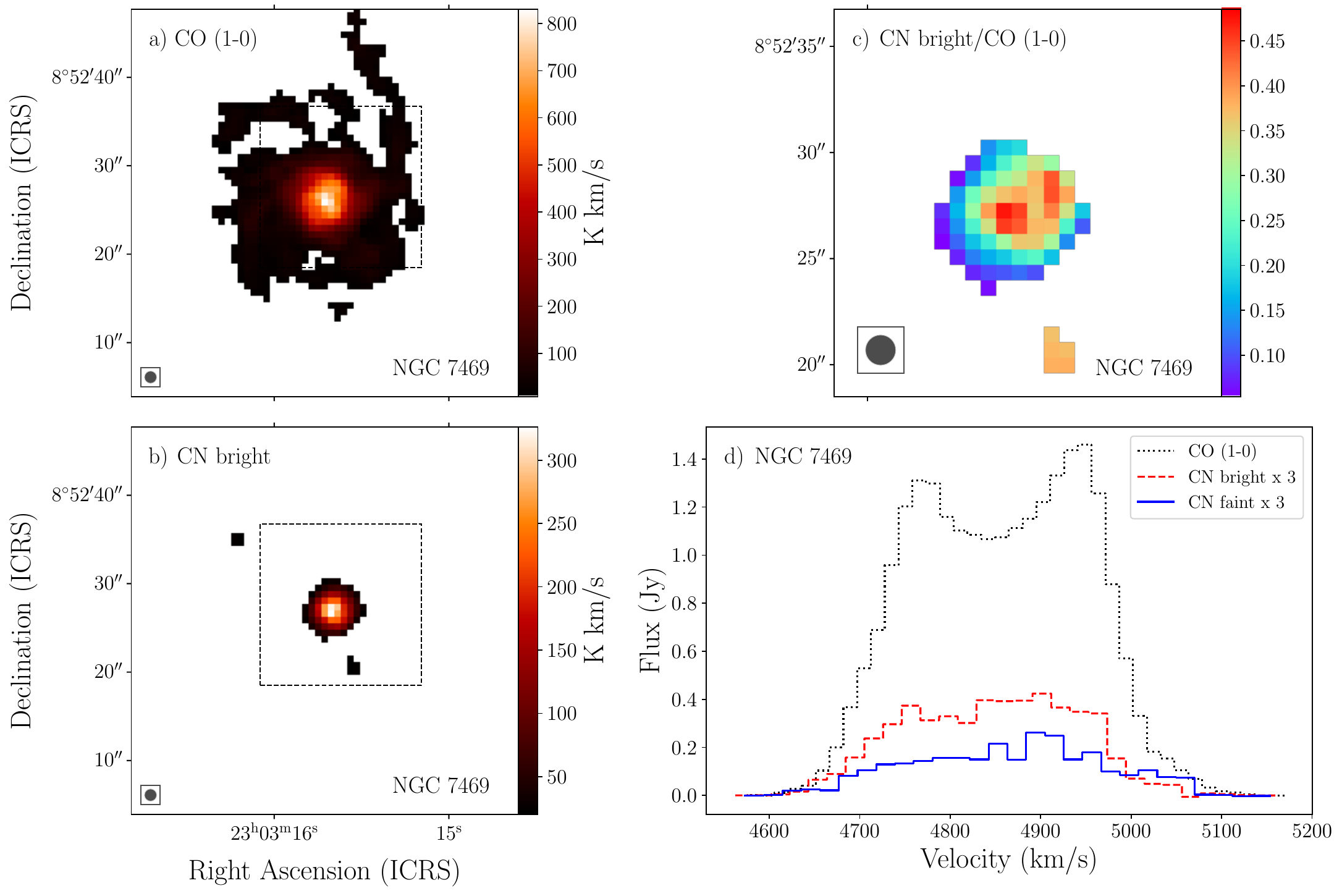}
    \caption{Moment maps, ratio maps, and spectra for NGC 7469. See Figure \ref{fig:arp220_4panel} for more details.}
    \label{fig:ngc7469_4panel}
\end{figure*}

\begin{figure*}
	\includegraphics[width=0.9\textwidth]{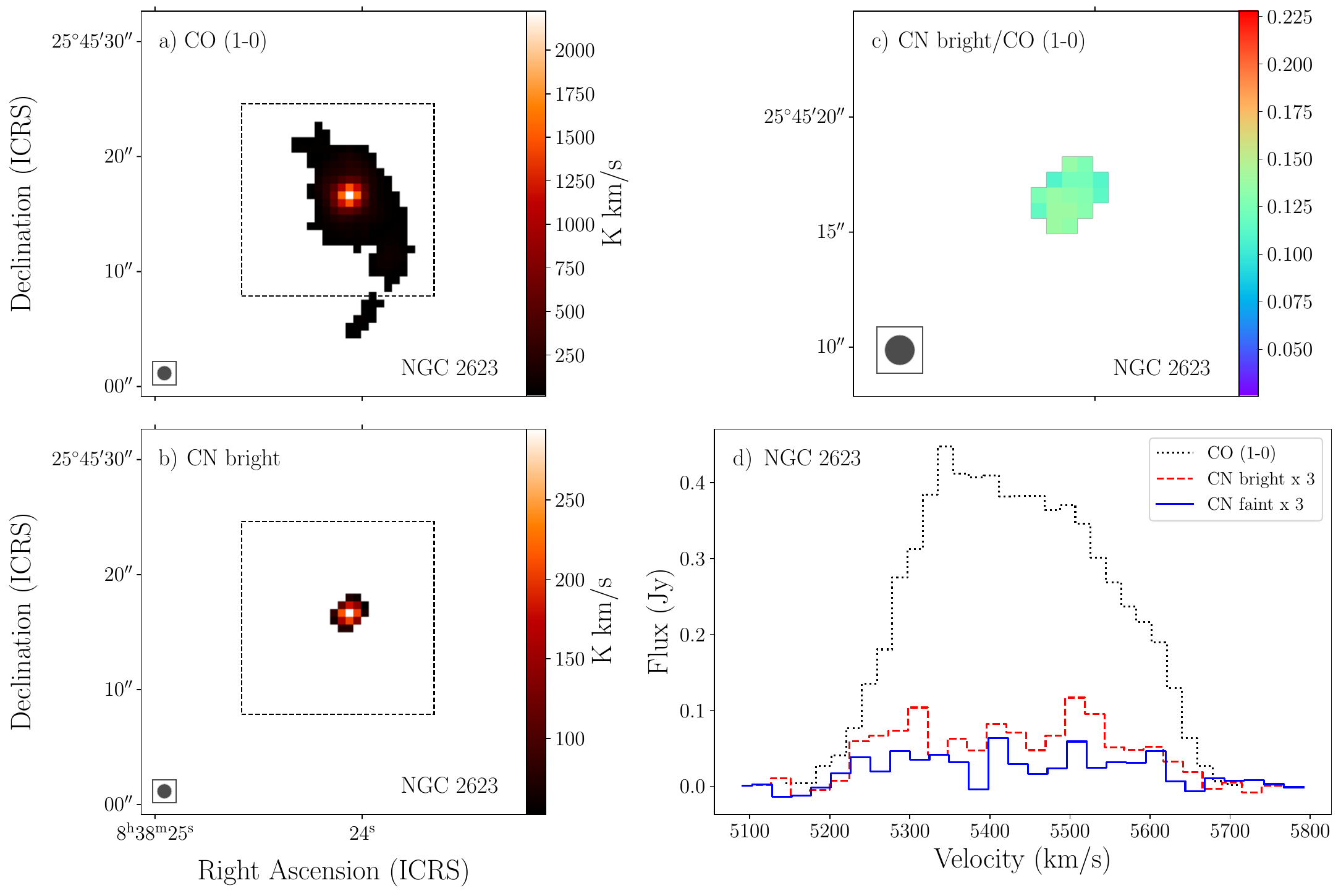}
    \caption{Moment maps, ratio maps, and spectra for NGC 2623. See Figure \ref{fig:arp220_4panel} for more details.}
    \label{fig:ngc2623_4panel}
\end{figure*}

\begin{figure*}
	\includegraphics[width=0.9\textwidth]{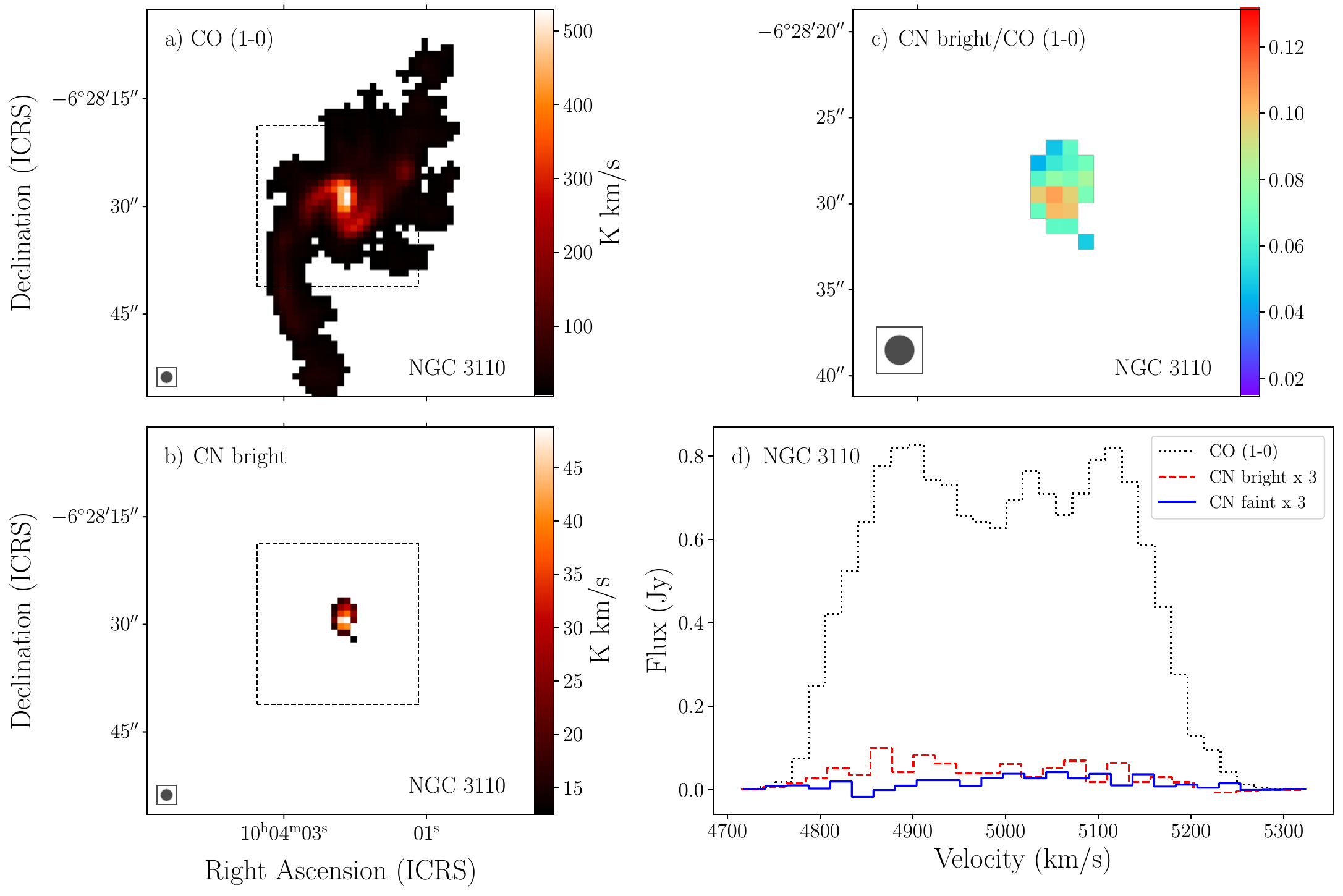}
    \caption{Moment maps, ratio maps, and spectra for NGC 3110. See Figure \ref{fig:arp220_4panel} for more details.}
    \label{fig:ngc3110_4panel}
\end{figure*}

\begin{figure*}
	\includegraphics[width=0.9\textwidth]{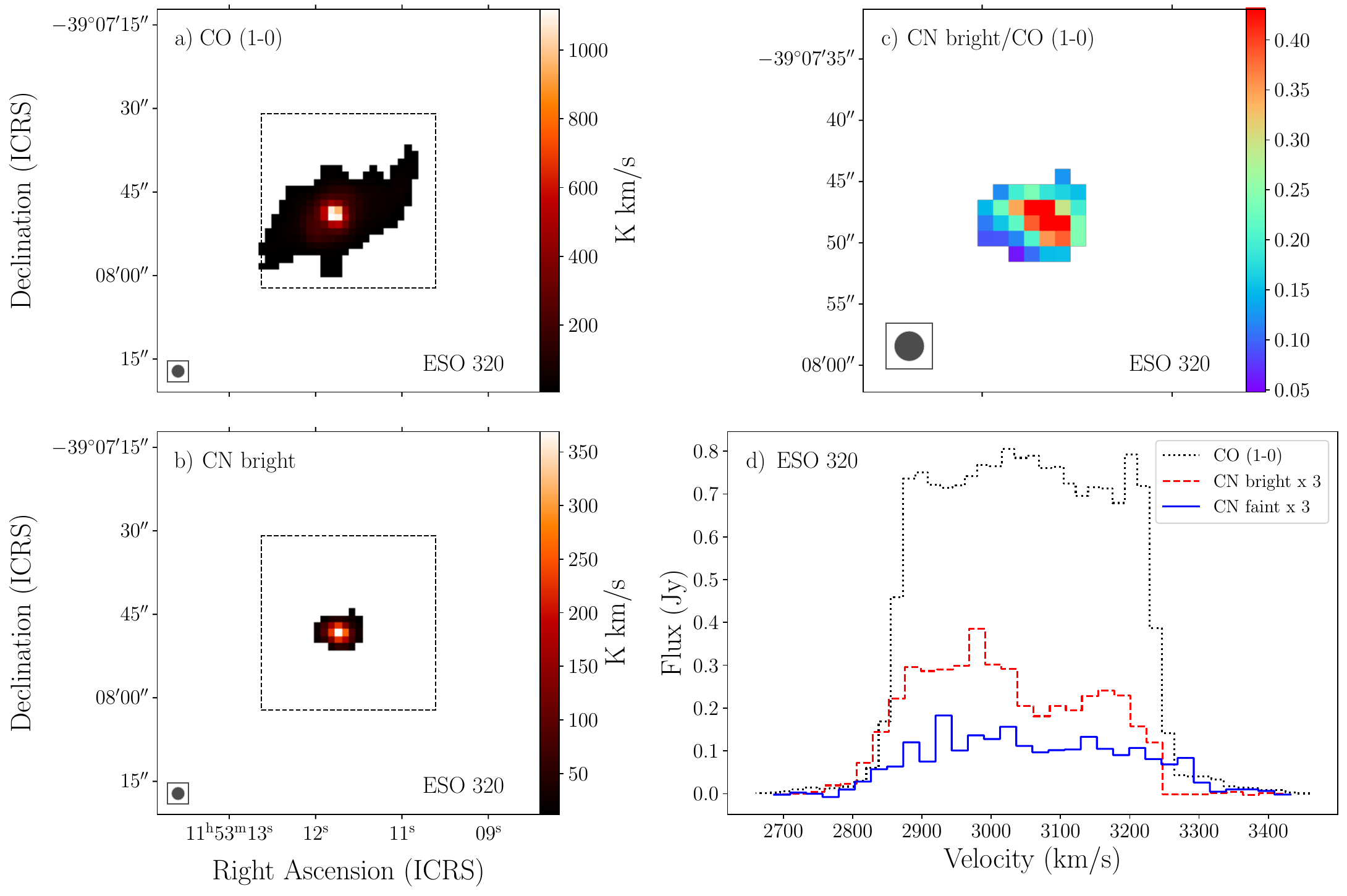}
    \caption{Moment maps, ratio maps, and spectra for ESO 320-G030. See Figure \ref{fig:arp220_4panel} for more details.}
    \label{fig:eso320_4panel}
\end{figure*}

\begin{figure*}
	\includegraphics[width=0.9\textwidth]{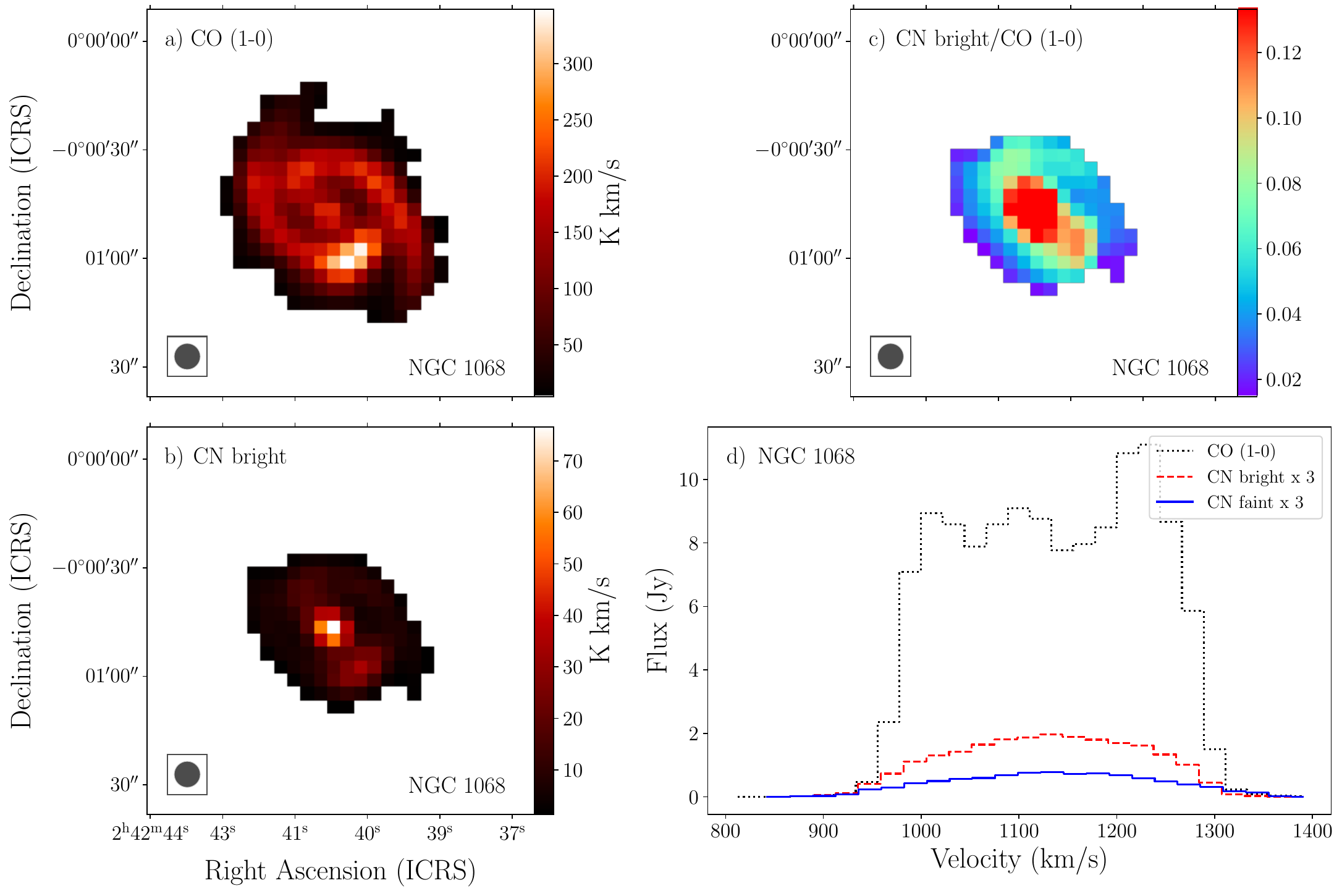}
    \caption{Moment maps, ratio maps, and spectra for NGC 1068. See Figure \ref{fig:arp220_4panel} for more details.}
    \label{fig:ngc1068_4panel}
\end{figure*}

\begin{figure*}
	\includegraphics[width=0.9\textwidth]{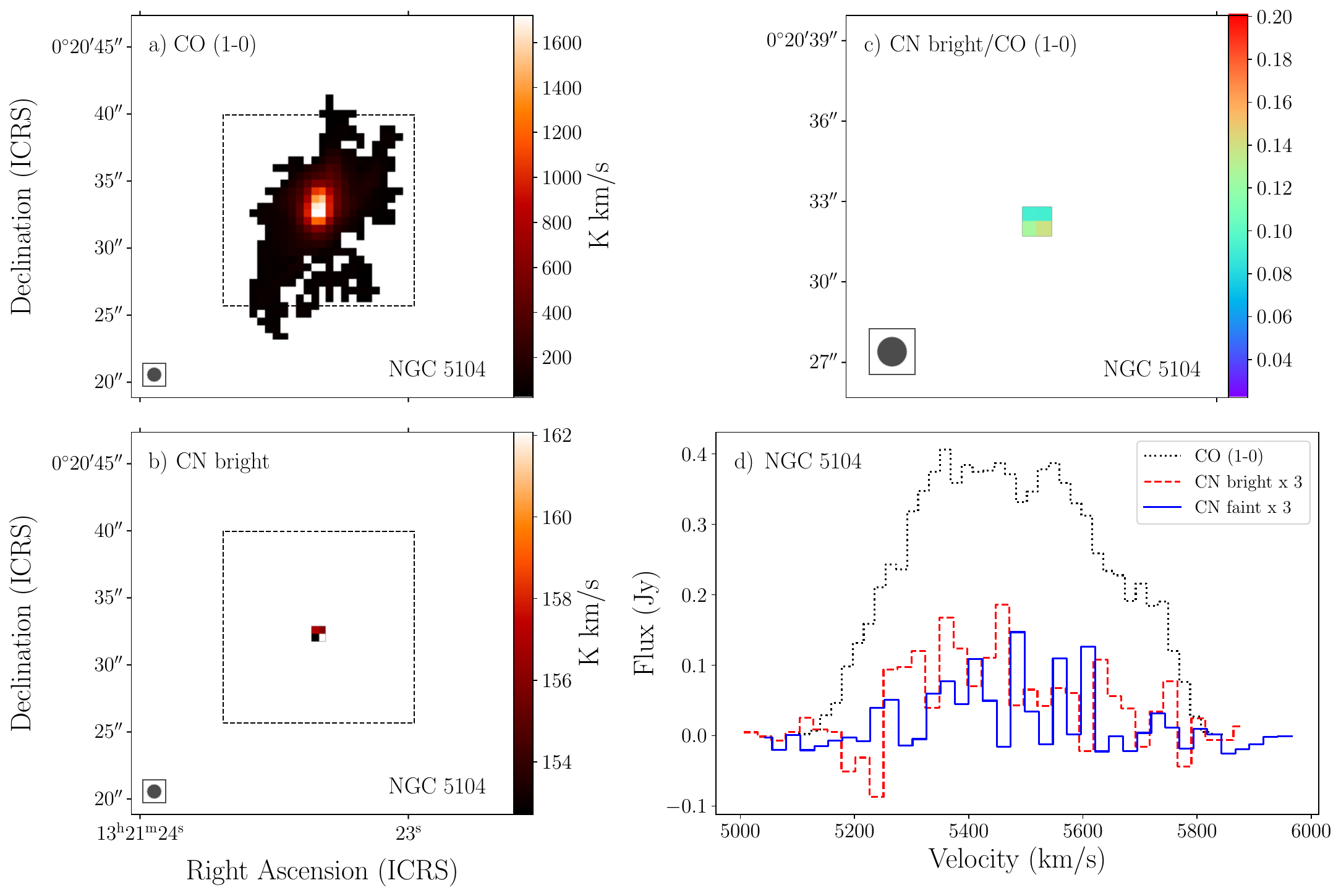}
    \caption{Moment maps, ratio maps, and spectra for NGC 5104. See Figure \ref{fig:arp220_4panel} for more details.}
    \label{fig:ngc5104_4panel}
\end{figure*}

\begin{figure*}
	\includegraphics[width=0.9\textwidth]{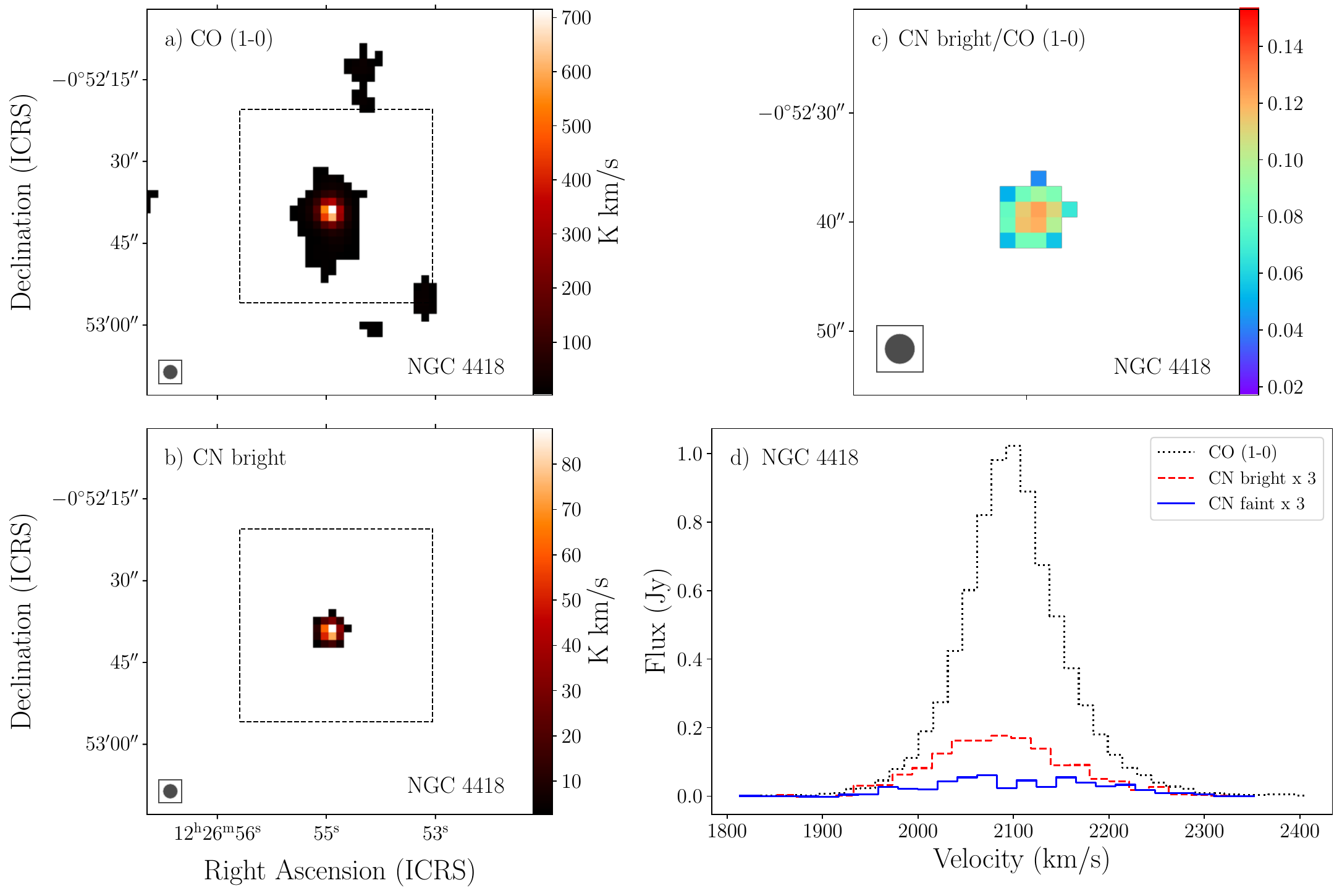}
    \caption{Moment maps, ratio maps, and spectra for NGC 4418. See Figure \ref{fig:arp220_4panel} for more details.}
    \label{fig:ngc4418_4panel}
\end{figure*}

\begin{figure*}
	\includegraphics[width=0.9\textwidth]{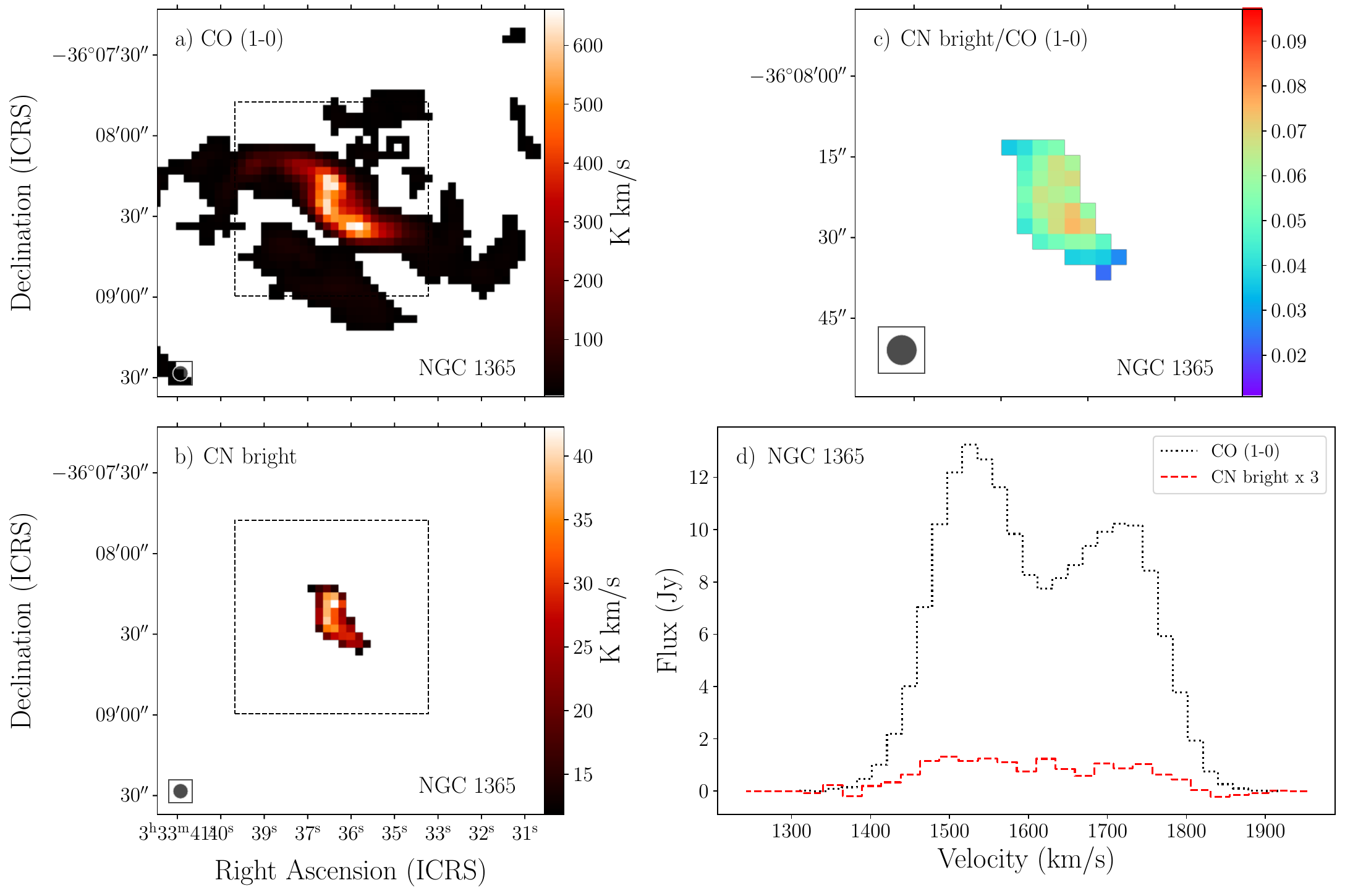}
    \caption{Moment maps, ratio maps, and spectra for NGC 1365. See Figure \ref{fig:arp220_4panel} for more details.}
    \label{fig:ngc1365_4panel}
\end{figure*}

\section{(CN bright)/(CN faint) ratio maps}
\label{append:cn_bright_faint_ratio_maps}
\begin{figure*}
	\includegraphics[width=\textwidth]{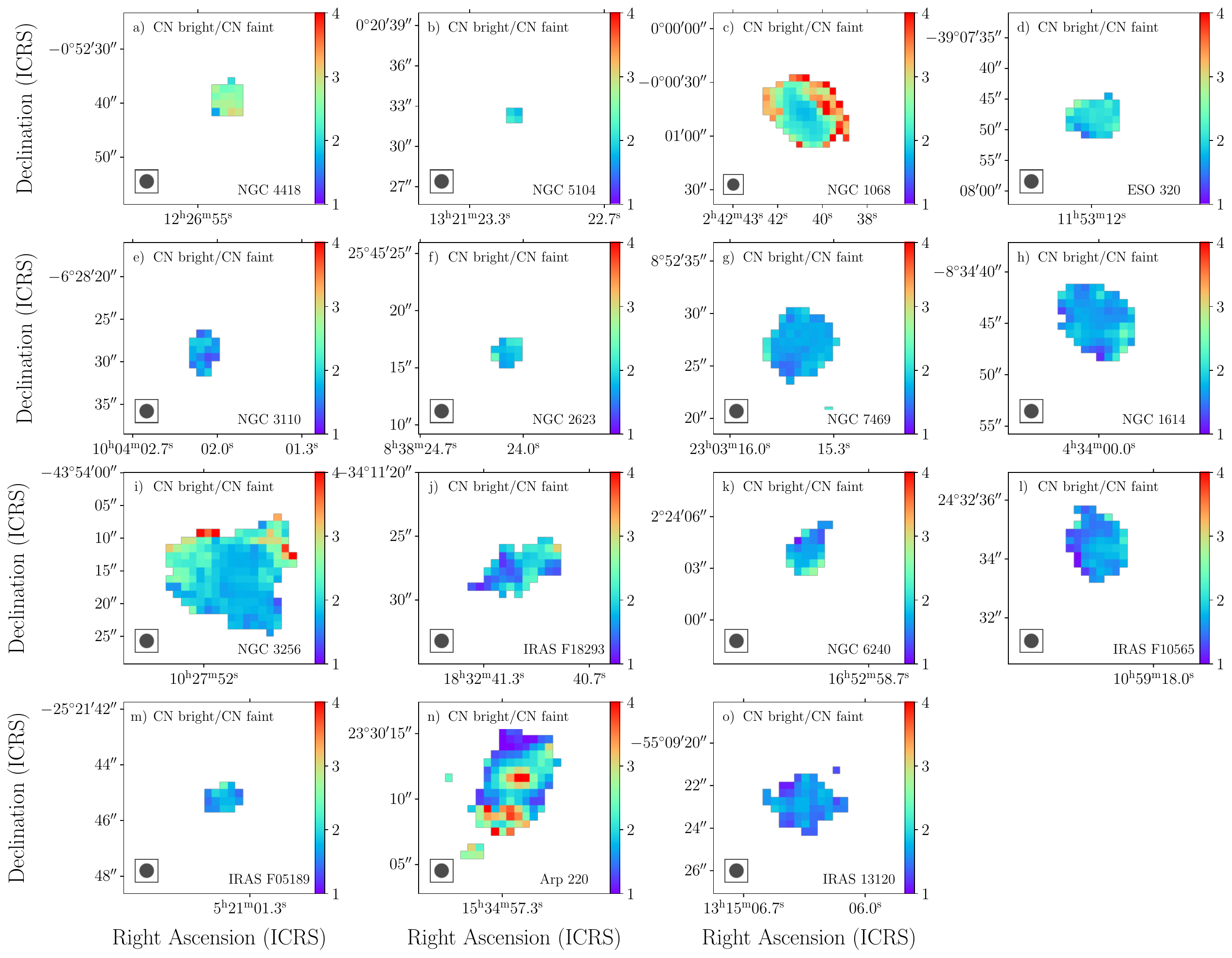}
    \caption{This figure shows the (CN bright)/(CN faint) intensity ratio in K km s$^{-1}$ units. The circle in the bottom left corner is the size of the beam smoothed to 500 pc. The colour bar in each panel is matched for easy comparison between galaxies. The pixels correspond to the CN bright and CN faint lines detected at $>6\sigma$ and $>3\sigma$, respectively. NGC 1365 is not shown because the observed spectral window did not cover the CN faint line.}
    \label{fig:cn_bright_faint_maps}
\end{figure*}

Figure \ref{fig:cn_bright_faint_maps} shows a compilation of the (CN bright)/(CN faint) ratio maps in each galaxy in our sample. Each ratio map includes pixels with $>6\sigma$ and $>3\sigma$ detections in the CN bright and CN faint lines, respectively. The maps included in this appendix section were used when measuring the spatially averaged (CN bright)/(CN faint) intensity ratios (Table \ref{tab:ratios}).

%\section{(CN bright)/CO binned ratios}
%\label{append:cn_co_bins}
%Figure \ref{fig:cn_co_bins_all_gals} shows a compilation of the (CN bright)/CO intensity ratios versus CO luminosity in each galaxy in our sample. Section \ref{subsubsec:CO_binning_method} describes the binning method and Figure \ref{fig:cn_co_bins_example} (right) compares the binned intensity ratios for all galaxies.

%\begin{figure*}
%	\includegraphics[width=\textwidth]{CN_CO_bins_all_galaxies.pdf}
%    \caption{This figure shows the log-scale intensity (CN bright)/CO intensity ratios versus CO intensity (K km s$^{-1}$ units). The large round connected symbols represent the binned values in the CO intensity bins with $>3\sigma$ detections in the CN stacked spectra (see Section \ref{subsubsec:CO_binning_method}). The open triangles represent the CO intensity binned pixels which correspond to $<3\sigma$ detections in the stacked CN bright spectra. The detected bins are connected by solid lines, while the non-detected binned data points are connected by dotted lines. The uncertainties on the binned data points are from Equation \ref{eqn:uncertainty}. The faint blue scattered crosses represent the individual pixels detected in CN with $>3\sigma$, while the grey open diamonds are the pixels with $<3\sigma$.}
%    \label{fig:cn_co_bins_all_gals}
%\end{figure*}

\section{Figures including Arp 220}
\label{append:arp220}

\begin{figure*}
	\includegraphics[width=\textwidth]{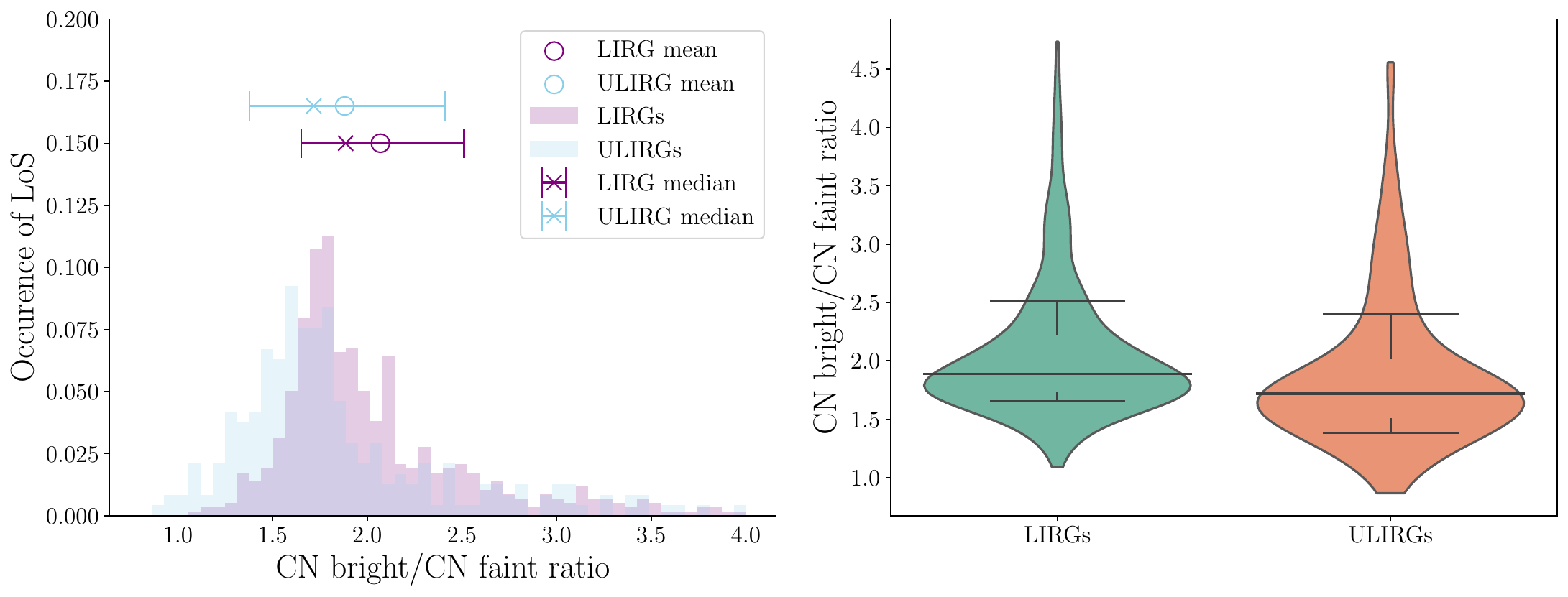}
    \caption{This figure shows the histograms of the (CN bright)/(CN faint) ratio after splitting the sample into pixels from ULIRGs and LIRGs. Data points from Arp 220 are included. We note that without the data from Arp 220, the ULIRG distribution only extends to a value of 2.5. \textit{Left:} The y-axis shows the number of pixels with the specific ratio value, and the x-axis gives the ratio in a linear scale. The blue and violet bars correspond to the ULIRG and LIRG data points, respectively. The open circles are the mean values of each distribution. The cross represents the median value, while the error bars extend to the 16\textsuperscript{th} and 84\textsuperscript{th} percentiles. A decreasing (CN bright)/(CN faint) ratio corresponds to an increasing optical depth. \textit{Right:} Violin plots of the (CN bright)/(CN faint) intensity ratio in ULIRGs (orange) compared to LIRGs (green). The black lines correspond to the 16\textsuperscript{th}, 50\textsuperscript{th}, and 84\textsuperscript{th} percentiles. The y-axis gives the (CN bright)/(CN faint) ratio in a linear scale.}
    \label{fig:cn_hist_violin_arp220}
\end{figure*}

\begin{figure*}
	\includegraphics[width=\textwidth]{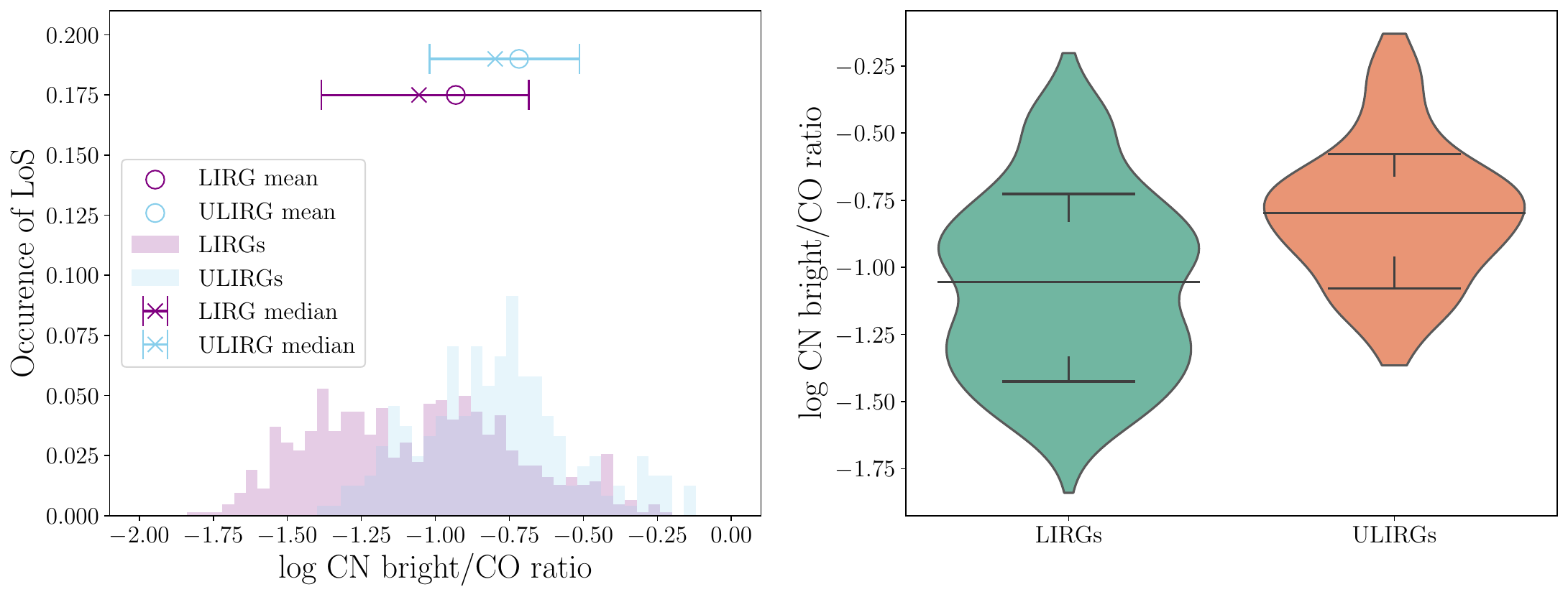}
    \caption{This figure shows the histograms of the (CN bright)/(CN faint) ratio after splitting the sample into pixels from ULIRGs and LIRGs. Data points from Arp 220 are included. We note that without the data from Arp 220, the spread in the ULIRG distribution is nearly a factor of 2 smaller. \textit{Left:} The y-axis shows the number of pixels with the specific ratio value, and the x-axis gives the (CN bright)/CO ratio on a logarithmic scale. The blue and violet bars correspond to the ULIRG and LIRG data points, respectively. The open circles are the mean values of each distribution. The cross represents the median value, while the error bars extend to the 16\textsuperscript{th} and 84\textsuperscript{th} percentiles. \textit{Right:} Violin plots of the (CN bright)/CO intensity ratio in ULIRGs (orange) compared to LIRGs (green). The black lines correspond to the 16\textsuperscript{th}, 50\textsuperscript{th}, and 84\textsuperscript{th} percentiles. The y-axis gives the (CN bright)/CO ratio on a logarithmic scale.}
    \label{fig:cn_co_hist_arp220}
\end{figure*}

\begin{figure}
	\includegraphics[width=\columnwidth]{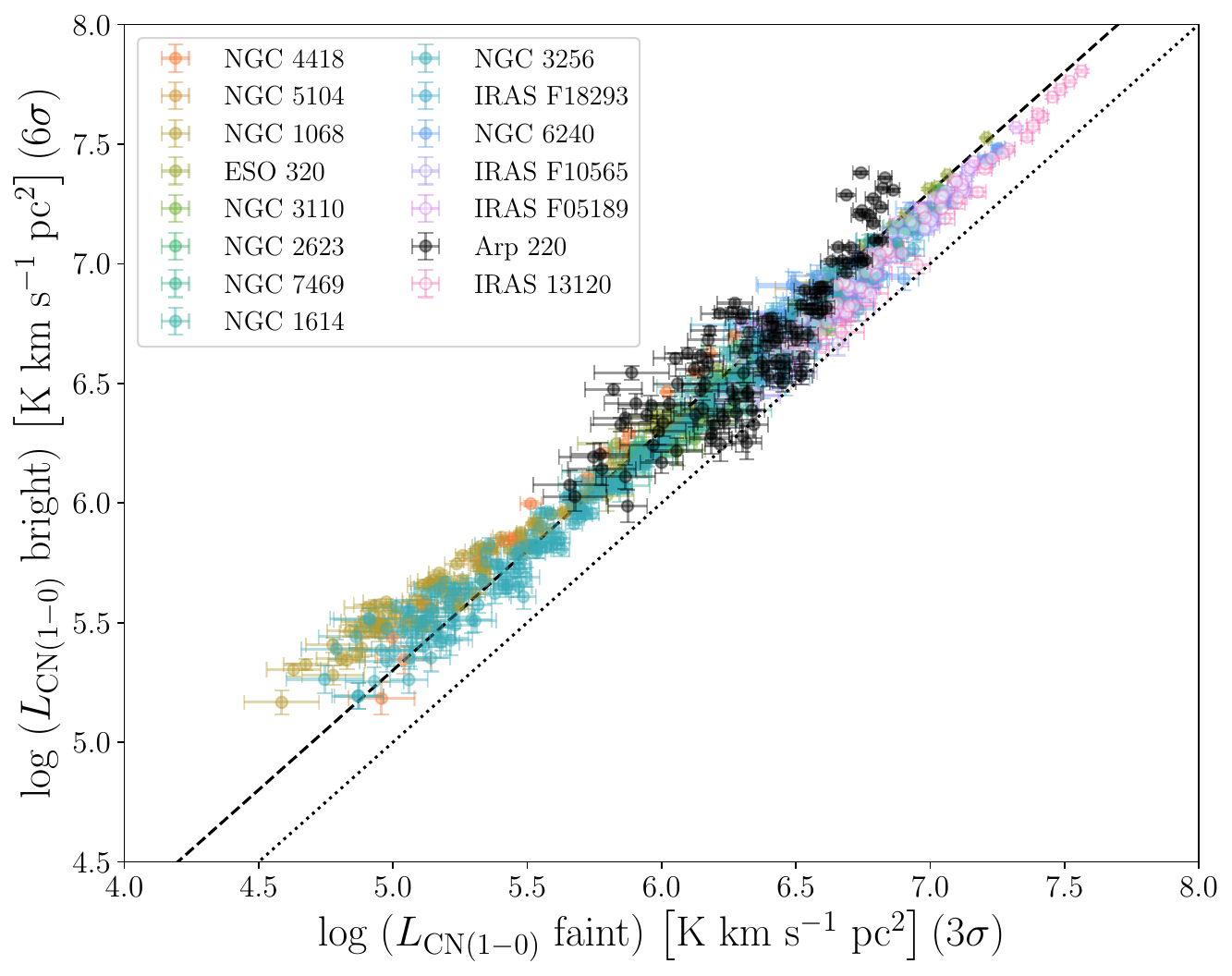}
    \caption{This figure compares the CN bright and CN faint lines on a pixel-by-pixel basis for 15 of the 16 galaxies in our sample. Uncertainties on individual pixels are calculated from Equation \ref{eqn:uncertainty}. Both pixel values and their uncertainties have been converted from fluxes to luminosities using Equation \ref{eqn:lum}. The pixels for the CN bright and CN faint lines are show with S/N cuts of $>6\sigma$ and $>3\sigma$, respectively. Pixels have been colourized by galaxy. Open circles are ULIRG galaxies. Closed circles are LIRG galaxies. The black-dotted and black-dashed lines represent 1:1 and 2:1 luminosity ratios, respectively. Data points from Arp 220 are highlighted in this figure as the black circles.}
    \label{fig:cn_optical_depth_arp220}
\end{figure}

This appendix shows the effect of including Arp 220 in Figures \ref{fig:cn_optical_depth}, \ref{fig:cn_hist_violin} and \ref{fig:cn_co_hist}. Individual pixels from Arp 220 were deemed untrustworthy due to variations in the noise throughout the cube, and as such we removed this galaxy from our results discussion when considering individual pixels within each galaxy. Figure \ref{fig:cn_hist_violin_arp220} includes the pixels from Arp 220 in the histogram of the (CN bright)/(CN faint) intensity ratio. The extension in the violin plot with combined ULIRG pixels to high ratio values is a result of the scatter of individual pixels within Arp 220. Figure \ref{fig:cn_co_hist_arp220} includes the pixels from Arp 220 in the histogram of the (CN bright)/CO intensity ratio. The violin plot with combined ULIRG pixels has a larger spread because of the scatter of individual pixels within Arp 220. Figure \ref{fig:cn_optical_depth_arp220} describes the (CN bright)/(CN faint) luminosity ratio with the Arp 220 pixels overplotted in black. This figure clearly demonstrates the scatter of individual pixels within Arp 220. 

\bsp	% typesetting comment
\label{lastpage}
\end{document}